\documentclass[msomc,NONBLINDREV]{informsArxiv}

\usepackage{color}
\usepackage{xcolor}
\usepackage{mathtools}
\usepackage{amssymb}

\usepackage{amsmath}
\usepackage{blkarray}
\usepackage{caption}
\usepackage{subcaption}
\usepackage{amsfonts}
\usepackage{dsfont}
\usepackage{mathrsfs}
\usepackage{float}
\usepackage{bbold}
\usepackage{graphicx}
\usepackage{listings}
\usepackage{framed}
\usepackage{multirow}
\usepackage{bbding}
\usepackage{fancyhdr}
\usepackage{booktabs}
\usepackage{capt-of}
\usepackage[inline]{enumitem}
\usepackage{lineno}
\usepackage{pbox}
\SetLabelAlign{parright}{\parbox[t]{\labelwidth}{\raggedleft#1}}
\usepackage{tabularx}
\def\BState{\State\hskip-\ALG@thistlm}

\usepackage{tabularx}
\usepackage{bbm}

\usepackage{algorithm}
\usepackage{algpseudocode}
\usepackage[toc,title,page]{appendix}

\usepackage{tikz}
\usetikzlibrary{shapes,arrows}
\usepackage{caption}
\usepackage{subcaption}

\definecolor{lightgray}{gray}{0.9}

\usepackage{natbib}
\bibpunct[, ]{(}{)}{,}{a}{}{,}
\def\bibfont{\small}

\TheoremsNumberedThrough
\ECRepeatTheorems

\EquationsNumberedThrough

\newif\ifpfapp  
\pfappfalse     

\begin{document}
	
\ARTICLEAUTHORS{
	\AUTHOR{Sanyukta Deshpande}
	\AFF{Industrial and Enterprise Systems Engineering, University of Illinois at Urbana-Champaign, Urbana, IL 61801, \EMAIL{spd4@illinois.edu}}
	\AUTHOR{Lavanya Marla}
	\AFF{Industrial and Enterprise Systems Engineering, University of Illinois at Urbana-Champaign, Urbana, IL 61801, \EMAIL{lavanyam@illinois.edu}}
	\AUTHOR{Alan Scheller-Wolf}
	\AFF{Tepper School of Business, Carnegie Mellon University, Pittsburgh, PA, 15213, \EMAIL{awolf@andrew.cmu.edu}}
	\AUTHOR{Siddharth Prakash Singh}
	\AFF{University College London School of Management, London, UK, E14 5AA, \EMAIL{siddharth.singh@ucl.ac.uk}}
	}
	
	\RUNTITLE{Hospital Capacity Management in a Pandemic}
	\RUNAUTHOR{Deshpande et al}
	\TITLE{Capacity Management in a Pandemic with  Endogenous Patient Choices and Flows}
	
	\ABSTRACT{

Motivated  by the experiences of a healthcare service provider during the Covid-19 pandemic, we aim to study the decisions of a provider that operates both an Emergency Department (ED) and a medical Clinic. Patients contact the provider through a phone call or may present directly at the ED; patients can be COVID (suspected/confirmed) or non-COVID, and have different severities.  Depending on severity, patients who contact the provider may be directed to the ED (to be seen in a few hours), be offered an appointment at the Clinic (to be seen in a few days), or be treated via phone or telemedicine, avoiding a visit to a facility. All patients  make joining decisions based on comparing their own risk perceptions versus their anticipated benefits: They then choose to enter a facility only if it is beneficial enough. Also, after initial contact, their severities may evolve, which may change their decision.  The hospital system’s objective is to allocate service capacity across facilities so as to minimize costs from patients deaths or defections.  We model the system using a fluid approximation over multiple periods, possibly with different demand profiles. While the feasible space for this problem can be extremely complex, it is amenable to decomposition into different sub-regions that can be analyzed individually; the global optimal solution can be reached via provably parsimonious computational methods over a single period and over multiple periods with different demand rates. Our analytical and computational results indicate that endogeneity results in non-trivial and non-intuitive capacity allocations that do not always prioritize high severity patients, for both single and multi-period settings.
	}
	
	
	
	\maketitle
	\vspace{-26 pt}

\section{Introduction}\label{sec:motivation}


This paper is inspired by a collaboration with hospitals in Champaign county, IL during the COVID-19 pandemic. The authors conducted multiple interviews with Carle Foundation Hospital and OSF hospitals to understand the best practices adopted, and gathered data to understand the system dynamics as they evolved during the pandemic.
As with many hospital systems across the US, ED and hospital overcrowding led our collaborating  hospitals to advise patients to first call before visiting hospitals, in order to avoid both overcrowding and contagion risk. Thus, many patients would call the hospital call centers for advice on making appointments, visiting EDs or quarantining at home. The call center would categorize callers into three severity categories. Our collaborating hospitals adopted a threshold policy -- patients (regardless of COVID-19 status) whose symptoms, as diagnosed through an interview,  were found to be severe were directed to EDs where they could be quickly treated (in a few hours). Those diagnosed as mild cases of COVID-19 were asked to remain at home and call in if their symptoms worsened, and those with moderate symptoms of COVID-19 were provided with appointments at quarantined sections of the hospital Clinics (COVID-Clinic) during the next few days. Call centers directed non-COVID patients that exhibited symptoms from other diseases to non-COVID sections of hospital Clinics (if moderate) or stay at home (if mild). Some  patients who called to the call center would follow these directions, whereas others would make different choices based on their perception of risk (potential contagion in ED queues) and urgency of service demanded. Additionally, some patients would directly walk in to EDs without calling into the call center first.
 
The partnering hospitals faced challenges of estimating the demand arrivals and staffing each facility, to meet their service goals. These challenges were exacerbated due to patients' severities, possibly evolving between the day they presented to the system and the day they were offered service, patients modifying their choices of entering a facility based on perceived wait times and risk, and the dynamically changing demands over the course of the pandemic. 

The key question we explore, thus, is the following: How should hospitals manage capacity at multiple facilities they operate, given that patients present with multiple levels of severity at varying rates over time? To answer the question the hospital must take endogeneity into account -- patient severities may evolve endogenously with the service rates provided; patients' choices of entering or balking from facilities depend on their expected delay and the associated risk of contagion in the facilities. Both of these phenomenon are endogenously determined by the service rates provided.

\subsection{Related Literature and Contributions}\label{sec:contrib_literature}
We now discuss literature that connects to our work either methodologically or in application.

\textbf{Capacity management during the COVID-19 pandemic:}
Since the beginning of the COVID-19 outbreak, the healthcare operations management community has generated a significant body of work on various pandemic-related issues, including optimizing capacity provision and deployment, managing healthcare demand, curbing ED congestion, ICU operations, lockdown policies, vaccine distribution, locating testing facilities and managing medical resource supply chains \citep{cacciapaglia2021multiwave, nicola2020evidence, fan2022distributionally, chang2021mobility}.
Several data driven approaches also study contextual disease spread in multiple countries. \cite{bertsimas2021predictions} propose data-driven approaches using epidemiological and clinical data to  alleviate the impact of COVID-19 using predictive and prescriptive analyses, with some of these policies implemented in various US cities. \cite{donelli2022disruptive} use the case study of a hospital in Italy and study behavioral, cognitive and contextual responses to derive insights on crisis management and resilience. \cite{bekker2022modeling} use queueing models for predicting hospital and bed occupancy in the Netherlands; \cite{ehmann2021operational} (Maryland, US),  \cite{zimmerman2022queuing} (Canada), and  \cite{melman2021balancing} find optimal policies for ventilator/resource allocation under scarcity in the UK. Our work contributes to this stream by studying capacity management in the setting of \emph{multiple} facilities managed by the same hospital system with \emph{demands changing} during the course of a pandemic, through the use of a \emph{general} fluid model framework.

\textbf{Capacity management in the broader healthcare context:}
\cite{mccaughey2015improving} present an extensive review of the literature on improving the capacity management (CM) in the Emergency Department (ED), primarily from 2000-2012. It has been established that ED crowding not only decreases hospital revenue but also substantially reduces the quality of service and access to healthcare, increasing risks \citep{bayley2005financial, pines2011international}. \cite{saghafian2015operations} extensively review the literature on optimizing patient flows to EDs.  \cite{dai2020om}  review recent literature on multiple aspects of healthcare operations management,  including patient behavior, incentives, policymaking, innovation, and financing.
 
Particularly relevant to our work, the problem of resource allocation has been studied under multiple settings: \cite{angalakudati2014business} study resource allocation under random emergencies.  \cite{ata2017organjet} make use of fluid and diffusion approximations to derive efficient solutions for delivering organ donations, considering geographical disparities.  \cite{bertsimas2013fairness} have  proposed data driven methods for efficient ways of kidney allocation under multiple fairness constraints.   
Close to our work, \cite{natarajan2014inventory} study the problem of inventory management under capacity constraints over multiple periods.  \cite{armony2018critical} study the problem of critical care capacity management by offering step down units (SDUs), designed in order to prioritize the most critical patients in ICUs. 
\cite{deglise2018capacity} solve the problem of capacity allocation while minimizing urgent patient delay. Recently,  the work by  \cite{hu2021optimal} takes a stochastic approach to address the problem of efficient resource allocation for  multiserver queues for two severity classes, where transitions between the classes are allowed,  and proposes a metric that indicates the most cost effective policy.

Methodologically closer to our work, \cite{akan2012broader} make use of fluid models to design an optimal liver allocation system with patients undergoing disease severity evolution while waiting in the queues. \cite{sharma2020reducing}  study the effects of patients' imperfect perfections about health on the non-urgent ED visits, using flow models. \cite{armony2009impact} also take the fluid model approach for setting up equilibrium conditions for delay announcements given to patients making an online contact. Taking a different direction, our work provides insights on multi period capacity allocations in a multi facility healthcare system, making use of stationary analysis in each period. 

\textbf{Management in healthcare through strategic queueing/ patient behavior:}
\cite{dong2019impact} provide empirical evidence that delay announcements indeed do affect patients' decisions on choosing service providers, along with their sensitivity to waiting, and that such information can lead to  increased coordination in the hospital system. \cite{batt2015waiting} provide insights on linking ED patient balking with observable queue lengths and waiting times.  \cite{xu2021interplay} review another crucial part of the healthcare system, where patients make decisions online appointments based on the information about clinic services. 
\cite{li2021next} work with three severity classes of patients and demonstrate optimal policies for patient prioritization under ED blocking. \cite{liu2018waiting} examine patient preferences and choice behavior while discussing various trade-offs relating to speed, quality and risk.  \cite{zacharias2017managing} work on the  problem of achieving resource utilization and shorter wait times under no-shows. In our work, we incorporate three severity classes of patients (for walk-ins to the ED and callers), and model transitions between the classes. We also model the patient choice of joining, balking and reneging incorporating multiple factors including severity, perceived risk, offered wait times and severity evolution.
	
The key contributions of this work are as follows.
\begin{enumerate}
    \item From a \emph{modeling standpoint}, we model a pandemic as a series of periods such that exogenous parameters are constant in each period. We first provide a fluid model framework to solve for optimal capacity management in a multi-facility healthcare system, taking endogenous patient behavior into account.
    \item From an \emph{analytical standpoint}, we characterize the solutions to the stationary one-period problem by decomposing its solution space into multiple cases with physically meaningful as well as computationally tractable solution structures; such that the global optimal can be found by a simple choice of the best among these cases. We further prove that there is a strong ordering of progression of optimal solutions as a function of the total available capacity.   
    \item We model the pandemic progression as a \emph{multi-period setting} with stationarity achieved in each period and current decisions endogenously affecting future demands through carryover demand. With additional structural analysis, we analytically show that the structures and ordering discovered for the one-period problem can be exploited to generate a parsimonious and provably efficient way of computing the multi-period optimal solution, without having to solve a complex dynamic programming problem.  
    \item We generate \emph{managerial insights} by implementing our solution approach on numerical experiments that mimic real-world scenarios for single period as well as multi-period pandemics. First, even in single period problems, capacity allocations are non-intuitive because of endogeneity due to patient choices and evolving severities. That is, the prioritizing highest severity patients should depend also on the relative number of medium severity patients and their evolution rates. Second, in the multi-period setting, optimal policies account for carryovers and future demands and thus make capacity allocations based on effective loads resulting in non-trivial allocations. Third, greedy solutions may be near-optimal only when the effect of carryovers to the next time period is minimal compared to exogenously arriving load. 
\end{enumerate}

\section{Modeling Framework}\label{sec:framework}

\pfapptrue  
\def\pfnotationtable{
\begin{table}[h]
\footnotesize	
\begin{tabular}{|lcc|}
\toprule\toprule
Parameters    & & \\
\midrule
 & Rate of inflow of patients             & $\lambda$                             \\ \hline
     & Severity levels $\in (0,1)$            & $s_{1}, s_{2}, s_{3}$             \\ 
\hline
    & Inflow rates according to severity     & $\lambda_{1}, \lambda_{2}, \lambda_{3}$                     \\
    & (1= high, 2=medium, 3=low) & \\
\hline
     & Fraction of patients calling in        & $p$                                               \\ \hline
     & Fraction of COVID patients             & $\mathds{P}(covid)$                                        \\ \hline
     & Risk parameter (at the ED)             & $r$                                               \\  \hline
     & ED queue length threshold for unit capacity                      & $\tau_1^E, \tau_2^E,\tau_3^E$                                                \\ \hline
     & Clinics queue length threshold for unit capacity                    & $ \tau^C, \tau^N$                                         \\ \hline
     & Severity evolution rates                        & $\delta_{ij}$ for $i,j \in {0,1,2,3,4}$                    \\ \hline
    & Rates of patients re-entering from the repositories  & $\beta_{1}, \beta_{2}, \beta_{3}$                          \\ \hline
    & Rates of patients leaving the repositories          & $\sigma_{1}, \sigma_{2}, \sigma_{3}$                         \\
    \hline 
    & Total available capacity                     & $\Gamma$                                               \\ 
    \midrule\midrule
Decision Variables &                             &                                   \\  \hline
    & Service rates at various facilities    & $\mu^{E}_{1}, \mu^{E}_{2,3}, \mu^C, \mu^N$         \\ 
    \midrule\midrule
    Performance Metrics &                             &                                   \\  \hline
    & Service efficiencies of various queues         & $\alpha^{E}_{1}, 
    \alpha^{E}_{2}, \alpha^{E}_{3},  \alpha^{C}, \alpha^{N}$         \\  \hline
    & Queue lengths at various facilities    & $Q^{E}_{1}, Q^{E}_{2,3}, Q^{C}, Q^{N}, H_{1}, H_{2}^C, H_{2}^N, H_{3}$ \\ 

    \bottomrule\bottomrule
\end{tabular}
\vspace{0.5cm}
\caption{\footnotesize	}
\label{tab: notation}
\end{table}
\pfappfalse  
}
We now describe the modeling of our hospital system. The basic hospital system consists of various facilities: (i) an Emergency Department (ED) that serves all patients but prioritizes patients with high severity, (ii) a quarantined COVID Clinic (Clinic), and (iii) a normal Clinic (NClinic) for non-COVID patients. These facilities work with service rates $(\mu^E,\mu^C, \mu^N)$ respectively, and the hospital determines these service rates by allocating its total service capacity $\Gamma$ (analogous to allocating medical staff) among these three facilities.  The primary focus of our paper is to provide guidance on how $\Gamma$ should be allocated among these three facilities. Note that because we model a pandemic, the demands and/or total service capacity available may vary across time periods, but we assume they remain constant within a time period. We describe this mathematically in Section \ref{sec: FluidModel}.


Next, we explain how we model patient flows into and across the various facilities in the hospital system. 
In any period, patients approach the hospital system with one of three severities -- high $(s_{1})$, medium ($s_{2}$) or low ($s_{3}$), with rates $\lambda_{1}, \lambda_{2}$ and $\lambda_{3}$ respectively (these rates will typically differ in different periods). Parameters $s_i, \forall i \in \{1, 2, 3\}$ represent the per unit time dis-utility of being sick for each severity level $i$.  Incoming patients also have a disease indicator which can take values C (COVID) or N (non-COVID). We assume that patients of all severity levels have disease indicator  C (respectively, N), with probability $\mathds{P}(covid))$ (respectively $1-\mathds{P}(covid))$). Incoming patients call in to the Callbank with probability $p$, or present directly to the Emergency department (ED), with probability $1-p$. Parameter $p$ is independent of severity and COVID status, although this can be generalized.  Patient severities are confirmed after consulting with the hospital and are then assumed to be accurate. The ED serves everyone; it has a High Priority queue (HPED) for severity $s_1$ patients and a Low Priority queue (LPED) for severity $s_2$ and $s_3$ patients, as indicated by our collaborating hospital. Callers to the Callbank with severity $s_1$ are directed to the ED. Severity $s_2$  callers are directed to the COVID Clinic or the non-Covid Clinic depending on their disease indicator. Severity $s_3$ callers are asked to stay at home and call the hospital again if they get worse. Note that the hospital may decide to keep one or more of these facilities non-functional (i.e., set $\mu^E$, $\mu^C$, or $\mu^N$ to zero); patients are aware of facilities' service rates. 

Although patients may be directed to specific facilities by the hospital, they are strategic, and decide whether they want to comply, switch to another facility in the system, or balk the hospital system altogether. These decisions are functions of the patients' severities and the congestion and contagion risk associated with each facility, which we assume are common knowledge. Specifically, the flowcharts in Figure \ref{fig:basic system} explicitly map the decision functions of each severity type.
\color{black}

We also incorporate the evolution of patients' severities over time. The waiting times at the ED are assumed to be of the scale of a few hours and thus, patients at the ED do not have their conditions evolve. However, both the Clinics work by appointments and have waiting times of the order of a few days, allowing patients to undergo severity evolution. Patients who balk the hospital system or asked to stay at home, ``keeping an eye on their symptoms", may also undergo severity evolution. To track such patients, we consider separate repositories $H_{1}$, $H_{2}^C$, $H_{2}^N$ and $H_{3}$, that hold patients who have balked with severity $s_1$, $s_2$ with COVID, $s_2$ without COVID, and $s_3$ respectively. These patients may re-enter the system  or ultimately leave the repositories due to recovery or death.  The complete hospital system, that includes both the hospital's facilities and the repositories, is given in Figure \ref{fig: completeHS} in the following subsection. 

Given this system with multiple types of flows, the hospital aims to optimally allocate total capacity $\Gamma$ as $(\mu^E, \mu^C, \mu^N): \mu^E + \mu^C + \mu^N \leq \Gamma$, so as to meet its service goals. In the next section, we formalize these service goals and illustrate all our modeling features within a fluid framework.

\tikzstyle{decision} = [diamond, draw, 
    text width=4.5em, text badly centered, node distance=3cm, inner sep=0pt]
\tikzstyle{block} = [rectangle, draw, 
    text width=5em, text centered, rounded corners, minimum height=4em]
\tikzstyle{line} = [draw, -latex']
\tikzstyle{cloud} = [draw, ellipse, text width=4em, text badly centered, node distance=3cm,  minimum height=2em]
 
\begin{figure}   
\captionsetup[subfigure]{font=footnotesize}
\subcaptionbox{Decisions for $s_1$ patients}[.5\textwidth]{
\begin{tikzpicture}[node distance = 2cm, auto, scale=0.7, every node/.style={transform shape}]
    \node [cloud] (s1walk) {$s_1$ Walk-ins};
    \node [cloud, xshift=15em] (s1call) {$s_1$ Callers};
    \node [decision, xshift=6.5em, yshift = -6 em] (isEDwaitOK) {Is ED Wait Time Acceptable?};
    \node [block, xshift=0em, yshift =-10em] (Home) {Home};
    \node [block, xshift=15em, yshift =-10em] (HPED) {HPED};
    \node [cloud, xshift=15em, yshift =-15em] (served) {Receive service};

    \path [line] (s1walk) -| (isEDwaitOK);
    \path [line]  (s1call) -| node [near start, above] {Directed} node [near start, below] {to HPED} (isEDwaitOK);
    \path [line]  (isEDwaitOK) -| node [near start, above] {No} (Home);
    \path [line]  (isEDwaitOK) -| node [near start, above] {Yes} (HPED);
    \path [line] (HPED) -- (served);
\end{tikzpicture}
}
\subcaptionbox{Decisions for $s_3$ patients}[.5\textwidth]{
\begin{tikzpicture}[node distance = 2cm, auto, scale=0.7, every node/.style={transform shape}]
    \node [cloud] (s3walk) {$s_3$ Walk-ins};
    \node [cloud, xshift=18em] (s3call) {$s_3$ Callers};
    \node [decision, xshift=6.5em, yshift = -6 em] (isEDwaitOK) {Is ED Wait Time Acceptable?};
    \node [block, xshift=0em, yshift =-10em] (Home) {Home};
    \node [block, xshift=18em, yshift =-10em] (LPED) {LPED};
    \node [cloud, xshift=18em, yshift =-15em] (served) {Receive service};

    \path [line] (s3walk) -| (isEDwaitOK);
    \path [line]  (s3call) -| node [near start, above] {Directed to Home} node [near start, below] {but check LPED} (isEDwaitOK);
    \path [line]  (isEDwaitOK) -| node [near start, above] {No} (Home);
    \path [line]  (isEDwaitOK) -| node [near start, above] {Yes} (LPED);
    \path [line] (LPED) -- (served);
\end{tikzpicture}
}
\par\medskip
\subcaptionbox{Decisions for $s_2$ walk-ins}[.5\textwidth]{
\begin{tikzpicture}[node distance = 2cm, auto, scale=0.7, every node/.style={transform shape}]
    \node [cloud] (s2walk) {$s_2$ Walk-ins};
    \node [decision, yshift = -7 em] (isEDwaitOK) {Is ED Wait Time Acceptable?};
    \node [block, xshift=-10em, yshift =-13em] (LPED) {LPED};
    \node [decision, xshift=10em, yshift = -13 em] (isClinicwaitOK) {Is Clinic/ NClinic Wait Time Acceptable?};
    \node [block, xshift=0em, yshift =-18em] (Home) {Home};
    \node [block, xshift=10em, yshift =-22em] (Clinics) {Clinic/ NClinic};
    \node [cloud, xshift=-10em, yshift =-22em] (served) {Receive service};
    
    \path [line] (s2walk) -- (isEDwaitOK);
    \path [line]  (isEDwaitOK) -| node [near start, above] {Yes} (LPED);
    \path [line]  (isEDwaitOK) -| node [near start, above] {No} (isClinicwaitOK);
    \path [line]  (isClinicwaitOK) -| node [near start, above] {No} (Home);
    \path [line]  (isClinicwaitOK) -- node [near start, right] {Yes} (Clinics);
    \path [line]  (LPED) -- (served);
    \path [line]  (Clinics) -- (served);
\end{tikzpicture}
}
\subcaptionbox{Decisions for $s_2$ callers}[.5\textwidth]{
\begin{tikzpicture}[node distance = 2cm, auto, scale=0.7, every node/.style={transform shape}]
    \node [cloud] (s2call) {$s_2$ Callers};
    \node [decision, yshift = -7.5 em] (isClinicwaitOK) {Is Clinic/ NClinic Wait Time Acceptable?};
    \node [block, xshift=-10em, yshift =-13em] (Clinics) {Clinic/ NClinic};
    \node [decision, xshift=10em, yshift = -13 em] (isEDwaitOK) {Is ED Wait Time Acceptable?};
    \node [block, xshift=0em, yshift =-18em] (Home) {Home};
    \node [block, xshift=10em, yshift =-22em] (LPED) {LPED};
    \node [cloud, xshift=-10em, yshift =-22em] (served) {Receive service};
    \path [line] (s2call) -- (isClinicwaitOK);
    \path [line]  (isClinicwaitOK) -| node [near start, above] {No} (isEDwaitOK);
    \path [line]  (isClinicwaitOK) -| node [near start, above] {Yes} (Clinics);
    \path [line]  (isEDwaitOK) -| node [near start, above] {No} (Home);
    \path [line]  (isEDwaitOK) -- node [near start, right] {Yes} (LPED);
    \path [line]  (LPED) -- (served);
    \path [line]  (Clinics) -- (served);
\end{tikzpicture}
}
\par \medskip
\caption{Flowcharts for patient decisions}
\label{fig:basic system}
\end{figure}
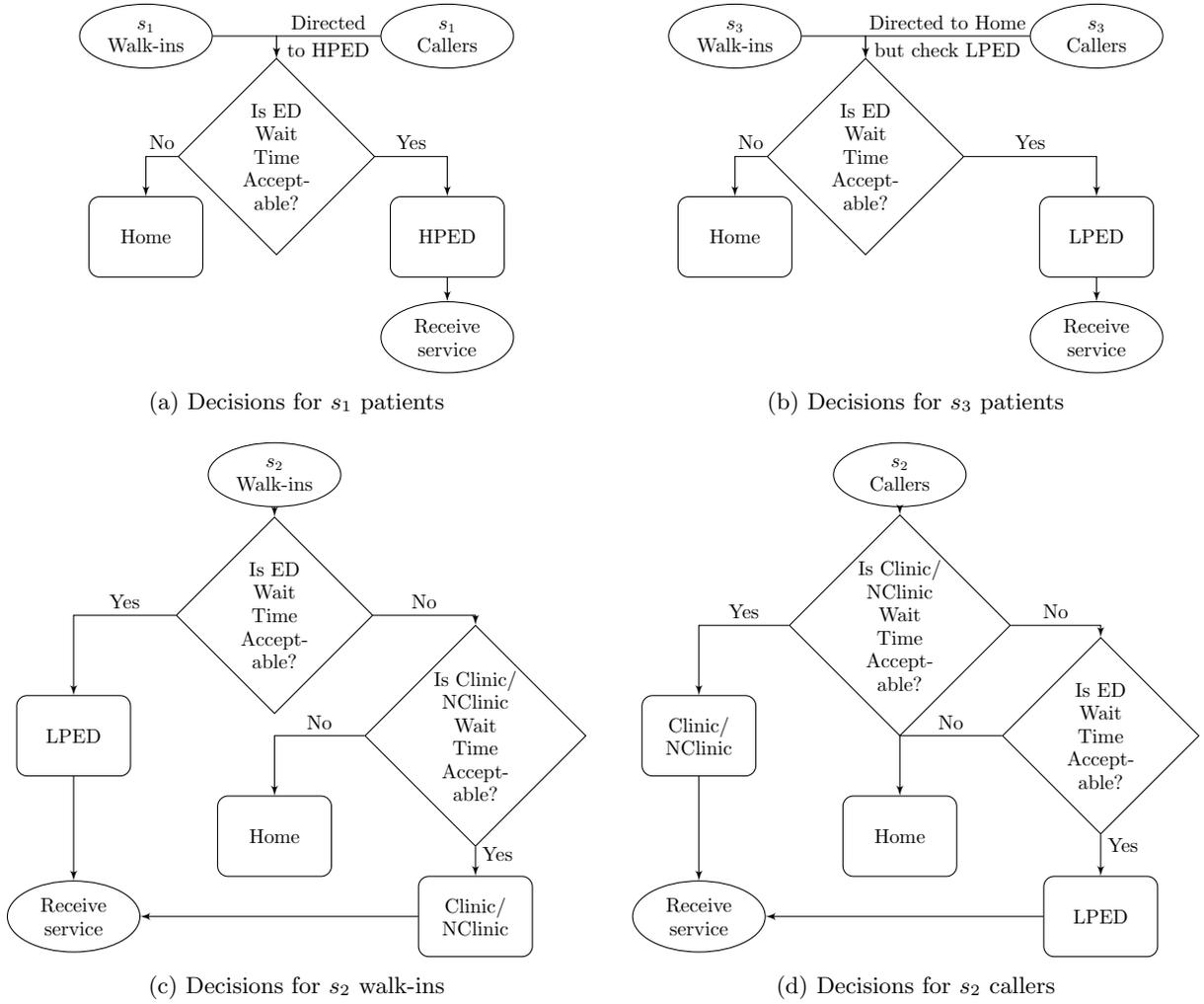

\subsection{Flow Modeling in the Hospital System}\label{sec:flow_chart}
We employ a fluid model in order to tractably capture the flows of patients into and between facilities, and across various severity levels. Since customers' decisions, as outlined in Figure \ref{fig:basic system}, depend on the wait times they encounter at each facility, we first explain how wait times are captured through our fluid model. We then use these to explain how we capture patients' decision functions. Subsequently, we discuss the modeling of patient severity evolution and finally, the repositories for customers who balk. In this work, we model and analyze the hospital system at stationarity, once equilibrium is reached. 

\textbf{Wait times:}
Each facility's wait time is a function of its queue length and its service rate. In our fluid setting, a facility working at rate $\mu$ that has $Q$ patients in the queue is associated with a deterministic wait time $\frac{Q}{\mu}$ for a newly joined patient, assuming FCFS service. We define $Q^E_1$, $Q^E_{2, 3}$, $Q^C$, and $Q^N$ as the queue lengths at the HPED, the LPED, the Clinic, and the NClinic respectively. Given that the HPED and LPED queues share service capacity $\mu^E$ and that severity $s_1$ patients have priority over severity $s_2$ and $s_3$ patients, there is service at the LPED queue only if the HPED queue is empty: The LPED offers service to severity $s_2$ and $s_3$ patients only if $Q^{E}_{1}=0$. The split of total ED capacity $\mu^E$ between capacity that serves the HPED queue (denoted as $\mu^E_1$) and the LPED queue (denoted as $\mu^E_{2,3}$) is endogenously determined by this prioritization. Accordingly, the wait times for HPED, LPED, Clinic and are $\frac{Q^{E}_{1}}{\mu^{E}_1}$, $\frac{Q^{E}_{2, 3}}{\mu^{E}_{2, 3}}$, $\frac{Q^{C}}{\mu^{C}}$, and  $\frac{Q^{N}}{\mu^{N}}$ respectively. We thus obtain the following general formula for all facilities:
\begin{align}
    \text{Wait time at facility } f \text{ for a patient with severity }s_i = \frac{Q^{f}_i}{\mu^{f}_i} \label{eq: wait time}.
\end{align}
For brevity, we drop the subscript $i$ for the Clinic and NClinic facilities because they only serve severity $s_2$ patients.

\textbf{Patients' Joining Decisions:}
Given the above wait time specification, we now utilize these wait times to explain how patients make decisions according to Figure \ref{fig:basic system}. Patients' decisions on whether to join or balk from a considered facility depend on the wait time they will experience, their severity and the contagion risk.

Let the patients receive a reward $B$ from receiving service, and let $r$ represent the contagion risk associated with waiting in the ED: We assume that patients waiting at the ED incur waiting costs at a faster rate. Similarly, we assume that more severely ill patients incur waiting costs at a faster rate. We also assume that waiting for a clinic appointment incurs costs at a lower rate, because patients can wait at home. To account for this, we multiply wait times at the clinic by an \textit{indifference factor} $\phi\leq 1$. Putting these together, we assume that the waiting cost for a severity $s_i$ patient waiting at the ED is $r+s_i$ per unit time, and the waiting cost for a severity $s_i$ patient waiting for a clinic appointment is $\phi s_i$ per unit time.

Patients who consider joining the ED do so if and only if their resulting utility is positive:
\begin{equation*}
    B - (r+s_i)\times \text{ wait time (ED)} \geq 0
\end{equation*}

For ease of exposition, we define parameter $\tau^{E}_i \equiv \frac{B}{r+s_i}$ so that patients with severity $i \in \{s_1,s_2,s_3\}$ join the ED if and only if:
\begin{align}
    \text{ wait time (ED) } & \leq \tau_i^{E},
    \label{eq: EDdecision}
\end{align}
with $\tau_1^{E} \leq \tau_2^{E} \leq \tau_3^{E} $. 

Similarly, severity $s_2$ patients who consider joining the Clinic or NClinic (recall that these facilities only serve patients with severity $s_{2}$) do so if and only if their resulting utility is positive:
\begin{equation*}
    B - \phi s_2\times \text{ wait time (Clinic/NClinic)} \geq 0.
\end{equation*}

Define parameter $\tau^{C}_2=\tau^{N}_2 \equiv \frac{B}{\phi s_2}$ for the Clinic and NClinic; a severity $s_2$ patient who considers joining the Clinic/NClinic queue does so if and only if:
\begin{align}
 \text{ wait time (Clinic/NClinic) } & \leq \tau^{C}_2=\tau^{N}_2.
   \label{eq: cdecision}
\end{align}
Using equations \eqref{eq: wait time}-\eqref{eq: cdecision}, we can find bounds on queue lengths that make queues attractive for patients of specific severities. Accordingly, a patient with severity $s_{i}$ will join facility $f$ if and only if the queue length $Q^{f}_{i}$ satisfies:
\begin{align}
 Q^{f}_{i} \leq \mu^{f}_i \tau_i^{f}. \label{eq: joiningrule}
\end{align}
Again, we drop the subscript $i=2$ for the clinic queues, i.e., for $f = N$ or $f=C$, and $\tau^C_2$ and $\tau^N_2$ are understood to be $\tau^{C}$ and $\tau^{N}$ respectively.

\textbf{Severity Evolution:}
As described earlier, we consider dynamic evolution of patient severity levels to model progression of the disease and recovery. In particular, we model severity evolution for patients waiting to be served in the Clinic and the NClinic; these patients will attempt to join the ED (and will do so or balk in accordance with the decision rule in equation \eqref{eq: joiningrule}) if their severity worsens to $s_1$, or leave the system (and monitor at home) if their severity level improves to $s_3$. We assume that such patients' appointments are left empty, i.e., that they cannot be filled at short notice. Once the patients are treated, we denote their severity level by $s_4$, and they leave the system permanently. Additionally, we consider that patients who are in the repositories may also undergo severity evolution, either becoming better or worse.  These patients, upon evolution, move to the repository corresponding to their updated severity level, from which they may seek to reenter the system.  Patients in the repositories who die are considered as reaching severity level $s_0$.

We define the rate of evolution $\delta_{ij}$ as the rate with which a patient with severity level $s_i$ reaches another level $s_j$. The associated transition matrix is given in \eqref{eq: transition}.

 \begin{equation}
\begin{blockarray}{cccccc}
s_0 & s_1 & s_2 & s_3 & s_4 \\
\begin{block}{(ccccc)c}
 1 & 0 & 0 & 0 & 0 & s_0 \\ 
 \delta_{10}& 1-\delta_{10}-\delta_{12}  & \delta_{12} & 0 &0 & s_1 \\ 
 0 & \delta_{21} & 1-\delta_{21}-\delta_{23} & \delta_{23} & 0 & s_2\\ 
0  & 0 & \delta_{32} & 1-\delta_{32}-\delta_{34} & \delta_{34} & s_3\\ 
 0 & 0 & 0  & 0 & 1 & s_4 \\
\end{block}
\end{blockarray} \label{eq: transition}
\end{equation}

The rates given in \eqref{eq: transition} can be understood as the instantaneous rates of transition into a different severity level, representing the per unit flow of the expected amount of fluid evolving at any instant. 

\textbf{Repositories:}
Patients who do not join any facility based on their decision functions are added to the repositories, which are considered as being outside the ``basic hospital system.'' In particular, severity $s_i$ patients who leave from the system are added to repository $H_i$ (severity $s_2$ patients are added to repository $H^C_2$ or $H^N_2$ depending on their disease indicator). As these patients are not being served in any facility, they may also transition to neighbouring severity states with rates given in \eqref{eq: transition}. 
Additionally, from each of the repositories $H_i$, we consider a  flow of $H_{i} \beta_{i}$ back to the basic hospital system: this flow represents patients who re-attempt to receive service. Similarly, we consider a flow $H_i \sigma_i$ leaving permanently, either to receive service at another hospital system, or to ignore getting treated altogether. We assume that the hospital aims to curb such flows, because they represent a loss of revenue to the hospital and/or poor health outcomes. 

Given these flows, in steady state, the total incoming rate of severity $i$ patients to the hospital is $\lambda_i + H_{i}\beta_i$. Similar to `new' patients, these patients also approach ED and the Callbank with proportions $p$ and $1-p$.

\begin{figure}
\centering
\hspace{-3cm}
\begin{subfigure}[b]{0.3\textwidth}
         \centering        \includegraphics[width=8cm]{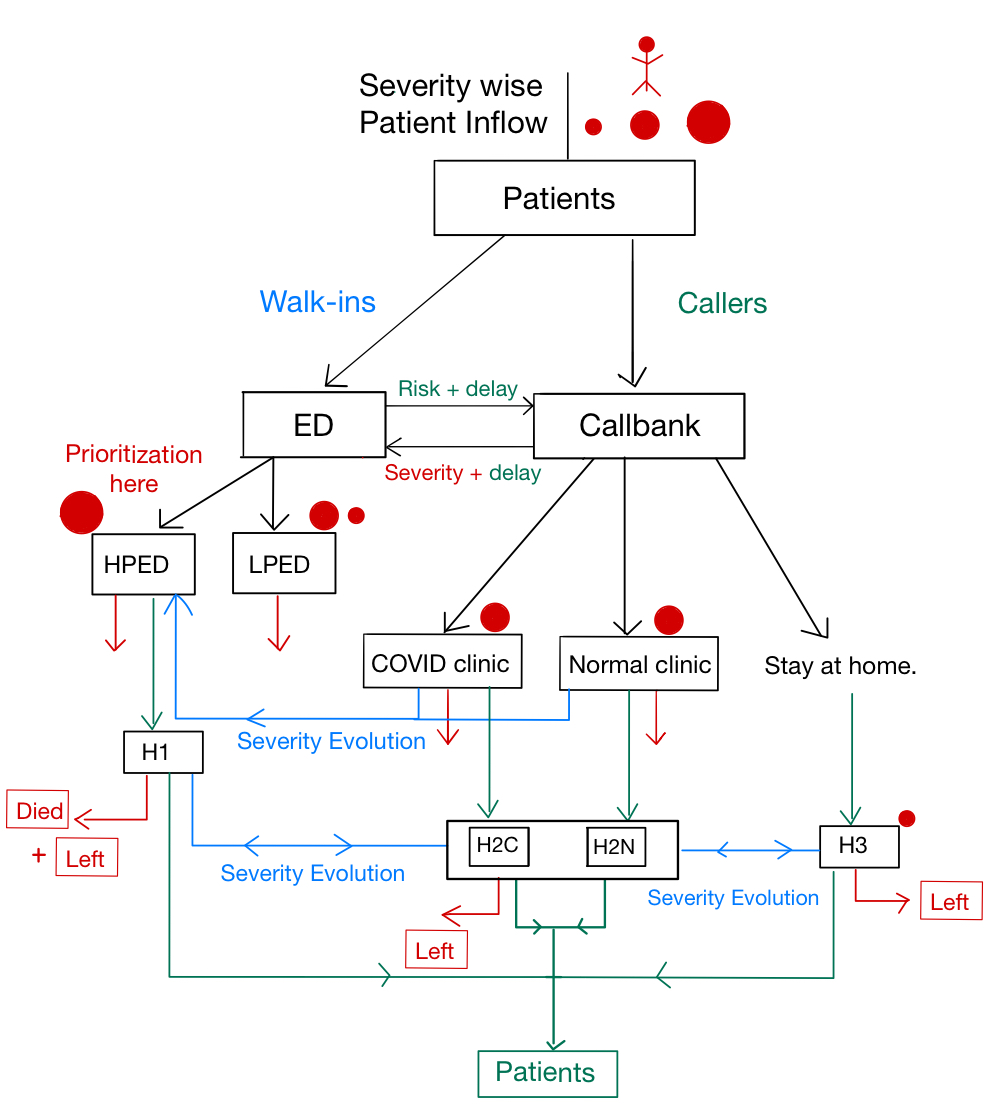}
         \caption{Complete hospital system}
         \label{fig: completeHS}
     \end{subfigure}
     \hspace{3cm}
     \begin{subfigure}[b]{0.3\textwidth}
         \centering
  \includegraphics[width=8cm]{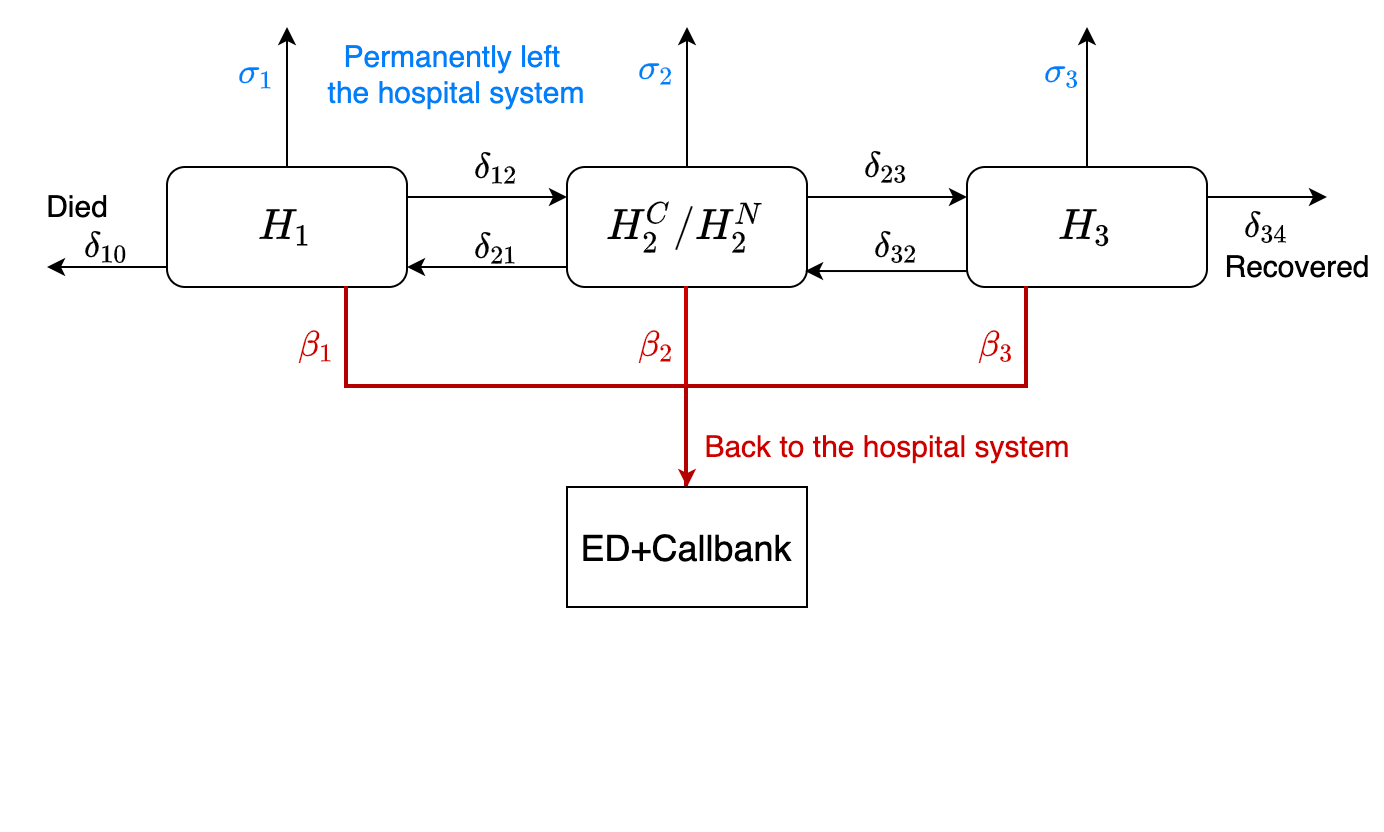}
       \caption{Flows across repositories}
         \label{fig: repoflows}
     \end{subfigure}
\caption{Patient flows in the hospital system}     
\label{fig: flow details1}
\end{figure}

By explicitly modeling the repositories,  we are able to keep track of patients who can  return to the treatment facilities, accounting for their possible severity evolution. This also allows us to track the possible evolution of disease, enabling us to track how  public health changes depending on the hospital's decisions. Figure \ref{fig: repoflows} presents a schematic representation of these flows.
 
This completes the description of our modeling framework. As we see from Figure \ref{fig: completeHS}, the facilities and repositories are connected to each other through multiple flows. These flows are governed by patients' decisions and their severity evolution, which in turn are functions of the chosen service rates. Naturally, the capacity allocation decision must account for patients' response to the hospital's capacity allocation and any corresponding feedback flows. As we see in the flowchart, the capacity at one facility can influence all other facilities. For instance, if the Clinic works with very high efficiency such that every patient entering gets an immediate appointment, there are secondary effects on HPED and LPED as well -- no patient waits long enough to undergo severity evolution and reach HPED, and callers with severity $s_{2}$ do not switch from the Clinic to LPED. Thus, as the service capacity at the Clinic changes, the considerations at the ED change as well.

\section{Fluid Model Framework}\label{sec: FluidModel}

In this section, we formally derive the relations between the fluid capacities at all facilities, their corresponding fluid queue lengths and various feedback flows in the hospital system.

During a pandemic, we expect that the incoming demand will fluctuate over time. For ease of modeling, we consider the time horizon to be divided into multiple time periods with demand being constant within each period. That is, we consider a period to have constant parameters (including $\lambda_i, \forall i \in \{1, 2, 3\}$,  $\mathds{P}(covid)$, $p$, and severity evolution rates).

We now formulate the single-period problem assuming period lengths are sufficient for a fluid equilibrium to be reached within each time period, using the fluid modeling framework introduced in Section~\ref{sec:framework}. In Section~\ref{sec:single_period_solutions} we will analyze our fluid model to arrive at an optimal allocation of capacity $\Gamma$ among the ED and the Clinics. 


\subsection{Fluid Formulations}\label{sec:formulations}
At equilibrium, fluid enters each facility at a constant rate; a part of the fluid may spill out. More precisely, if the rate of incoming patients $\lambda^f$ to facility $f$ is higher than its service rate $\mu^f$, flow spills out at rate $\lambda^f -\mu^f$.  Recall equations \eqref{eq: wait time}, \eqref{eq: EDdecision}, and \eqref{eq: cdecision} which imply that  patients with severity level  $s_i$ join facility $f$ only if the queue length $Q^f_i$ is smaller than a constant, say $c^f_i = \mu^f_i \tau^f_i$. Thus, a fraction of patients entering the system, if they see a queue of length $c^f_i$, see that the queue is congested, and they do not join it. But a fraction sees `free' space in the queue because of the constant service rate, and they join the queue. In the fluid sense, we use $\alpha^f_i =  E[\mathds{I}(Q^f_i<c^f_i)]$ to  denote the efficiency of the queue; $\alpha^f_i$ can be interpreted as the proportion of time queue $Q^f_i$ is not congested {from the perspective of a severity $i$ patient. Note that a queue could be congested for one type of severity and free for another. This may specifically be seen in the case of the ED which serves patients of all severities. We discuss this case in detail shortly. 
 
Thus, a queue at each facility could be: non-functional (i.e. the facility could be closed), congested with a non-zero length, or working with full efficiency (i.e. having no queue or spillage)\footnote{Technically, could have a positive, stable queue with $\lambda=\mu$}. We define $\alpha^{E}_{1}$ for (ED, $s_{1}$), $\alpha^{E}_{2}$ for (ED, $s_{2}$), $\alpha^{E}_{3}$ for (ED, $s_{3}$), $\alpha^{C}$ for (Clinic, $s_{2}$), and $\alpha^{N} $ for (NClinic, $s_{2}$)
 as the efficiencies of the facilities. We have  $\alpha_i^f\in [0,1]$ with 0 implying non-functionality and 1 implying full efficiency. If $\alpha_i^f \in (0,1)$, then the queue is congested with a non-zero efficiency, and serves only part of the incoming flow. In this case, the stationary queue length is fixed at the threshold $c^f_i$.

To find these efficiencies, we solve the fluid balance equations 
\[\frac{dQ}{dt} =0 \]  for $Q \in \{Q^{E}_{1}, \ Q^{E}_{2,3} , Q^{C}, \ Q^{N}, H_{1}, H_{2}^C, H_{2}^N, H_{3}\}$.  These equations establish the equilibrium at all queues.  We now illustrate how to derive these fluid balance equations for the various facilities. We describe the procedure for the HPED and LPED in detail and defer the rest of the discussion about the Clinics and the repositories to Appendix~\ref{appndx: ForFluidFormulations}.

\textbf{HPED:} We derive the fluid balance equation for HPED by adding the flow of incoming patients who join the queue and receive service and subtracting the outgoing flow of HPED customers who receive service at the ED. If  $P(s=s_i, U^{f}>0)$ is  the proportion of severity $i$ patients who join the queue at facility $f$ (i.e. who find their utility of joining to be positive), we can define $\alpha^{E}_{1}$ as \footnote[2]{All entering flows with severity $s_1$ have the same $\alpha_1^E$}
 \begin{equation*}
      P(s = s_{1}, U^{ED} >0) = E[\mathds{I}(\{Q^{E}_{1} <\mu^{E}_{1} \tau_1^{E} \})] =  \alpha^{E}_{1}
    \end{equation*}
\noindent Accordingly, we have:
\begin{align}
    \frac{dQ^{E}_{1}}{dt} = [\lambda_{1} + \beta_{1}H_{1} + (Q^{C}+Q^{N}) \delta_{21}]\alpha^{E}_{1} -  \mu^{E}_1 = 0 \label{eq: flowHP}
\end{align}
In the equation above, the flow that is served is the total incoming $s_1$ flow arriving directly to ED and the callbank, the feedback flow from $H_{1}$ (including any $H_2$ flow transitioning to $H_1$), and the flow from patients undergoing severity evolution from the Clinic and NClinic queues, multiplied by the fraction of patients who actually join the ED queue.

Furthermore, we have: 
\begin{align}
    (1-\alpha^{E}_{1}) \left( Q^{E}_{1} - \mu^{E}_{1} \tau_1^{E}\right) = 0,
    \label{eq: IndicatorHP}
\end{align}
to impose the condition that either the queue is congested with a fixed queue length $\mu^{E}_{1} \tau_1^{E}$ or it is working with full efficiency i.e. $\alpha^{E}_{1} = 1$. Note that $\alpha^{E}_{1} = 1$  indicates that  $Q^{E}_{1}$ is a free variable, but since this implies that we have enough capacity to serve everyone, we may fix $Q^{E}_{1} = 0< \mu^{E}_{1} \tau_1^{E}$.

\textbf{LPED:}
In order to formulate the fluid balance equations for the LPED, we first discuss the prioritization effects: LPED patients cannot enter and get served if the HPED queue is congested, i.e., if some part of the entering $s_1$ fluid spills over. The HPED is congested if the service rate $\mu^E$ is less than the incoming flow to the HPED.  Conversely, if $\mu^E$ is greater than or equal to the incoming rate, then prioritization ensures that the fluid level in HPED is zero (high priority patients seeking service at the ED are all served), and the LPED is served with any remaining capacity. In this case, the HPED fluid will not spill over as its queue has full efficiency. The LPED queue spills over if the remaining capacity is less than the incoming rate of $s_2$ and $s_3$ patients to the LPED; accordingly, the LPED may be congested with some efficiency. As a result, the possible configurations of the stationary queue lengths at the ED are: (i) both ED queues are empty; (ii) the HPED queue is empty and the LPED queue is working with some partial efficiency; and (iii) the HPED queue is congested and the LPED is not serving (or equivalently, serving at zero efficiency). 

The flow balance equation for the LPED queue includes incoming severity $s_2$ and $s_3$ flows to the ED, the feedback flows from $H_{2}^C$, $H_{2}^N$, and $H_{3}$ and incoming flows of unsatisfied patients from the Clinic and the NClinic. 
Now, we compute the proportions of $s_{2}$ and $s_{3}$ patients coming to the LPED. This expression uses two indicator functions: the first relates to the current queue length at the LPED and the second indicates if the HPED is congested. 
\begin{align*}
         P(s = s_{2}, U^{ED} > 0) & =
          E[\mathds{I}( Q^{E}_{2,3} \leq \mu^{E}_{2,3} \tau_2^{E})] \mathds{I}(Q^{E}_{1}< \mu^{E}_1 \tau_1^{E}) =  \alpha^{E}_{2} \mathds{I}(Q^{E}_{1}< \mu^{E}_1 \tau_1^{E} )\\
         P(s = s_{3}, U^{ED} > 0) & =
         E[\mathds{I}( Q^{E}_{2,3} \leq \mu^{E}_{2,3} \tau_3^{E})] \mathds{I}(Q^{E}_{1}< \mu^{E}_1 \tau_1^{E})   = \alpha^{E}_{3} \mathds{I}(Q^{E}_{1}< \mu^{E}_1 \tau_1^{E} ) \end{align*}
    
Therefore, the fraction of $s_2$ patients unhappy with the Clinic and entering the LPED is: 
    \begin{align*}
      P(s = s_{2}, U^{ED} >0, \ U^{Clinic } <0)  & =  E[\mathds{I}( Q^{E}_{2,3} \leq \mu^{E}_{2,3} \tau_2^{E})] E[\mathds{I}(Q^{C}\geq \mu^C \tau^{C} )] \times \mathds{I}(Q^{E}_{1}< \mu^{E}_1 \tau_1^{E} ) \notag \\ 
     & =  \alpha^{E}_{2} (1-\alpha^{C}) \mathds{I}(Q^{E}_{1}< \mu^{E}_1 \tau_1^{E} )
    \end{align*}  
    The flow from the NClinic and for $s_3$ patients is computed similarly. In addition, the indicator  condition $\mathds{I}(Q^{E}_{1}< \mu^{E}_1\tau_1^E$) can be represented as 
\begin{align}
 Q^{E}_{1} \times \mu^{E}_{2,3} = 0 \label{eq: IndicatorLP1} 
\end{align}
This equation suffices since (i) if $Q_1^E = \mu^{E}_1 \tau_1^{E}$, then $\mu^{E}_{2,3} = 0$, and (ii) if $Q_1^E < \mu^{E}_1 \tau_1^{E}$, then the queue is non-congested at stationarity, and  $Q_{1}^{E} = 0$. Putting all this together, we can simplify the flow equation to  \eqref{eq:flowLP}:
\begin{align}
    \frac{dQ^{E}_{2,3}}{dt} & = (1-p)[\lambda_{2}+ \beta_{2}H_{2}^C+ \beta_{2}H_{2}^N] \alpha^{E}_{2} + [\lambda_{3} + \beta_{3}H_{3}] \alpha^{E}_{3} +   [p\lambda_{2} \mathds{P}(covid) + p\beta_{2}H_{2}^C ]  (1-\alpha^{C})\alpha^{E}_{2} \notag\\
     & + \   [p\lambda_{2}(1-\mathds{P}(covid))+ p\beta_{2}H_{2}^N]   (1-\alpha^{N})\alpha^{E}_{2}  - \mu^{E}_{2,3}  =0
     \label{eq:flowLP} 
\end{align}

In addition, we get two separate indicator function equations for $s_{2}$ and $s_{3}$ patients in the LPED. 
\begin{align}
 (1-\alpha^{E}_{2})\left(Q^{E}_{2,3}- \mu^{E}_{2,3} \tau_2^{E} \right) = 0 \label{eq: IndicatorLP2} \\
   (1-\alpha^{E}_{3})\left( Q^{E}_{2,3} - \mu^{E}_{2,3} \tau_3^{E}\right)= 0 
  \label{eq: IndicatorLP3}
\end{align}

The queue length threshold which makes a queue congested is given by the last term of these equations. 
The maximum acceptable queue length for $s_{2}$ patients is less than that for $s_{3}$, as $s_3$ arrivals are less patient. Thus, if the queue length for the LPED is $\mu^{E}_{2,3} \tau_2^{E}$, then $s_{3}$ patients will always join. If the incoming rate of $s_{3}$ patients exceeds the available capacity, then the queue length will exceed  the maximum acceptable queue length for $s_{2}$ patients. In this case, only $s_{3}$ patients join and the equilibrium length is $\mu^{E}_{2,3} \tau_3^{E}$. To capture this, we add the following constraint:
\begin{align}
    \alpha_{2}^{E} < \alpha_{3}^{E} \label{eq: LPEDextracases}
\end{align}
Accordingly, all these equations ensure that exactly one of these four cases occurs with regards to the LPED: (i) The LPED is working with full efficiency for severities $s_2$ and $s_3$; (ii) the LPED is congested only for $s_2$; (iii) the LPED is congested for $s_3$ and is effectively closed for $s_2$; and (iv) the LPED is not serving any patients, as the HPED is congested.

\pfapptrue  
\def\pfClinics{

\textbf{Clinic and NClinic.}
Similar to the earlier fluid equations, we set up equations for the Clinic and NClinic, where we add flows into these clinics, feedback flows from $H_{2}^C$ and $H_{2}^N$, flows of unsatisfied patients from the ED, and finally subtract the flow of served patients, as in Fig \ref{fig: flow details1}. The equations below represent the equations for Clinic, with the first term capturing the inbound flows, the second terms capturing the feedback flows and the flow of unsatisfied patients from the ED, and the third term capturing the served patients.   
\begin{align*} \frac{dQ^{C}}{dt} =      p[\lambda_{2}+  \beta_{2}H_{2}^C]  \times \mathds{P}(covid) \times P( s =  s_{2}, \ U^{Clinic}>0)    \notag \\ +(1-p)[\lambda_{2}+ \beta_{2}H_{2}^C]  \times \mathds{P}(covid) \times P( s= s_{2}, \ U^{Clinic}>0, U^{ED}<0)  - \mu^C 
  \end{align*}
where
\begin{align*} P( s = s_{2}, \ U^{Clinic}>0) & =  E[\mathds{I}(Q^{C}\leq \mu^C \tau^{C} ) ] =  \alpha^{C} \\
  P( s= s_{2}, \ U^{Clinic}>0, U^{ED}<0)  \notag 
 & =  E[\mathds{I}( Q^{E}_{2,3} \geq \mu^{E}_{2,3} \tau_2^{E} )] \times E[\mathds{I}(Q^{C}\leq \mu^C \tau^{C} ])] \notag \\
  & = (1- \alpha^{E}_{2}) \alpha^{C}
  \end{align*} 
We can write similar flow equations for the NClinic. Then, we simplify both Clinic and NClinic equations to write
\begin{align}
    \frac{dQ^{C}}{dt} & = \alpha^{C}[   p\lambda_{2}\mathds{P}(covid) + p\beta_{2}H_{2}^C +  \notag\\
    &  [(1-p) \lambda_{2} \mathds{P}(covid)+ (1-p) \beta_{2}H_{2}^C] (1-\alpha^{E}_{2}) ] - \mu^C = 0 \label{eq: flowC}\\
    \frac{dQ^{N}}{dt} & = \alpha^{N}[ p  \lambda_{2}(1-\mathds{P}(covid)) + p\beta_{2}H_{2}^N + \notag\\
    &  [(1-p) \lambda_{2}(1-\mathds{P}(covid))+ (1-p)\beta_{2}H_{2}^N] (1-\alpha^{E}_{2}) ] - \mu^N = 0 \label{eq: flowN}
    \end{align}
and the indicator functions reduce to:
\begin{align}
  (1-\alpha^{C}) \left(Q^{C}- \mu^C \tau^{C}\right)=0  \label{eq: IndicatorC} \\
 (1-\alpha^{N})\left(Q^{N}- \mu^N \tau^{N}\right)=0 
 \label{eq: IndicatorN}
 \end{align}
 These equations indicate that a queue can either be congested with a fixed queue length $(Q^{f} = \mu^f \tau^{f}) $ or work with full efficiency i.e. $\alpha^{f} = 1$ for $f \in \{C, N\}$.

\textbf{Repositories.}
Finally, the number of patients in repositories $H_1, H_2^C, H_2^N, H_3$ follows equations \eqref{eq: flowH1}-\eqref{eq: flowH3} (see in conjunction with Figure \ref{fig: flow details1}).
\begin{align}
\frac{dH_{1}}{dt} & = [\lambda_{1} + H_{1}\beta_{1} + (Q^{C}+Q^{N})\delta_{21}] (1-\alpha^{E}_{1}) \notag\\
    & + H_{2}^C\delta_{21}+H_{2}^N \delta_{21} -H_{1}[  \sigma_{1}  +\delta_{12} + \delta_{10} + \beta_{1}]= 0  \label{eq: flowH1}\\
  \frac{dH_{2}^C}{dt} & = [\mathds{P}(covid)\lambda_{2} + H_{2}^C\beta_{2}  ](1-\alpha^{E}_{2})(1-\alpha^{C})\notag\\
    &  +(H_{3}\delta_{32}+ H_{1}\delta_{12})\mathds{P}(covid)- H_{2}^C[\sigma_{2} +\delta_{23} + \delta_{21} +\beta_{2}] = 0 \label{eq: flowH2C} \\
    \frac{dH_{2}^N}{dt} & =  [(1-\mathds{P}(covid))\lambda_{2} + H_{2}^N \beta_{2} ](1-\alpha^{E}_{2})(1-\alpha^{N}) \notag\\
    &  +(H_{3}\delta_{32}+ H_{1}\delta_{12}) (1-\mathds{P}(covid))- H_{2}^N[\sigma_{2} +\delta_{23} + \delta_{21} +\beta_{2}] = 0 \label{eq: flowH2N} \\
   \frac{dH_{3}}{dt} & = [\lambda_{3}+H_{3}\beta_{3}] (1-\alpha^{E}_{3}) + \delta_{23}H_{2}^C+ \delta_{23}H_{2}^N - H_{3} [\delta_{32}+\delta_{34} +  \sigma_{3}+\beta_{3}] = 0 \label{eq: flowH3}
\end{align}
 Equation \eqref{eq: flowH1} captures the flow balance equation for the repository of unsatisfied $s_1$ patients. The incoming flows include the unsatisfied HPED flow and the severity evolution flows from $H_{2}^C$ and $H_{2}^N$. The outgoing flows include those going into the hospital system, out of the system and severity evolution. Equation \eqref{eq: flowH2C} captures the flow balance of COVID $s_2$ patients reneging from the Clinic as well as the LPED. Since the inflow to Clinic includes the flow from $H_{2}^C$, the inflow to $H_{2}^C$ also includes the doubly rejected flow of $H_{2}^C$. Similarly, we derive $\eqref{eq: flowH2N}$ for non-COVID $s_2$ patients. Finally,  $H_{3}$ includes the flow of unsatisfied $s_3$ patients. The outgoing, as usual, includes flow back to the system, out of the system and severity evolution, as given in \eqref{eq: flowH3}.
 \pfappfalse  
}

\subsection{Mathematical Formulation}\label{subsection: optprogram}
Penalizing the weighted stationary flow of unserved patients in our objective, we seek to find the optimal capacity allocation. The constraints include all fluid balance equations for facilities and repositories, indicator functions for queue length and the efficiency and total capacity constraints.  We aim to determine the capacities, efficiencies and queue lengths at all facilities, and queue lengths at repositories through this program. (Note that since we solve the stationary problem, any transient and terminal effects are immaterial.)

The objective function, determined in conjunction with our hospital partner, minimizes the number of people exiting the system due to dissatisfaction or mortality, weighted by severity. 
\textbf{Optimization Problem $\mathcal{P}$:}
\begin{align}
   \min  & \ 1 \times H_{1}\delta_{10} +s_{1} \sigma_{1} H_{1}+ s_{2} \sigma_{2} H_{2}^C+ s_{2} \sigma_{2} H_{2}^N + s_{3} \sigma_{3}H_{3} \label{eq: MainObj} \\
    \text{ s.t. } 
   & \frac{dQ}{dt} =0, \ Q \in \{Q^{E}_{1}, \ Q^{E}_{2,3} , Q^{C}, \ Q^{N}, H_{1}, H_{2}^C, H_{2}^N, H_{3} \},  \label{eq: dqdtfinal} \\
   & \eqref{eq: IndicatorLP1}, \eqref{eq: IndicatorLP2}, \eqref{eq: IndicatorLP3}, \eqref{eq: LPEDextracases} \ \text{ (For LPED) }, \\
   & (1-\alpha^{f})\left(Q^{f}- \mu^{f} \tau^f \right) = 0, \quad \ f \in \{C, N\}, \label{eq: IndicatorClinicsfinal} \\
  & \sum_f \mu^f \leq \Gamma.  \label{eq: Capacityconstraintfinal} 
 \end{align}

\section{Analytical Characterization of the Single Period Problem}\label{sec:single_period_solutions}
The optimization problem $\mathcal{P}$ consists of multiple complex and nonlinear constraints, which are computationally challenging to solve. Rather than solving this problem directly, we decompose its feasible space into a finite number of cases, which we call `\textit{combinations},' solve each combination independently, and then choose the best one. The resulting problems for each combination are computationally tractable and each combination has a concrete physical meaning, allowing us to better interpret our obtained solutions. 

We explain and prove the validity of our decomposition approach in Section~\ref{sec: Decomposition}. Next, in Section~\ref{sec: computationaltractability}, we show that the decomposed problems are computationally tractable. Then, in Section~\ref{sec: preferenceorder}, we provide further insights by showing that as the total capacity increases, the combinations become optimal in a specific, non-trivial order; Section~\ref{sec: preferenceorder} we also discuss the implications of this result on which facilities should be prioritized as the total capacity increases.   




\begin{table}[htbp]
\centering
\footnotesize	
\begin{tabular}{|c|c|c|c|c|c|}
\hline
Combination & \textbf{$\alpha^{E}_{1}$} & \textbf{$\alpha^{E}_{2}$ } & \textbf{ $\alpha^{E}_{3}$ } & \textbf{ $\alpha^{C}$ } & \textbf{ $\alpha^{N}$ } \\ \hline
1     & 1                & 1                   & 1                   & 1                    & 1                 \\ \hline

2     & 1               & 1                   & 1                   & $[0,1)$                & 1                    \\ \hline
3     & 1              & 1                   & 1                  & 1                    & $[0,1)$                   \\ \hline

4     & 1               & 1                  & 1                   & $[0,1)$                   & $[0,1)$                  \\ \hline

5     & 1                & $[0,1)$                 & 1                  & 1                   & 1           \\ \hline
6     & 1                & $[0,1)$                  & 1                   & $[0,1)$                   & 1                     \\ \hline
7     & 1                & $[0,1)$                  & 1                   & 1                    & $[0,1)$                    \\ \hline
8     & 1                & $[0,1)$                  & 1                   & $[0,1)$                   & $[0,1)$                   \\ \hline
9     & 1                & 0                   & $[0,1)$                  & 1                    & 1                     \\ \hline
10    & 1                & 0                   & $[0,1)$                  & $[0,1)$                   & 1                     \\ \hline
11    & 1                & 0                   & $[0,1)$                  & 1                    & $[0,1)$                    \\ \hline
12    & 1                & 0                   & $[0,1)$                  & $[0,1)$                   & $[0,1)$                   \\ \hline
13    & $[0,1)$               & 0                   & 0                   & 1                    & 1                     \\ \hline
14    & $[0,1)$               & 0                   & 0                   & $[0,1)$                   & 1                     \\ \hline
15    & $[0,1)$               & 0                   & 0                   & 1                    & $[0,1)$                    \\ \hline
16    & $[0,1)$               & 0                   &     0                & $[0,1)$                   & $[0,1)$                   \\ \hline
\end{tabular}
\caption{Combination specifications}
\label{table : StarSolutions}
\end{table}



\subsection{Decomposition Framework} \label{sec: Decomposition}
To decompose the solution space, we observe that each solution can be characterized by the congestion status of all (facility, severity) pairs, i.e., $\alpha^f_i$ values: each (facility, severity) pair can be served with (i) zero, (ii) partial or (iii) full efficiency. Accordingly, we decompose the feasible space into sub-spaces (our \textit{combinations}), each of which represents a particular combinations of  congestion statuses. 
Crucially, this decomposition does not involve exhaustively enumerating all $3^5$ possible efficiency configurations, as some combinations are inconsistent with our prioritization rules (e.g. $\alpha^E_1=0, \alpha_{2,3}^E>0$). Specifically, our decomposition yields 16 valid combinations, as listed in Table~\ref{table : StarSolutions}. In what follows, we show that the combinations taken together are mutually exclusive and collectively exhaustive sub-spaces of the feasible space. Therefore, we can solve our optimization problem by solving the problem associated with each combination, giving us a set of candidate solutions, and then choosing the best candidate. 
\begin{theorem} \label{thm: existence}
Optimization Problem $\mathcal{P}$ can be written as a set of 16 mutually exclusive and collectively exhaustive optimization problems,  indexed in Table \ref{table : StarSolutions}. For any set of parameters, an optimal solution always exists and can be computed as the best of these 16 combinations. Of these, combinations 1-5 are all feasible (and optimal) only when capacity $\Gamma$ exceeds the incoming flow $\lambda$.
\end{theorem}

\proof{Proof of Theorem \ref{thm: existence}.}
We prove this theorem through a series of arguments and lemmas. The details of the proof are in Appendix~\ref{appndx: ForOnePeriodSection}. Briefly, 
\begin{enumerate}
    \item First, we note that the combinations are mutually exclusive by their definition in Table~\ref{table : StarSolutions}.
    \item Next, we prove exhaustiveness, by showing that every feasible solution to $\mathcal{P}$ belongs to one of the 16 combinations (Lemma~\ref{lem: exhaustive}).
    \item We next show that at least one combination is always feasible for each parameter setting (Lemma~\ref{lem: existencelemma}).
    \item Finally, we show that combinations 1-5 are feasible (and optimal because they achieve an objective value of 0) only when $ \Gamma > \lambda$ (Lemma~\ref{lem: 1to5feasible}). Thus, unless we have enough capacity to serve all flow, the optimal combination is one of combination 6-16. \halmos \endproof
\end{enumerate}  

\pfapptrue  
\def\pfexhaustive{
\begin{lemma} \label{lem: exhaustive}
 Every feasible solution to problem $\mathcal{P}$ belongs to exactly one of the 16 combinations listed in Table \ref{table : StarSolutions}.
\end{lemma}


\proof{Proof of Lemma \ref{lem: exhaustive}.}  We characterize the combinations by finding the cross product of the only possible cases for the ED and the callbank. Note that all variables $\alpha^f_i$ indicate efficiencies, with their values constrained between 0 and 1.  We first observe that the absolute priority to HPED allows LPED to serve patients only if $Q^{E}_{1}=0$. Thus, when $Q^{E}_{1}>0$ i.e. $\alpha^{E}_{1}<1$, we have $\alpha^{E}_{2} = \alpha^{E}_{3} = 0$.  Next, considering the dynamics of LPED, we can refer to the discussion in Section \ref{sec:formulations}, where we have the following types of solutions for $(\alpha^{E}_{1},\alpha^{E}_{2}, \alpha^{E}_{3})$ : $(1,1,1), (1,c,1), (1,0,c), (c,0,0)$, where $c$ is a constant $\in [0,1)$.
  
  Considering the callbank part now, we see that since the systems Clinic and NClinic are independent, and we have four types of solutions for $(Clinic, NClinic) $ : $(1,1), (1,c), (c,1), (c,c')$. Thus, taking the cross product of the ED and callbank possibilities, we see that these 16 combinations are the only possible solutions. 
 \halmos \endproof
 
 \pfappfalse  
}

\pfapptrue  
\def\pfcontiguityoffeasible{
Next, we make a technical observation about the contiguity of the feasible combinations in Lemma~\ref{lem: contiguity of feasible}.  
 \begin{lemma} \label{lem: contiguity of feasible}
  For any fixed set of parameters, every combination is feasible for a continuous range of $\Gamma$.
  \end{lemma}

  \proof{Proof of Lemma \ref{lem: contiguity of feasible}.}
    We just need to prove that if a solution is feasible at $\Gamma=m$, it is feasible for all $\Gamma>m$. This comes from the understanding that we can always choose $ \mu_f$ such that $\sum_f\mu^f =m <\Gamma$, where the combination is feasible.
  \halmos \endproof
  \pfappfalse  
}

 \pfapptrue  
\def\pfexistencelemma{
\begin{lemma} \label{lem: existencelemma} An optimal solution always exists to problem $\mathcal{P}$ and it can be computed as the best among the candidate solutions.
 \end{lemma}

 \proof{Proof of Lemma \ref{lem: existencelemma}.} 
 We prove the first part by showing that combination 16 which indicates non-negative spillage from all facilities, is always feasible.
 
For the feasibility of combination 16, we only need to prove that it is feasible for capacity $\Gamma = \epsilon$, where $\epsilon$ is small enough. Here, $\epsilon \leq \lambda_1$ suffices. Then, Lemma \ref{lem: contiguity of feasible} ensures the feasibility for all values of $\Gamma$. Combination 16 is characterized by  $\alpha_1^E, \alpha^{C}, \alpha^{N} \in [0,1)$ and the rest $\alpha_2^E, \alpha_3^E = 0$. For $\epsilon$ capacity, we create a feasible instance of combination 16 as follows: let  $\mu^C, \mu^N = 0$, forcing $Q^C, Q^N = \alpha^{C}, \alpha^{N}  = 0$, and we satisfy the fluid balance equations for Clinic and Nclinic  \eqref{eq: flowC}- \eqref{eq: IndicatorN}. We also trivially satisfy the LPED equations  \eqref{eq:flowLP}- \eqref{eq: IndicatorLP3}. Let $\mu_1^E = \epsilon$ and define $\alpha_1^E$ and $Q_1^E$ satisfying the linear constraints in \eqref{eq: flowHP}-\eqref{eq: IndicatorHP}. The choice of small enough epsilon ensures  $\alpha_1^E \in [0,1)$. It can be easily checked that equations \eqref{eq: flowH1}-\eqref{eq: flowH3} result in $H_i>0$ for our choice of $\alpha_1^E$. Thus, we see that our construction satisfies all constraints within combination 16, proving its feasibility for all $\Gamma\geq\epsilon$. 

After having established that space can be decomposed into only 16 possible combinations, we now want to show that we can find the optimal solution as the best amongst these candidate solutions. We observe that while solving a particular combination,  its candidate solution (optimal within that combination) may go beyond the specifications of the corresponding combination. For instance, combination 16 requires all facilities to be congested with fixed queue lengths but if we have enough capacity to make HPED full efficient, the corresponding candidate solution will have $\alpha^E_1 = 1$. This is because we cannot force strict inequalities $\eta^i \in [0,1)$ even if we want a particular facility to be congested. Going by the strict definition, although the region of combination 16 is feasible, the corresponding candidate solution does not exist within it.  We denote such candidate solutions as \textit{boundary} solutions. 

Luckily, even if the boundary candidate solutions exist, our process of choosing an optimal solution by picking their best circumvents this issue.  In other words, even if we discard all boundary candidate solutions to find the best out of the rest, our optimum doesn't change.  We need to give a little more explanation to say that if we discard the boundary solutions, the corresponding combination type may still be feasible but no solution within that combination can be optimal. Lemma \ref{lem: Degeneracy} establishes that we can always consistently find the optimal solution  through the decomposition of the solution space into 16 combinations. 
  \begin{lemma} \label{lem: Degeneracy}
    If a candidate solution is a boundary solution,  there exists another combination which dominates all feasible solutions in the discarded combination.
  \end{lemma}

   \proof{Proof of Lemma \ref{lem: Degeneracy}.}
   We discard a candidate solution when it belongs to a different combination (say 2) than the corresponding one (say 1). This happens when both parts of one or more indicator constraints $eqn1 \  \times \  eqn2 =0$ wiz., $eqn1 = 0$ and $eqn2 = 0$  hold true.  In that case, we compute the candidate capacity allocation for combination 1 and feed it into combination 2. It is easy to see that this solution is feasible for combination 2 as all constraints are trivially satisfied. The objective either remains the same or decreases as we make the relevant queue lengths equal to zero, thereby keeping the same constraints and variable assignments.  Since we can find a feasible solution for combination 2 that has at least the same objective, we conclude by noting that combination 2 dominates all feasible solutions of combination 1.
  \halmos \endproof

\pfappfalse  
}
 \begin{figure}
\centering
\begin{subfigure}[b]{0.3\textwidth}
         \centering
         \includegraphics[width=6cm]{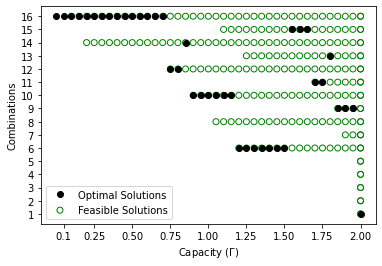} 
         \caption{Feasibility vs Capacity}
         \label{fig:feasibility vs capacity_Ex1}
     \end{subfigure}
     \hspace{1cm}
     \begin{subfigure}[b]{0.3\textwidth}
         \centering
         \includegraphics[width=6.1cm]{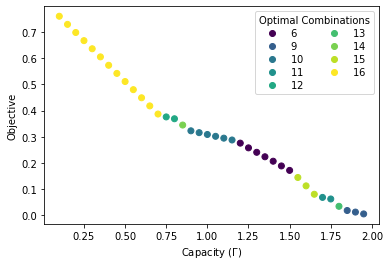}
         \caption{Objective vs Capacity}
         \label{fig:obj vs capacity}
     \end{subfigure}
\caption{The progression of optimal (and feasible) solutions with varying capacity, for our example. The green and black circles represent feasibility and optimality, respectively, of a combination at the specified capacity} \label{fig: sec4examplePrefOrder}
\end{figure}
 
\pfapptrue  
\def\pfOnetoFivefeasible{
  
 \begin{lemma} \label{lem: 1to5feasible}
   The first 5 combinations are feasible (or optimal) only at full capacity.  
   \end{lemma}
   
   \proof{Proof of Lemma \ref{lem: 1to5feasible}.}
   Consider the first five solutions in Table \ref{table : StarSolutions}. In each of these solutions, all $s_{1}, s_{2}, s_{3}$ patients can join at least one queue that works with full efficiency. For example, combination 4 has ED working with full efficiency. Thus, $s_{1},s_{3}$ patients can directly join the ED queues. $s_{2}$ patients who don't join Clinic and NClinic due to congestion, join LPED, which works with full efficiency. Thus, every entering patient joins a facility, which is possible only if $\sum_f \mu^f = \Gamma= \lambda$ i.e. when the total capacity equals the total incoming rate.  More formally, we check the constraints in section $\ref{subsection: optprogram}$ and see that the equations $\frac{dH_i}{dt} = 0$ reduce to homogeneous equations in $H_i$. This is because there is no incoming flow to the $H_i$'s. The solution to this homogeneous system is $H_i = 0$. This leads to zero objective value, as the objective penalizes the lengths of $H_i$'s.  Moreover, we can now check the fluid balance equations for each facility, with $H_i = 0$. Summing the total incoming rate for each queue and equating it to the corresponding  $\mu_f$ results in $\sum_f \mu^f  =\Gamma= \lambda$.  We then conclude that the first five combinations are feasible only when $\lambda  = \Gamma$. Moreover, with this objective function, we note that the first five combinations are feasible and optimal if and only if $\lambda = \Gamma$. They are simultaneously optimal since all solutions correspond to zero number of patients that we lose or $H_i = 0$ and objective $= 0$.
   \Halmos
\endproof
Considering the objective function in \eqref{eq: MainObj}, which is bounded below by 0, we see that an optimal solution always exists and gives the minimum objective value amongst these 16 math programs.   With the technical result in Lemma \ref{lem: 1to5feasible}, we have proved that the optimal solution  can always be computed as the best amongst combinations 6-16. Thus, we conclude the proof of Theorem \ref{thm: existence}. 
\Halmos
\pfappfalse  
}


We illustrate this section's results using a representative example. (The detailed parameter settings for this example are provided as Example 1 in Table \ref{tab: paramsforoneperiod} in Section \ref{sec: NumericalSection}, where we discuss our numerical results in deeper detail.) Figure \ref{fig:feasibility vs capacity_Ex1} shows each combination's feasibility and optimality as a function of $\Gamma$, and illustrates that multiple combinations can be feasible at a particular capacity. It also illustrates that each combination is feasible for a continuous range of $\Gamma$ values (we formalize this result in Lemma~\ref{lem: contiguity of feasible}). The figure also shows that each combination is \textit{optimal} for a continuous range of $\Gamma$ values, hinting at a deeper structural connection between capacity and the optimal combination; we explore this connection in detail in Section~\ref{sec: preferenceorder}. The objective values are shown in Figure \ref{fig:obj vs capacity}; we show that the objective value is piecewise linear in $\Gamma$ in Section~\ref{sec: computationaltractability}.

\subsection{Computational Tractability of the Decomposed Problems} \label{sec: computationaltractability}
In Section~\ref{sec: Decomposition}, we showed that $\mathcal{P}$ can be solved by solving 16 separate optimization problems -- one for each combination. We now show through Theorems~\ref{thm: 9to16space} and \ref{thm: 6to8space} that these problems are computationally tractable. To this end, it is helpful to separate the space of all combinations into three groups: (i) Combinations 1-5, which by Theorem~\ref{thm: existence}, are not feasible unless we have enough capacity to serve all incoming flow; (ii) Combinations 6-8 which have non-linear feasible spaces; and (iii) Combinations 9-16, which have linear feasible spaces. 

We note first that group (i) is feasible and optimal only when $\Gamma \geq \lambda$, and that their solution is trivial: Assign sufficient capacity to match all the flows, which makes it feasible (as $\mu\geq \lambda$) and optimal (as it yields an objective equal to zero). We now establish the tractability of combinations in groups (ii) and (iii). With this aim, we first show that we always use all available capacity in any optimal solution when $\Gamma < \lambda$, and therefore, we can replace the constraint $\sum_f \mu^f \leq \Gamma $ by $\sum_f \mu^f = \Gamma $ for combinations 6-16. We formalize this result in Lemma~\ref{lem: fullcapacity} (Appendix~\ref{appndx: ForOnePeriodSection}). With this modified constraint, we now state Theorems~\ref{thm: 9to16space} and \ref{thm: 6to8space}, that establish tractability.


 \pfapptrue  
\def\pffullcapacity{

 \begin{lemma} \label{lem: fullcapacity}
   For any optimal solution with $\Gamma < \lambda$, $\sum_f \mu^f  =\Gamma$.
 \end{lemma} 
 
\proof{Proof of Lemma \ref{lem: fullcapacity}.}
    Let us assume that an optimal solution exists where $\sum_f \mu^f <\Gamma$. Since $M< \lambda$, the optimal solution belongs to the set of the last 6-16 combinations, where at least one facility queue is congested. Without loss of generality, let us choose a facility $f$ with corresponding $\alpha^f<1$. Since $\sum_f \mu^f <\Gamma$, we can increase $\mu^f$ by epsilon amount. This increase is followed by the increase in $\alpha^f$ in its corresponding  equation $\frac{dQ^f}{dt} = 0$. Since the term $1-\alpha^f$ appears in the rest of the equations $\frac{dQ^g}{dt} = 0$, we can increase $\alpha^{g}$ in these equations, so as to satisfy the equalities. Then, the total number of patients who enter $H_i$  decreases, since all $\alpha^{g}$ either increase or remain the same. This leads to an overall lower objective as $H_i$ are smaller. This contradicts the optimality of the current solution. Therefore, $\sum_f \mu^f =\Gamma$ in every optimal solution. 
  \halmos \endproof
\pfappfalse  
}


\begin{table}[h]
\footnotesize	
\centering
\begin{tabular}{|c|c|c|c|}
\hline
Combination & Equations & Variables & \multicolumn{1}{l|}{Extreme Points} \\
\hline
1-5          & -         & -         & 1                                   \\
\hline
6-8          & -         & -         & 3                                   \\
\hline
9           & 8         & 8         & 1                                   \\ \hline
10          & 7         & 8         & 2                                   \\ \hline
11          & 7         & 8         & 2                                   \\ \hline
12          & 6         & 8         & 3                                   \\ \hline
13          & 7         & 7         & 1                                   \\ \hline
14          & 6         & 7         & 2                                   \\ \hline
15          & 6         & 7         & 2                                   \\ \hline
16          & 5         & 7         & 3                                   \\ \hline
\end{tabular}
\caption{Properties of  combinations 1 through 16}
\label{tab: simplifiedcombs}
\end{table}



\begin{theorem} \label{thm: 9to16space}
  The feasible region of each combination in  9-16 is a polytope, resulting in enumerable extreme points that span the space of optimal solutions. Thus, the objective values and flows are piecewise linear with respect to the inputs. 
  \end{theorem} 
    \proof{Proof of Theorem \ref{thm: 9to16space}.} 
    
     
    Each combination's polytope is specified by enforcing the constraints corresponding to its $\alpha^f_i$ values from Table~\ref{table : StarSolutions}. Adding these constraints shows that combinations 9-16 reduce to linear programs. The remaining details of the proof are presented in Appendix \ref{appndx: ForOnePeriodSection}.     \halmos \endproof

 \pfapptrue  
\def\pfthmtwo{

\proof{Details of the Proof to Theorem \ref{thm: 9to16space}.}
Consider the simplified constraints for combination 12, which has $\alpha^{E}_{1} = 1$, $\alpha^{E}_{2}= 0$, $\alpha^{E}_{3}, \alpha^{C}, \alpha^{N} \in (0,1)$.
\begin{align}
   \mu^{E}_1 & = \left( H_{1}\beta_{1} + \lambda P(s_{1}) + (Q^{C}+Q^{N})\delta_{21}\right)  \label{eq: com$12_b$eq1}\\
    \alpha^{E}_{3} & = \frac{\mu^{E}_{2,3}}{[\lambda P(s_{3}) + \beta_{3}H_{3}]}   \label{eq: com$12_b$eq2}\\
    \alpha^{C} & = \frac{\mu^C}{\lambda P(s_{2})P(covid) + \beta_{2}H_{2}^C} \label{eq: com$12_b$eq3}\\
    \alpha^{N} & = \frac{\mu^N}{\lambda P(s_{2})(1-P(covid)) + \beta_{2}H_{2}^N}  \label{eq: com$12_b$eq4}\\
   \frac{dH_{1}}{dt} & =  H_{2}^C\delta_{21}+H_{2}^N \delta_{21} -H_{1}[  f_{H}^{out}  +\delta_{12} + \delta_{10} + \beta_{1}]= 0  \label{eq: com$12_b$eq5} \\
  \frac{dH_{2}^C}{dt} & = [P(covid)\lambda P(s_{2}) + H_{2}^C\beta_{2}  ](1-\alpha^{C}) \notag\\
    &  +(H_{3}\delta_{32}+ H_{1}\delta_{12})P(covid)- H_{2}^C[\sigma_{2} +\delta_{23} + \delta_{21} +\beta_{2}] = 0  \label{eq: com$12_b$eq6}\\
    \frac{dH_{2}^N}{dt} & =  [(1-P(covid))\lambda P(s_{2}) + H_{2}^N \beta_{2} ](1-\alpha^{N}) \notag\\
    &  +(H_{3}\delta_{32}+ H_{1}\delta_{12})(1-P(covid))- H_{2}^N[\sigma_{2} +\delta_{23} + \delta_{21} +\beta_{2}] = 0 \label{eq: com$12_b$eq7}\\
   \frac{dH_{3}}{dt} & = [\lambda P(s_{3})+H_{3}\beta_{3}] (1-\alpha^{E}_{3}) + \delta_{23}H_{2}^C+ \delta_{23}H_{2}^N - H_{3} [\delta_{32}+\delta_{34} +  \sigma_{3}+\beta_{3}] = 0 \label{eq: com$12_b$eq8}
   \end{align}
   \begin{align}
Q^{E}_{2,3}  = \mu^E_{2,3} \tau^{E}_2 \\
   Q^{C}  = \mu^{C} \tau^{C}  \\
 Q^{N} = \mu^{N} \tau^{N} \\
   \Gamma  =  \mu^{E}_1+ \mu^{E}_{2,3} + \mu^C+\mu^N \label{eq: com$12_b$eq14}
\end{align}
Notice that if we substitute the expressions for $\alpha^{E}_{2},\alpha^{C}$ and $\alpha^{N}$ in the equations for $H_i$, we get linear equations. For example, the equation for $H_{2}^C$ reduces to 
\begin{align}
  \frac{dH_{2}^C}{dt} & = [P(covid)\lambda P(s_{2}) + H_{2}^C\beta_{2} -\mu^{C}] \notag\\
    &  +(H_{3}\delta_{32}+ H_{1}\delta_{12})P(covid)- H_{2}^C[\sigma_{2} +\delta_{23} + \delta_{21} +\beta_{2}] = 0 \label{eq: simplified H2C}
    \end{align}
similarly, $H_{2}^N$ and $H_{3}$ can be reduced to linear equations. Thus, all equations are linear, along with a linear objective. Considering equations \eqref{eq: com$12_b$eq1}, \eqref{eq: com$12_b$eq5}-\eqref{eq: com$12_b$eq8}, \eqref{eq: com$12_b$eq14} as the 6 linear constraints and $\mu^{E}_1, \mu^{E}_{2,3}, \mu^N, \mu^{C}, H_{1}, H_{2}^C, H_{2}^N, H_{3}$ as the 8 variables, we can solve a linear optimization problem. We also have 3 inequality constraints coming from \eqref{eq: com$12_b$eq2}-\eqref{eq: com$12_b$eq4}, along with the positivity constraints on all variables. Note that all variables except $\alpha^{E}_{3}, \alpha^{C}, \alpha^{N}$ vary linearly with small changes in any of the inputs. These values are computed once we have the values for the 8 variables.  Even if the efficiencies $\alpha^i$ are not linearly dependent on the inputs, the costs and queue lengths are linear, making costs linear. Thus, we establish that combination 12 is reduced to a linear problem. 



 With this, we also prove that the objective and variables for this combination change piecewise linearly with total capacity $\Gamma$. Similarly, we can check that all other combinations 9-16 give piecewise linear formulations, with the number of variables and equations listed in Table \ref{tab: simplifiedcombs}.  In Lemma \ref{lem: 9to16extremepts}, we enumerate the number of possible solutions in each combination.
\begin{lemma} \label{lem: 9to16extremepts}
  Each combination in 9-16 has an enumerable number of possible optimal solutions, computed as the extreme points as given in Table \ref{tab: simplifiedcombs}.
\end{lemma}

\proof{Proof. of Lemma \ref{lem: 9to16extremepts}}
As given in Table \ref{tab: simplifiedcombs}, the simplified space for combination may be of  $n\times (n+2)$ equations and variables (combination 12, 16), $n\times (n+1)$ equations and variables  (combination 10, 11, 14, 15) or that of $n\times n$ (combination 9, 13). The number of extreme points in a $n\times n$ system is 1. That for a $n\times (n+1)$ system is 2, since it's essentially a single dimensional space. Thus, under the assumption of a convex/linear objective, the solution to this optimization problem is always going to be one of these extreme points. 

We now analyse a $n\times (n+2)$ system, through combination 12. We know that the space is of 8 variables and 6 linear equality constraints. There are positivity constraints for each of the variables, and three inequality constraints coming from the efficiencies being less than 1. Thus, we can easily compute the upper bound on the total number of extreme points. Further, we may do some analysis to find the exact number of extreme points. We can deduce that no $H_i$ can equal zero, because if so, then the corresponding efficiency of the queue becomes becomes greater than 1. Then, any extreme point involves 6 tight constraints from the equalities and 2 tight constraints come either from the positivity constraints or the inequality constraints from \eqref{eq: com$12_b$eq2}- \eqref{eq: com$12_b$eq4} (inequality constraints of the form $\alpha^f \leq 1$). We also know that although the extreme points may occur at efficiencies being equal to 1, those points are already covered in other combinations. Then, the only possibilities are through the positivity constraints on $\mu^C, \mu^N$ and $\mu^{E}_{2,3}$. There are 3 ways to select 2 efficiencies and thus the number of extreme points equals 3. Similarly, we can compute the number of extreme points in combination 16.   
\halmos \endproof

With this lemma, we conclude the proof to Theorem \ref{thm: 9to16space}, with a  tractable number of possible solutions from combinations 9-16.
 \halmos \endproof
\pfappfalse  
}
 
\begin{figure}
\centering
\begin{subfigure}[b]{0.3\textwidth}
         \centering
         \includegraphics[width=6cm]{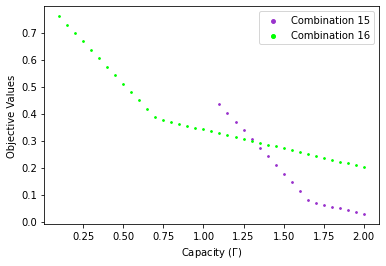}
         \caption{Combinations 15,16}
         \label{fig:combs15_16}
     \end{subfigure}
     \hspace{1cm}
     \begin{subfigure}[b]{0.3\textwidth}
         \centering
         \includegraphics[width=6.1cm]{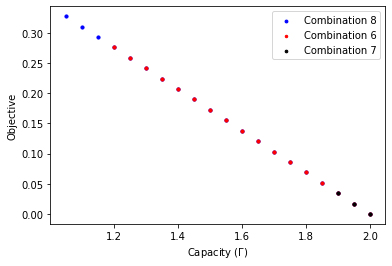}
         \caption{Combinations 6, 7, 8}
         \label{fig:combs6_7_8}
     \end{subfigure}
\caption{Piece-wise linear objectives corresponding to various combinations}
\label{fig: sec4exampleLinearCombs}
\end{figure}

\begin{theorem}  \label{thm: 6to8space}
  The feasible regions of combinations 6-8 are nonlinear spaces. However, for each combination, the optimal solution lies in a polytope contained within the non-linear feasible region. Accordingly, the candidate solutions for each combination are extreme points that can be enumerated, and the flows and objectives are linear in the inputs.  
  \end{theorem}
  \proof{Proof of Theorem \ref{thm: 6to8space}.}
 This proof is similar to that of Theorem \ref{thm: 9to16space}. We consider optimization problem $\mathcal{P}$ restricted to combinations 6-8 and study the simplified constraints. We prove that the optimal solution is contained within a linear subspace of the original nonlinear regions, by finding equivalent linearizations of the  non-linear terms in the objective and constraints using the structure of the objective function. The details of the proof can be found in  Appendix \ref{appndx: ForOnePeriodSection}. \halmos \endproof
 
\pfapptrue  
\def\pfthmthree{
\proof{Details of the Proof to Theorem \ref{thm: 6to8space}.}
 We start with discussing the case for combination 6.
 This combination has HPED, NClinic and LPED for $s_3$ working with full efficiency with $\alpha^{E}_{1} = \alpha^{E}_{3} = \alpha^{N}  = 1$. Since LPED is congested for $s_2$, we define  $\alpha^{E}_{2} = 1- x_1$, $\alpha^{C} = 1- x_2$ as our variables. As we see later, $H_{2}^C = x_3 $ is another explicit variable we use while formulating the simplified optimization problem for combination 6. We write the simplified constraints from Section \ref{subsection: optprogram}.
  \begin{align}
   \mu^{E}_1 & =  H_{1}\beta_{1} + \lambda_{1} + Q^{C}\delta_{21} \\ 
     \mu^{E}_{2,3} & = (1-p)[\lambda_{2}+ \beta_{2}H_{2}^C+ \beta_{2}H_{2}^N] \alpha^{E}_{2} \ + \ [\lambda_{3} + \beta_{3}H_{3}]  \notag\\
     & + \   [p\lambda_{2} \mathds{P}(covid) + p\beta_{2}H_{2}^C ]  (1-\alpha^{C})\alpha^{E}_{2} \\
     \mu^C & = \alpha^{C}[   p\lambda_{2}\mathds{P}(covid) + p\beta_{2}H_{2}^C +  \notag\\
    &  [(1-p) \lambda_{2} \mathds{P}(covid)+ (1-p) \beta_{2}H_{2}^C] (1-\alpha^{E}_{2}) ]  \\
   \mu^N & =  p  \lambda_{2}(1-\mathds{P}(covid)) + p\beta_{2}H_{2}^N + \notag\\
    &  [(1-p) \lambda_{2}(1-\mathds{P}(covid))+ (1-p)\beta_{2}H_{2}^N] (1-\alpha^{E}_{2})   \\
    \frac{dH_{1}}{dt} & =  H_{2}^C\delta_{21}+H_{2}^N \delta_{21} -H_{1}[  \sigma_{1}  +\delta_{12} + \delta_{10} + \beta_{1}]= 0\\
  \frac{dH_{2}^C}{dt} & = [\mathds{P}(covid)\lambda_{2} + H_{2}^C\beta_{2}  ](1-\alpha^{E}_{2})(1-\alpha^{C})\notag \\
    &  +(H_{3}\delta_{32}+ H_{1}\delta_{12})\mathds{P}(covid)- H_{2}^C[\sigma_{2} +\delta_{23} + \delta_{21} +\beta_{2}] = 0  \label{eq: thm3pf1}  \\
   \frac{dH_{2}^N}{dt}  & =  (H_{3}\delta_{32}+ H_{1}\delta_{12})(1-\mathds{P}(covid))- H_{2}^N[\sigma_{2} +\delta_{23} + \delta_{21} +\beta_{2}] = 0 \\
   \frac{dH_{3}}{dt} & = \delta_{23}H_{2}^C+ \delta_{23}H_{2}^N - H_{3} [\delta_{32}+\delta_{34} +  \sigma_{3}+\beta_{3}] = 0
\end{align}
\begin{align}
Q^{E}_{2,3}= \mu^{E}_{2,3} \tau_2^{E} \\
Q^{C}= \mu^C \tau^{C}\\
    \mu^{E}_1+ \mu^{E}_{2,3} = \mu^E \\
    0 \leq \alpha^{E}_{2} \leq \alpha^{E}_{3} \leq \alpha^{E}_{1} \leq 1\\
 \mu^{E}_1+ \mu^{E}_{2,3} + \mu^C+\mu^N = \Gamma \label{eq: thm3pf2}
\end{align}

Note that we skipped certain equations that yield $\alpha^{N} = 1, Q^{N} = 0, \alpha^{E}_{1}= 1, Q^{E}_{1} = 0, \alpha^{E}_{3} = 1$ etc, as they are part of the specifications of Combination 6.  To simplify the remaining set of constraints, note that all equations except \eqref{eq: thm3pf1}, \eqref{eq: thm3pf2}  are independent and can be solved uniquely once we compute the variables $\alpha^{C}, \alpha^{E}_{2}$ and $H_{2}^C$. 
We can write variables $H_{1}, H_{3}$ in terms of $H_{2}^C$,  $H_{2}^N$ and further express $H_{2}^N$ in terms of $H_{2}^C$. 
The equations involving $\mu^f$ can be used to independently compute the $\mu^f$. We know the values for $\mu^f$ would be positive being sums of positive numbers. There are no other constraints on our variables. 

We now explain the non-linear formulation of combination 6. To solve this set of equations, we put $H_{2}^C = x_3, 1-\alpha^{E}_{2} = x_2$, and $1-\alpha^{C} = x_1$. 
We also simplify the objective function by representing it as a constant multiple of $H_{2}^C$. Then, our optimization problem can be rewritten as 
\begin{equation}
\begin{aligned}
\min_{x_1,x_2,x_3} \quad & x_3\\
\textrm{s.t.} \quad & k_1 x_2x_1 + k_2x_1x_2x_3  - k_3 x_3 + k_4 = 0\\
&  k_5 x_3 + k_6 x_1 + k_7 x_1x_3+ k_8x_2 + k_9x_3x_2 + k_{10} x_1x_2x_3 = k_{11}\\
& x_2,x_1 \in (0,1]\ x_3 \geq 0
\end{aligned} \label{eqn: simplifiedoptprogfor6}
\end{equation}
where $k_i$ are some positive constants, giving a non-linear formulation for combination 6.
\\
 \textbf{Combination 7 and 8}
Since  combination 6 and 7 differ only in the treatment of  $p(covid)$, combination 7 can be simplified to a similar program like 6. Combination 8 has $\alpha^{C}, \alpha^{N}, \alpha^{E}_{2}, H^C_2, H_{2}^N$ as variables. 
We consider three equations wiz. the fluid balance equations for $H_{2}^C, H_{2}^N$ while optimizing. The remaining constraints assign values to the other variables, while also satisfying the positivity constraints. We have one extra constraint for $H_{2}^N$. Rewriting the optimization program for combination 8, we have 
\begin{equation}
\begin{aligned}
\min_{x_1,x_2,x_3, x_4} \quad & c_1x_1+c_2x_2\\
\textrm{s.t.} \quad & k_1 x_2x_3 + k_2x_1x_2x_3 + k_3 x_2x_4 + k_4x_1x_2x_4 - k_4 x_1 = 0\\
& k_6 x_1x_3 + k_7x_1x_2x_3 + k_8 x_1x_4 + k_9x_1x_2x_4 - k_{10} x_2 = 0\\
&  k_{11}x_1 + k_{12} x_3 + k_{13} x_1x_3+  k_{14}x_2 + k_{15}x_1x_2 + k_{16} x_1x_2x_3 \\
& + k_{17} x_3 + k_{18} x_1x_4+ k_{19} x_1x_2x_4= k_{20}\\
& x_3,x_4, x_5 \in (0,1)\ x_1, x_2 \geq 0 \label{eqn: simplifiedoptprogfor8}
\end{aligned} 
\end{equation} Looking at this simplified formulation, we again conclude that the cost and flows may vary non-linearly with inputs.  
 
Although the feasible regions for combinations 6-8 are not polytopes, we see that the candidate solutions always belong to polytope like regions, with the optimality of the objective enforcing an extra hyperplane.   Thus, the feasible spaces for combinations 6-8 are also linear, if restricted to optimality of our objective function. We formally prove this in Lemma \ref{lem: 6to8extremepts} and and also compute the number of extreme points for each of these combinations.  
\begin{lemma} \label{lem: 6to8extremepts}
Although their feasible regions are described by nonlinear constraints, the optimal solutions to combinations 6-8 can always be reduced to linear polytopes. Moreover, the optimal solutions for each are extreme points that we can enumerate, as given in Table \ref{tab: simplifiedcombs}.  
\end{lemma}

\proof{Proof of Lemma \ref{lem: 6to8extremepts}.}
We prove the theorem for combination 6,8 and note that combination 7 follows the exact same argument as combination 6. Since combination 6 is characterized by $\alpha^E_1 =\alpha^{N} = \alpha^E_3= 1, \alpha^E_2, \alpha^{C} \in [0,1)$, we just claim that its candidate solution always has $\alpha^{C}  = 0 = \mu^C$. With this, the nonlinear equations get reduced to linear equations, leading to extreme point optimal solutions. We first assume that the converse holds, i.e. the candidate solution involves $\alpha^{C} \in (0,1)$ and $\mu^C >0$. Then, $Q^C = \mu^C\tau^C \neq 0$. The incoming flow at the LPED involves $s_2$ flow to the ED and rejected Covid $s_2$ flow from the clinic. Moreover, the HPED has a positive flow $Q^C(\delta_{21})$ as the evolution flow from the Clinic.

Now, if we shift the Clinic capacity to LPED, the total $s_2$ flow served by LPED+Clinic remains the same. However, the evolution flow to the HPED now equals 0,  reducing the capacity needed at HPED to serve with full efficiency. We now serve more $s_2$ flow, thereby strictly decreasing the flow to $H_2^C$.   Thus, we strictly improve the efficiency of the system, contradicting the optimality. Hence, the optimal solution always has $\alpha^{C} = 0$ and combination 6 as well as combination 7 have linear frontiers.  The only extreme point for each is given as the unique solution to the two linear constraints with two variables, after substituting $x_1 = 1$ in equations \eqref{eqn: simplifiedoptprogfor6}.

We observe that a similar argument holds for combination 8. Assuming any solution with  $\mu^C, \mu^N \in (0,1)$, we see that there's a constant severity evolution flow to the HPED. We can always improve the objective by shifting the capacity to LPED. Thus, the optimal region of combination 8 also reduces to a polytope. We can then find the only corresponding extreme point by solving 3 linear constraints with 3 variables. Thus, a total of three extreme points can be derived for combinations 6-8.
\halmos
\endproof

With this, we conclude the proof to Theorem \ref{thm: 6to8space} by presenting an enumerable number of possible optimal solutions to combinations 6-8.   \halmos
\pfappfalse  
}

Table \ref{tab: simplifiedcombs} presents the outcomes of Theorems \ref{thm: 9to16space} and \ref{thm: 6to8space}, and establishes that solving $\mathcal{P}$ can be reduced to evaluating these enumerable extreme points. Figure \ref{fig: sec4exampleLinearCombs} shows the piecewise linear objectives corresponding to combinations 15-16 and 6-8 for our running example, in line with Theorems \ref{thm: 9to16space} and \ref{thm: 6to8space}. In Figure \ref{fig:combs6_7_8},  combinations 6, 7 and 8 lead to the same objective value for a given capacity, although they are feasible for different ranges of $\Gamma$. Additional capacity allows the hospital system to reduce the number of patients exiting the system due to dissatisfaction or mortality, and consequently, the objective value. Corollary \ref{cor: globalobj} shows that this decrease is piecewise linear in $\Gamma$; see Figure \ref{fig:obj vs capacity} for an illustration.

 
\begin{corollary} \label{cor: globalobj}
The global objective function is piecewise linear in the inputs, including in the total available capacity $\Gamma$.
\end{corollary}
\proof{Proof of Corollary  \ref{cor: globalobj}.}
Due to the linearity of the objective, the global objective can be written as the minimum over all the candidate solutions, each of which is the outcome of solving a linear program. Thus, it follows that the global objective is piecewise linear.
\halmos
\endproof

\subsection{Preference Order} \label{sec: preferenceorder}
In Theorem~\ref{thm: preference order}, we characterize the link between the capacity level and the optimal combination.
\begin{theorem} \label{thm: preference order}
  Depending on whether more patients are Covid or non-Covid, the optimal combinations always follow one of the two ``preference orders''  in Table \ref{tab: preforder}, as we increase $\Gamma$.
  \end{theorem}
  \pfapptrue  
\def\pfthmfour{
  \proof{Proof of Theorem \ref{thm: preference order}}
To prove this statement, we rely on using two arguments A1 and A2 repeatedly, which are as follows: \\
(A1): For a pair of combinations $(i,j)$, we say combination $j$ strictly dominates $i$, if the feasibility of $j$ implies that $i$ is feasible and the objective of $j$ is strictly better than that of $i$. In such cases, combinations $i$ and $j$ are same for all facilities except the ones where $j$ is strictly better. Even if $i$ is currently optimal and $j$ suddenly becomes feasible, then $i$ can never become optimal again. The validity of this argument can be established by comparing the extreme points of both $i$ and $j$, and it ensures a permanent dominance.\\
(A2): For a pair of combinations $(i,j)$, we say combination $j$ loosely dominates $i$, if the current optimality of $j$ implies that with increasing capacity $j$ dominates $i$. In other words, $i$ cannot become optimal right after $j$ has become optimal. In such cases, we compare the facilities we choose to invest on, as we increase the capacity $\Gamma$ for each of the combinations $i$ and $j$. If $j$ directs this new capacity to serve severity $s_j$ and $i$ sends to  $s_i$ with $s_j> s_i$, then the objective decreases faster for $j$. This implies that the current optimality of $j$ ensures future dominance of $j$ over $i$. The validity of this argument can also be established by comparing the extreme points of both $i$ and $j$, and it ensures a temporary dominance.

We first prove for the Covid majority order and through symmetry, the non Covid majority order is clear. In Lemma \ref{lem: existencelemma}, we've proved that combination 16 is always feasible by noting that for $\epsilon$ capacity, only 16 is optimal. We next see that if 12 is feasible for some $\Gamma$, 16 can never become optimal as we increase $\Gamma$, as 12 is better in serving $s_1$ patients, the rest being same for both the combinations. Every extreme point of combination 16 $\{(c, 0,0,0,0), (0,0,0,c,0), (0,0,0,0,c)\}$ is dominated by 12 $\{(1,0,0,0,c), (1,0,0,c,0), (1,0,c,0,0)\}$. In other words, combination 12 strictly dominates 16 through (A1).  

After combination 12, we see that combination 8 dominates the previous combinations. The optimality of 8 implies the feasibility of 12 and 16. Moreover, it dominates 12 through (A2) and 16 through (A1). 

We then see that combination 14 dominates 16, 12 and 8 through (A2). Out of the two extreme points of 14 $\{(c,0,0,0,1), (0,0,0,c,1)\}$, the first is strictly better as it prioritizes $s_1$ over $s_2$. If 14 is optimal,  16, 12 and 8 cannot become optimal since the objective for 14 decreases faster with increasing capacity. Thus, combination 14 dominates all extreme points of previous combinations. One may argue that 12 may suddenly become feasible while 14 is still optimal, and may have a better objective. But, the infeasibility of 12 until then contradicts the current optimality of 14, as 16 is better than 14 in that case.  Next, it can be observed that 10 dominates  8, 12, 14 and 16 through (A1). It is also easy to see that the current  optimality of 14 ensures that the only next possible optimal combination is 10.  Note that by our definition of combination 8, it is optimal when combination 6 cannot be made feasible.  Next, we claim that 6 dominates all the previous combinations through (A1), except 10.  It is easy to see this through the dominating extreme point of combination 6, which is $(1, c, 1, 0, 1)$. Optimality of 6 also implies the feasibility of 10 and further it dominates 10 through (A2).    With this, we prove the first six levels of the preference order, as no previous combination can become optimal once a current combination is optimal.

The dominance of the next three combinations is a consequence of the Covid majority, as apparent from the order.  Similar to the case of (14,10), combination 15 dominates all the previous combinations through (A2) and its current optimality also ensures that the next optimal combination is 11. We see that 11 dominates 16, 12, 8, 6 through (A1) and 14, 10 using the fact that (A1) is applicable through the Covid majority. Further, combination 7 dominates all previous combinations through (A1) and Covid majority, except 11. The current optimality of combination 7 implies the feasibility of 11, which it then dominates through (A2). 

Next, it is easy to see that combination 13 dominates all previous combinations through (A2) and ensures that the next optimal combination can only be 9. Combination 9 clearly dominates all previous combinations through (A1), and dictates that further capacity addition will let us directly reach combination 1. With this, we complete the proof of the preference order for Covid majority. We note that the preference order for non- Covid majority follows the same proof through symmetry, as we exchange the place of the block of combinations (14, 10, 6) with the block of combinations (15, 11, 7). 
\halmos
\endproof
\pfappfalse  
}
Theorem~\ref{thm: preference order} asserts that the order in which combinations may become optimal as we increase the total capacity from 0 to $\lambda$ is independent of the particular parameter setting; it is dependent only on whether Covid patients outnumber non-Covid or vice versa. We prove this result by showing that once a later combination in the order is optimal, a previous combination can never dominate its objective value. All details are in Appendix~\ref{appndx: ForOnePeriodSection}.
\begin{table}[H]
\footnotesize	
\centering
\begin{tabular}{|c|c|c|c|c|c|}
\hline
Combination & \textbf{$\alpha^{E}_{1}$} & \textbf{$\alpha^{E}_{2}$ } & \textbf{$\alpha^{E}_{3}$ } & \textbf{$\alpha^{C}$} & \textbf{$\alpha^{N}$} \\ \hline
16    & $(0,1)$                & 0                   &     0                & $(0,1)$                    & $(0,1)$                   \\ \hline
12    & 1                & 0                   & $(0,1)$                   & $(0,1)$                    & $(0,1)$                   \\ \hline
8     & 1                & $(0,1)$                   & 1                   & $(0,1)$                    & $(0,1)$                   \\ \hline
\color{red} 14    &\color{red} $(0,1)$                & \color{red}0                   & \color{red}0                   & \color{red}$(0,1)$                    & \color{red}1                     \\ \hline
\color{red}10    & \color{red}1                & \color{red}0                   & \color{red}$(0,1)$                   &\color{red} $(0,1)$                    & \color{red}1                     \\ \hline
\color{red}6     & \color{red}1                & \color{red}$(0,1)$                   & \color{red}1                   & \color{red}$(0,1)$                    & \color{red}1                     \\ \hline
\color{red} 15    & \color{red} $(0,1)$                &\color{red} 0                   & \color{red}0                   & \color{red}1                    & \color{red}$(0,1)$                     \\ \hline
\color{red}11    & \color{red}1                & \color{red}0                   & \color{red}$(0,1)$                   & \color{red}1                    & \color{red}$(0,1)$                     \\ \hline
\color{red}7     & \color{red}1                & \color{red}$(0,1)$                   & \color{red}1                   & \color{red}1                    & \color{red} $(0,1)$                    \\ \hline
13    & $(0,1)$                & 0                   & 0                   & 1                    & 1                     \\ \hline
9     & 1                & 0                   & $(0,1)$                   & 1                    & 1                     \\ \hline
1     & 1                & 1                   & 1                   & 1                    & 1                     \\ \hline
\end{tabular} 
\quad 
\begin{tabular}{|c|c|c|c|c|c|}
\hline
Combination & \textbf{$\alpha^{E}_{1}$} & \textbf{$\alpha^{E}_{2}$ } & \textbf{$\alpha^{E}_{3}$ } & \textbf{$\alpha^{C}$} & \textbf{$\alpha^{N}$} \\ \hline
16    & $(0,1)$                & 0                   &     0                & $(0,1)$                    & $(0,1)$                   \\ \hline
12    & 1                & 0                   & $(0,1)$                   & $(0,1)$                    & $(0,1)$                   \\ \hline
8     & 1                & $(0,1)$                   & 1                   & $(0,1)$                    & $(0,1)$                   \\ 
\hline
\color{red}15    & \color{red}$(0,1)$                & \color{red}0                   & \color{red}0                   & \color{red}1                    & \color{red}$(0,1)$                     \\ \hline
\color{red}11    & \color{red}1                &\color{red} 0                   & \color{red}$(0,1)$                   &\color{red} 1                    & \color{red}$(0,1)$                     \\ \hline
\color{red}7     & \color{red}1                &\color{red} $(0,1)$                   & \color{red}1                   & \color{red}1                    & \color{red}$(0,1)$                    \\
\hline
\color{red}14    & \color{red}$(0,1)$                &\color{red} 0                   & \color{red}0                   & \color{red}$(0,1)$                    & \color{red}1                     \\ \hline
\color{red}10    & \color{red}1                & \color{red}0                   & \color{red}$(0,1)$                   & \color{red}$(0,1)$                    & \color{red}1                     \\ \hline
\color{red}6     &\color{red} 1                &\color{red} $(0,1)$                   &\color{red} 1                   &\color{red} $(0,1)$                    & \color{red}1                     \\  \hline
13    & $(0,1)$                & 0                   & 0                   & 1                    & 1                     \\ \hline
9     & 1                & 0                   & $(0,1)$                   & 1                    & 1                     \\ \hline
1     & 1                & 1                   & 1                   & 1                    & 1                     \\ \hline
\end{tabular}
\caption{Preference orders for $\mathds{P}(covid)>0.5$ (left) and $\mathds{P}(covid)<0.5$ (right), differences in red}
\label{tab: preforder}
\end{table}
This order is significant and non-trivial, providing insights on how the allocations should change as the total available capacity $\Gamma$ changes. Observe that these preference orders are not ``greedy'' in the sense of first giving capacity to the ED, and then to the Clinics (also see Fig.~\ref{fig:feasibility vs capacity_Ex1}). For example, consider the order for the case of $\mathds{P}(covid)>0.5$: Combinations 16, 12, 8 first prioritize serving patients in the ED, as increasing $\mu^E$ allows us to serve more $s_1$ patients. However, in going from Combination 8 to 14, we stop serving $s_1$ patients with full efficiency; instead, when we have sufficient capacity, we prefer to make the NClinic fully efficient, as this prevents $s_2$ patients from evolving to worse health in the NClinic queue. The block 14, 10, 6 proceeds similarly to the block 16, 12, 8, awarding additional capacity to the ED. Subsequently in block 15, 11, 7, we divert our attention to the Clinic, which we now have sufficient capacity to serve with full efficiency. Finally, in block 13, 9, 1, we serve both clinics with full efficiency, and use any additional capacity in the ED.

We note that for all settings the optimal combinations always satisfy the ordering, but all combinations in the order need not become optimal. For example, Figure \ref{fig:feasibility vs capacity_Ex1} shows one set of optimal combinations for Covid majority,  where combinations appear in order (16, 12, 14, 10, 6,  15, 11, 13, 9, 1), skipping combinations  7 and 8\footnote{
We note that whenever combination 6  is optimal, 8 is too, but not vice versa. This is because we can shift the entire capacity of a fully  working  NClinic in combination 6 to LPED so that the resulting solution is in combination 8, without changing our objective. In such a case, we say that combination 6 is optimal as it has more facilities working than 8. Thus, when combination 8 is optimal, we check if combination 6 is also optimal and if it is, we say combination 6 is the resulting optimal solution. This convention helps in establishing a definitive  preference order.  }. We examine this in greater detail, in Section \ref{sec: NumericalSection}.

This concludes our analysis of the one-period problem. Next, we extend our approach to a multi-period setting, where we allow parameters to change from period to period.  

\section{The Multi-Period Problem}\label{sec: multiperiodanalysis}
The analysis in the preceding sections applied to the case of a single period, during which exogenous parameters (such as the arrival and evolution rates) remained fixed. 
In this section we extend our approach to the case of multiple periods, allowing these exogenous parameters (and our capacity allocation) to change between periods. This enables us to model an evolving pandemic, characterized by changes in the distribution over severity levels (pandemic stage),  patients' evolution rates (possibly due to new variants), risk perceptions (pandemic fatigue) as well as the capacity available (changes in medical staff/equipment). In order to study the multi-period problem, we 
formulate a  dynamic programming problem over $n$ periods, in which we capture the interdependence between periods in a parsimonious fashion.  We prove structural properties of our formulation that allow us to continue to solve the problem efficiently.

Before formalizing the $n$ period problem, we intuitively demonstrate the importance of capturing the interdependence between successive periods. Consider a two-period setting, and let us focus on Covid $s_2$ patients. At the conclusion of Period 1, the Covid Clinic queue, LPED and $H_2^C$ 
may contain a fixed number of Covid $s_2$ patients; the precise number of such patients is indicated by the stationary solution at the end of Period 1.  With a different set of exogenous parameters in Period 2, the hospital may wish to change its capacity allocation, but the remaining $s_2$ patients from Period 1 must still be accounted for. 
Thus, the total flow of Covid $s_2$ patients in Period 2 depends on the Period 2 exogenous parameters, the Period 2 capacity allocation decisions, and the carryover flow of patients remaining from Period 1. To capture this carryover flow in a parsimonious fashion, we model these patients as arriving uniformly throughout Period 2. Accordingly, we account for COVID $s_2$ carryovers by adjusting the input flow $\lambda_{2}$ for Period 2. 

More generally, we adjust all input flows (``effective'' input flows, denoted $\lambda^e_s$ for severity level $s$), accounting for carryovers from the first period. This enables us to capture the interdependence between periods while continuing to leverage our one period analysis. We consequently assume that all buffers in Period 2 (like in Period 1) begin empty, and allow the system to reach stationarity. Our optimal multi-period capacity allocation decisions induce a sequence of optimal $\lambda^e$, given a sequence of $\lambda$. In other words, we wish to find $\mu^f$ in each period so that the resulting sequence of $\lambda^e$ is optimal in the sense of minimizing the total undiscounted loss. The natural way to formulate this problem is as a dynamic program.

\subsection{The Dynamic Programming Formulation}
We begin by explicitly describing how we account for carryovers from the previous period to calculate the effective input rates, $\lambda^e_s$. To illustrate this process, consider the buffer levels of the LPED queue, consisting of $s_2$ and $s_3$ patients (details in Section \ref{sec: FluidModel}). 
Recall that the incoming flow to the LPED is:
\begin{align}
 \text{LP Flow } = (1-p)[\lambda P(s_{2})+ \beta_{2}H_{2}^C+ \beta_{2}H_{2}^N] \alpha^{E}_{2} \ + \ [\lambda P(s_{3}) + \beta_{3}H_{3}] \alpha^{E}_{3} \notag\\
     + \   [p\lambda P(s_{2}) P(covid) + p\beta_{2}H_{2}^C ]  (1-\alpha^{C})\alpha^{E}_{2} +   [p\lambda P(s_{2})(1-P(covid))+ p\beta_{2}H_{2}^N]   (1-\alpha^{N})\alpha^{E}_{2}
\end{align}
Out of this total flow, we  separate the flows for $s_2$ Covid, $s_2$ non-Covid and $s_3$ patients: 
\begin{align*}
 s_3 \text{ LP Flow } & =  [\lambda P(s_{3}) + \beta_{3}H_{3}]\alpha^{E}_{3} \\
   s_2 \text{ LP covid Flow } & = (1-p)[\lambda P(s_{2}) P(covid)+ \beta_{2}H_{2}^C] \alpha^{E}_{2} \\
     & + \   [p\lambda P(s_{2}) P(covid) + p\beta_{2}H_{2}^C ]  (1-\alpha^{C})\alpha^{E}_{2}\\
  s_2 \text{ LP non-Covid Flow } & = (1-p)[\lambda P(s_{2}) (1-P(covid) + \beta_{2}H_{2}^N] \alpha^{E}_{2} \\ & +   [p\lambda P(s_{2})(1-P(covid))+ p\beta_{2}H_{2}^N]   (1-\alpha^{N})\alpha^{E}_{2}
\end{align*}
 The buffer level of $s_3$ patients is given by $Q^{E}_{2,3}\times \mathds{I}(\alpha_{3}^E < 1) $, i.e. the entire queue length if the LPED is partially efficient for $s_3$, as $\tau_2^E< \tau_3^E$.  We find the proportion  of $s_2$ Covid and non-Covid patients in the LPED queue as their flows divided by the total flow, if the LPED is partially efficient for $s_2$.

For the HPED and the Clinics, the queues only consist of one type of patients- $s_1$, and $s_2$ Covid and $s_2$ non-Covid  respectively- so their buffer levels are just the queue lengths. Similarly, the repositories $H_{1}, H_{2}^C, H_{2}^N, H_{3}$ add patients according to their categories. Therefore after we obtain a stationary solution, the total number of patients carrying over to the next period are given by:
\begin{align}
    \text{Buffer } s_1 & = H_{1} + Q^{E}_{1}\\
    \text{Buffer Covid } s_2 & = H_{2}^C+ Q^{E}_{2,3}\times \frac{s_2 \text{ Covid LP Flow }}{s_2 \text{ LP Flow }}\mathds{I}(\alpha_{3}^E = 1) + Q^C\\
     \text{Buffer non-Covid } s_2 & = H_{2}^N+ Q^{E}_{2,3}\times \frac{s_2  \text{ non-Covid LP Flow }}{s_2 \text{ LP Flow }}\mathds{I}(\alpha_{3}^E =  1)+ Q^N\\
     \text{ Buffer } s_3 & = H_{3} + Q^{E}_{2,3}\times \mathds{I}(\alpha_{3}^E < 1)
\end{align}
\begin{lemma} \label{lem: 2periodlemma}
  The buffers are continuous and linear functions of the decision variables corresponding to  optimal solutions.
\end{lemma}
\proof{Proof of Lemma \ref{lem: 2periodlemma}.} In Appendix \ref{appndx: ForTwoPeriodSection}.
\halmos \endproof
 \pfapptrue  
\def\pftwoperiodlemma{
\proof{Proof of Lemma \ref{lem: 2periodlemma}.}
We have already proved that combinations 9-16  reduce to polytopes with linear constraints, and have extreme points as candidate solutions. Moreover, combinations 6-8 also reduce to linear polytopes at optimality. Then, we see that for each combination, the optimal queue and repository lengths are also linear functions of the decision variables. Moreover, it is obvious to see that the buffer lengths are linear in the queue and the repository lengths. We note that the decomposition of the LPED flow into various flows is also linear, restricted to each extreme point. This is because we exactly know the flow entering each facility for every extreme point, linearizing the buffer flows from the LPED.  Thus, the buffers are also linear functions of the decisions for optimal solutions. 
\halmos \endproof
\pfappfalse  
}

Having computed the buffers, we now add them uniformly to the subsequent period's arrival rates. Denoting $\lambda_i^k$ as the original rate of incoming  severity $i$ patients in period $k$, the effective $\lambda^{k,e}$ is computed as follows:
\begin{align*}
    \lambda^{k,e} & = \lambda^k + 
    \frac{\text{Total buffer }}{t}, \text{ i.e.}\\
    \lambda^{k,e}_{1} &  = \lambda_{1}^k + \frac{\text{ Buffer } s_1}{t }\\
    \lambda^{k,e}_{2} \times P(covid)^{k,e} &  = \lambda_{2}^k \times  P(covid)^k + \frac{\text{ Buffer Covid } s_2}{\text{length } t }\\
    \lambda^{k,e}_{2} \times (1-P(covid)^{k,e}) &  = \lambda_{2}^k \times  (1-P(covid)^k) + \frac{\text{ Buffer non-Covid } s_2}{\text{length } t }\\
    \lambda^{k,e}_{3} &  = \lambda_{3}^k + \frac{\text{ Buffer } s_3}{t },
\end{align*}
where $t$  is the length of period $k$. Clearly, the optimal solution depends on $t$: As $t$ increases, the effect of the earlier period diminishes. This is natural as buffers capture the transient effects of moving from one period's stationary solution to another, and as $t$ grows the relative effect of the transient portion diminishes. With this framework, it is possible to study the effects of $t$ on decisions. Indeed, we could have different $t$ for different periods, allowing us to capture long-lasting versus rapidly evolving phases of pandemics.

\textbf{The Objective:} In our one period formulation our objective function penalizes the rate at which patients permanently leave the system, weighted by their severity. We follow the same high-level idea in the multi-period setting, while also weighing periods by their period lengths. Additionally, for the last period, we add a terminal penalty for patients remaining in the queues. (Note that for non-terminal periods, our carryover procedure accounts for patients remaining in the queues.) Accordingly, for a two period problem, our objective function can be written as
\begin{align}
    \text{Global objective }= \ &  [\text{(objective for the 1st period)} \times t_1 + \text{(objective for the 2nd period)} \times t_2 \notag \\
    & +  \sum_{i=1,2,3} \text{(2nd period queue and repositories lengths for severity } s_i) \times  s_i]  \times \frac{1}{t_1+t_2} \label{eqn: 2periodobjective}
\end{align}
where $s_i$ are appropriate weights for severities. This setting can easily be generalized to $n$ periods. 


\subsection{Solution characteristics of the dynamic program}
In this section we analyze the multi-period problem and explore the solution space associated with our dynamic programming formulation. Recall that in the one period setting,  the feasible space can be decomposed into 16 subspaces such that we can restrict our search to the extreme points of these subspaces. The dynamic programming formulation over multiple periods is fundamentally different, due to the interdependence between periods. The optimal solution for the first period will consider the carryover effect for the second and later periods and may accordingly choose a different combination than the myopically optimal one period solution. However, we show that there exists a similar structure in determining the optimal solution for the $n$ period problem, in Theorem \ref{thm: nperiods}. 

\begin{theorem} \label{thm: nperiods}
There exists an optimal solution to the $n$ period problem involving only the extreme points as solutions to all the periods. 
\end{theorem}

Before proving Theorem  \ref{thm: nperiods} formally, we discuss the structure of the optimal solutions in a two period problem, which forms the base case of our induction proof for the $n$ period problem. In one period, we know that the space of each combination can be simplified to systems of  $x$ constraints and $y$ variables and the extreme points within each combination can be enumerated as seen in Theorems \ref{thm: 9to16space}, \ref{thm: 6to8space}. Let $\mathcal{E}^k$ be the set of all extreme points in period $k$. 
\begin{proposition} \label{prop: 2periods} There exists an optimal solution to the 2-period problem involving  extreme points $\in \mathcal{E}^1$ as the solution to period 1 and extreme points $\in \mathcal{E}^2$ as the solution to period 2. 
\end{proposition}
\proof{Proof of Proposition \ref{prop: 2periods}.}
We prove this through a series of lemmas, which we state and explain here; please see Appendix \ref{appndx: ForTwoPeriodSection} for  details.

Let the optimal solution to the 2-period problem be $(sol1, sol2)$ where $sol1$ and $sol2$ are the optimal solutions for Period 1 and 2, respectively. 
 
\begin{lemma} \label{lem: extptperiod2}
  The optimal solution for Period 2, i.e. $sol2$ belongs to $\mathcal{E}^2$.
\end{lemma}
 \pfapptrue  
\def\pfextptperiodtwo{
\proof{Proof for Lemma \ref{lem: extptperiod2}.}
We note that for any type of solution (extreme or non-extreme point) $sol1$ coming from the first period, we can uniquely compute the buffers that are added as inputs to the second period. With these updated inputs, the problem in the second period can be viewed as a single period problem. To be precise, recall that the global objective function for the 2 period problem in equation \eqref{eqn: 2periodobjective} consists of the sum of objective functions of the two periods and the weighted queue lengths at the end of Period 2. After we compute the fixed Period 1 objective using $sol1$, we're left with solving only the second period problem with the remaining Period 2 linear objective. Since $(sol1, sol2)$ is an optimal solution, $sol2$ must be an optimal solution to this updated 2nd period problem. Otherwise, the optimality of $(sol1, sol2)$ is violated as with the same solution $sol1$ in Period 1, we are able to find a better solution in Period 2. Then, using the fact that the optimal solution belongs to $\mathcal{E}$  in a single period problem, we conclude that the solution to the second period $sol2$ is always in $\mathcal{E}^2$. Note that $sol2$ follows Theorem \ref{thm: 6to8space} and \ref{thm: 9to16space} as well. As discussed above, the solution $sol2$ can be located by using the updated inputs from $sol1$ to Period 2 and going over $\mathcal{E}^2$.
\halmos \endproof
\pfappfalse  
}

We next need to prove that $sol1$ also belongs to $\mathcal{E}^1$. We prove this by contradiction. Let us assume that the optimal solution from Period 1 i.e. $sol1 \not \in \mathcal{E}^1$, but $sol2$ is in $\mathcal{E}^2$ within a combination. Let  combination $x\in  \{6,\dots,16\}$ contain the optimal solution $sol1$ in Period 1 and let $sol2$ belong to a combination $y \in \{6, \dots, 16\}$. These assignments are well defined since all capacity allocations uniquely correspond to a combination i.e. the decomposition into 16 combinations holds for each period. We next claim that the  global objective function is linear in terms of $sol1$. Before we do so, we define the ``restricted'' global objective function for period 1 to be the objective function formed as $x$ progresses through combinations, restricted to those  which make $y$  feasible in period 2. Within $x$, some points may result in carryovers that make $y$ infeasible and make some other combination $y'$ optimal, while others may maintain the feasibility of $y$. 
\begin{lemma} \label{lem: linearperiod1}
  For every extreme point $sol2$ in Period 2, the restricted global objective function for Period 1 is linear in the Period 1 variables.
\end{lemma}
 \pfapptrue  
\def\pflinearperiodone{
\proof{Proof of Lemma \ref{lem: linearperiod1}.}
We have already established that $sol2$ is an extreme point of the polytope of $y$. The optimality of the extreme point $sol2$ in $y$ ensures that a small enough deviation in Period 2 inputs will not change its combinatorial type, but will change the objective and all variables in a linear fashion. The fixed combinatorial structure of $y$ then allows us to write the variables of the second period and hence its objective as linear functions of its inputs. This implies that the restricted global objective function is in fact a linear function of the second period inputs.  The second period inputs, moreover, are linear functions of the Period 1 solutions, as proved in Lemma \ref{lem: 2periodlemma}. To summarize, the existence of extreme point $sol2$ fixes the combinatorial structure for Period 2 and hence the restricted global objective for Period 1 can be written as a linear function of only Period 1 variables. 
\halmos \endproof
\pfappfalse  
}

With this linear restricted global objective function in hand, let us now consider the optimal solution $sol1$ in the first period. To further understand the objective, we may think of $y$ describing the rules which guide $sol1$ to form a linear global objective for Period 1. 
To be precise, the restricted problem for Period 1 (corresponding to combination $y$ being optimal in Period 2) should have a new restricted objective associated with $y$ in Period 2 and additional constraints for the feasibility of $y$, which may become tight for some interior (i.e. non-extreme) points in $x$. We refer to these interior points as `switching points' and collect these points in a set denoted as $\mathcal{S}^1$. The set consists of some points on edges of the period 1 polytope, each satisfying a unique set of feasibility constraints at equality, and corresponding to a specific combination $y$ in Period 2. We  now characterize the possible optimal solutions for Period 1:
\begin{lemma} \label{lem: candidateoptsolnspd1}
  The optimal solution for a combination in Period 1 belongs to $\mathcal{E}^1 \cup \mathcal{S}^1$. 
\end{lemma}
 \pfapptrue  
\def\pfcandidateoptsolnspdone{
\proof{Proof of Lemma \ref{lem: candidateoptsolnspd1}.}
The linear global objective makes sure that the optimal solution from Period 1 has full rank. If we prove that there exists a unique linear constraint that makes $y$ feasible, we can easily see that the candidate solutions are the $\mathcal{E}^1$ for $x$ as well as the `switching' points on the edges $\mathcal{S}^1$, which have a tight  feasibility constraint. This is supported by the fact that the switching points need to have rank $(n-1)$ i.e. edge for the polytope with $n$ variables, so that the linear feasibility constraint being tight implies full rank.

To see the uniqueness of this constraint, let us take $y=10$ as an example. Combination 10 is characterized by $(\alpha_1^E, \alpha_2^E, \alpha_3^E, \alpha^{C}, \alpha^{N}) = (1,0,c, c', 1)$, where $c, c' \in[0,1)$ are some constants. This combination is feasible only if HPED and the NClinic can function with full efficiencies. This happens only when (the  incoming flow to the facility) + (feedback flow from the corresponding repository) matches the capacity allocated to the facility. More formally, we go back to the linear formulation corresponding to combination 10, and observe that the constraints for $H_i$ are linear and can be simplified to contain only the parameters and $\mu^i$'s. We can easily check that the $H_i\geq0$ constraints are satisfied in these solutions as these contain positive sums of parameters. The feasibility of the combination then only depends on whether $\Gamma\geq \mu^N+\mu^E_1 \geq $ (total incoming flow to NClinic and HPED), which is a linear function of the parameters and the $\mu_i$'s. If this constraint is satisfied, then we have enough capacity to make combination 10 feasible. Intuitively, we can check if $\Gamma$ is sufficient to serve the facilities with full efficiency, by allocating all the capacity to those and if so, the combination is always feasible. The same argument can be extended to all combinations. This includes combinations 6-8 as well, since the feasibility only requires that some facilities be fully efficient, disregarding the non-linear partially efficient part. \halmos \endproof
\pfappfalse  
}

To complete the proof, we only need to argue that the switching points in $\mathcal{S}^1$   are always dominated by the extreme points $\mathcal{E}^1$. We prove that this is the case in Lemma~\ref{lem: thefinallemma}, through observing that either the combinations themselves are not optimal when a feasibility constraint is tight (such that we are at a switching point in period 1), or that moving in a direction from a feasible to an infeasible point (i.e. from an extreme point toward a switching point) does not improve the objective. 
\begin{lemma} \label{lem: thefinallemma}
 There is always an extreme point in  $\mathcal{E}^1$  that is the optimal solution in Period 1. 
\end{lemma}
 \pfapptrue  
\def\pfthefinallemma{
\proof{Proof of Lemma \ref{lem: thefinallemma}.} 
 Firstly, Lemma \ref{lem: candidateoptsolnspd1} says that for each combination in Period 2, there exists  a unique linear constraint for its feasibility. We need to consider the extreme points for combination $x'$ in Period 1, corresponding to a combination $y'$ in Period 2 if and only if $y'$ is feasible and infeasible for non empty sets of extreme points in $x'$: If all extreme points are infeasible for $y'$, then the entire combination $x'$ (polytope) is infeasible. If all extreme points within $x'$ are feasible for $y'$, then the linear constraint is irrelevant to the combination, and thus it suffices to check the extreme points in $\mathcal{E}^1$. The only interesting case is where the linear feasibility constraint intersects with the feasible region (polytope) of a combination.

We have already seen that the switching points $\mathcal{S}^1$  are  candidates in $x$ only when there exists at least one extreme point in $x$ that yields a carryover such that $y$ is feasible in Period 2 and at least one yielding a carryover that makes $y$ infeasible. Then the switching point lies on the edge connecting these two extreme points. We now argue that for every optimal combination $y$ in the second period,  the  global objective cannot decrease while moving from the feasible point to the infeasible point in combination $x$. This would then imply that the extreme point will have a better global objective and hence checking only the extreme points suffices, for every possible optimal combination in Period 2. Depending on various combinations for Period 2, we have two types of arguments establishing that the extreme points are better than the switching points.

Firstly, we consider the linear feasibility constraint for a combination, say 10. Let $e_f$ and $e_i$ denote the feasible and infeasible extreme points in combination $x$, corresponding to combination 10 for Period 2. While moving from $e_f$ to $e_i$, we will hit the feasibility constraint when the Period 1 carryovers are such that Period 2 can barely manage to serve HPED and the NClinic with full efficiency, after allocating all its capacity to these two facilities. This implies that while moving from the feasible $e_f$, we have increased the carryover of either $s_1$ and/or $s_2$ non-Covid patients by internally shifting the capacity allocation in Period 1. Increasing the $s_1$ carryover can never improve our global objective, as this indicates an increase in Period 1 and an increase or no change in the Period 2 objective, as it shifts its capacity deficit to treating the extra $s_1$ flow. The increase may be associated with the increase in some repository levels as we shift our capacity. Thus, keeping the same capacity allocation for $s_1$ i.e. HPED  in Period 1, we must have increased the $s_2$ non-Covid carryover by shifting its capacity to $s_2$ Covid or $s_3$. In both case, the objective for  Period 2 doesn't change or increases, while the objective for Period 1 doesn't change or strictly increases if $s_3$ carryover decreases. To summarize, changing the carryover type for Period 2 may not change its objective, when it shifts the capacity deficit to match with the increasing carryover, but doing so increases the Period 1 objective as we increase the higher severity carryovers.  Thus, we never improve while moving from $e_f$ to $e_i$. Similar arguments work for combinations 9, 11 and 12 as well, in Period 2. 

The second type of dominance argument rests on the fact that the combination $y$ itself cannot be optimal if we indeed reach such a switching point.  Thus, again, checking the extreme points is enough. Let's take combination 8 as an example, where its linear feasibility constraint being tight implies that all capacity is distributed towards $s_1$ and $s_3$ together in Period 2. In this case, we're prioritizing $s_3$ over $s_2$, thereby making $y$ not optimal in this case. A similar argument works for combinations 6, 7, 13, 14, 15 as well. As proved earlier in Lemma \ref{lem: existencelemma}, combination 16 is always feasible and thus we never have to worry its feasibility constraint being tight.
\halmos \endproof
\pfappfalse  
}

Thus, we have proved that only extreme points $\mathcal{E}^1$  can be optimal in Period 2 and every optimal Period 2 combination requires extreme points as optimal solutions in Period 1 as well.
\halmos \endproof

We have proved that the two period problem's structure enables us to find the optimal solution by checking over enumerable extreme points. Fortunately, we can prove that this structural result is not restricted to the two period problem, but can naturally be extended to the multi period problem. We use a simple induction argument to generalize this result, as shown in proof to Theorem \ref{thm: nperiods}.
\proof{Proof of Theorem \ref{thm: nperiods}.} 
We prove this theorem using an induction argument. We have already proved in Proposition \ref{prop: 2periods} that the optimal solution to a 2 period problem involves extreme points in $\mathcal{E}^1$ and $\mathcal{E}^2$  as solutions in period 1a and 2, respectively.  

We begin by assuming that there exists an optimal solution to any $n$ period problem, having only extreme points for all the $n$ periods. Now consider an $n+1$ period problem. For any type of optimal solution in the first period, we can add the corresponding carryovers to the 2nd period and now the problem is reduced to $n$ periods. Using the induction hypothesis,  for periods $2,..n+1$, 
the optimal solutions for each are extreme points. Next, using the knowledge of the 2nd period extreme point, we can take the Period 1 carryovers and find $H_i$ and $Q_i$ for Period 2, which are linear in the Period 1 carryovers, which in turn linearly impact the Period 2 carryovers and thereby Period 3 inputs. We can similarly convert Period 3 inputs to period 4 inputs and so on. Thus, for any period, we can form a linear objective involving parameters and the first period decisions (through period 1 carryovers, themselves linear in the decisions). Then, using Lemma \ref{lem: thefinallemma}, for any optimal combination in this chain, a switching point is never optimal.  It is the property of the later combinations, rather than the Period 1 combination, as the arguments in Lemma \ref{lem: thefinallemma} show.  Thus, we should not move from extreme points in the first period as well, thereby extending the result of $n$ periods to $n+1$ periods.   
\Halmos \endproof 

Thus, we prove that the dynamic programming formulation for any multi period problem is computationally tractable and does not require us to scan the entire space of each period.

\section{Numerical Results and Insights} \label{sec: NumericalSection}
In this section, we complement our analytical results with numerical experiments that illustrate the utility of our framework to shed insight on how capacity should be managed in different phases of a pandemic. In particular, we study:
\begin{enumerate}
    \item \textbf{Two one-period problems} that represent contrasting pandemics: Example 1 is an early wave of the pandemic (pre-vaccinations) with a majority of Covid patients and higher evolution rates. Example 2 is a later wave after vaccination of a significant part of the population, with the patient population being majority  non-Covid and skewed towards lower severities with health status improving faster, and lower Covid spread. 
    \item \textbf{Two three-period problems} that represent different pandemic evolution profiles, or equivalently, different stages of a single pandemic. The first setting (Example 3; see Figure~\ref{fig: set3set4}), represents the early stages (pre-vaccination) of a pandemic, with low initial load that increases and then gently tapers off as people take precautions. The second setting (Example 4; see Figure~\ref{fig: set3set4}) represents a later stage wave, starting with a high peak but good vaccination coverage and effective treatments reducing high severity incidence in following periods. While overall load is high in periods 1 and 2, it tapers off in the third period. 
\end{enumerate}

\subsection{One-period problems}

\begin{table}[]
\footnotesize	
\centering
\begin{tabular}{|c|c|c|}
\hline
Parameters                                                                    & One-Period Example 1         & One-Period Example 2         \\ \hline
$\lambda$                                                                     & 2                            & 2                            \\ \hline
$s_{1}, s_{2}, s_{3}$                                                         & 0.5, 0.25, 0.125             & 0.5, 0.25, 0.125             \\ \hline
$\lambda_{1}, \lambda_{2}, \lambda_{3}$                                       & 0.6, 1.2, 0.2                & 0.4, 1.2, 0.4                \\ \hline
$p$                                                                           & 0.7                          & 0.7                          \\ \hline
$\mathds{P}(covid)$                                                           & 0.85                         & 0.4                          \\ \hline
$r$                                                                           & 0.2                          & 0.2                          \\ \hline
$\delta_{10},  \delta_{21}, \delta_{32},\delta_{12},\delta_{23}, \delta_{34}$ & 0.3, 0.3, 0.2, 0.3, 0,3, 0.2 & 0.1, 0.1, 0.1, 0.4, 0,4, 0.4 \\ \hline
$\beta_{1}, \beta_{2}, \beta_{3}$                                             & 0.25, 0.25, 0.25             & 0.25, 0.25, 0.25             \\ \hline
$\sigma_{1}, \sigma_{2}, \sigma_{3}$                                          & 0.2, 0.2, 0.2                & 0.2, 0.2, 0.2                \\ \hline
$\Gamma$ range                                                                & (0.3, 2)                     & (0.3, 2)                     \\ \hline
\end{tabular}
\caption{Parameter settings for one-period problem instances} 
\label{tab: paramsforoneperiod}
\end{table}

Consider the two one-period problems with settings as described in Table \ref{tab: paramsforoneperiod}, which we refer to as Example 1 and Example 2. Example 1 (also seen in Section \ref{sec:single_period_solutions}) represents an early wave of a pandemic, with a majority of Covid patients and higher evolution rates, along with slightly higher disease intensity. By contrast, the majority of patients in Example 2 are non-Covid, and the patient population is skewed towards lower severities;  and health status improves faster. The former example corresponds to a setting before vaccinations are discovered; and the latter to a setting in which a large part of the population is vaccinated.

In both settings, we vary total capacity $\Gamma$ from 0.3 to 2 to understand the implications this change has for optimal capacity allocations. Figure \ref{fig: sec6ex2fig1} shows the sets of feasible and optimal solutions for the range of $\Gamma$ values. We show the variable values corresponding to the optimal policies in Figure \ref{fig: sec6comparisonsPD1} (efficiencies $\alpha^f_i$, capacity allocations $\mu^f_i$) and in Figure \ref{fig: sec6comparisonsAppndx} (facility queue $Q^f_i$, repository lengths $H_i^f$) in Appendix \ref{appndx: ForNumericalSection}. Observe that the capacities and efficiencies (Figure \ref{fig: sec6comparisonsPD1}) change non-linearly and non-monotonically with increasing $\Gamma$. Correspondingly, we also see facilities fluctuate between being non-functional, partially efficient and fully efficient. This pattern may be very different from that of the capacities. Further discussion on allocations and efficiencies is in Appendix \ref{appndx: ForNumericalSection}.

\begin{figure}
\centering
\begin{subfigure}[b]{0.3\textwidth}
         \centering        \includegraphics[width=6cm]{feasibility_vs_capacity_Ex1}
         \caption{Example 1 (pre-vaccination, Covid majority)}
         \label{fig: sec6ex1order}
     \end{subfigure}
     \hspace{1cm}
     \begin{subfigure}[b]{0.3\textwidth}
         \centering
         \includegraphics[width=6cm]{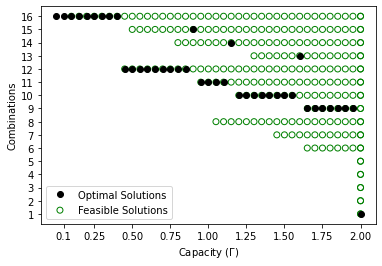}
         \caption{Example 2 (many vaccinated, non-Covid majority)}
         \label{fig: sec6ex2order}
     \end{subfigure}
\caption{The progression of optimal solutions with varying capacity} \label{fig: sec6ex2fig1}
\end{figure}

\textbf{Preference Order: } From Figure \ref{fig: sec6ex2fig1}, observe that all combinations are feasible at some value of capacity for Example 1, although 7 and 8 never become optimal. At the lowest capacity, which is insufficient for any facility operating at full efficiency, only combination 16 is feasible and hence optimal, in line with Lemma \ref{lem: existencelemma}. As capacity increases, combination 14 becomes feasible (but not immediately optimal as $s_2$ patients are not prioritized, and Covid majority implies larger capacity is required to fully serve Covid patients) and it now becomes possible to serve the NClinic at full efficiency. Also, combination 12 becomes optimal as soon as it becomes feasible, i.e., we serve HPED with full efficiency and then allocate remaining capacity to Clinics. Next, 14 briefly dominates 12 as the increasing evolution flows from the partially efficient NClinic become costlier than only a fraction of $s_1$ patients being served at HPED. Observe here that the capacity allocation changes non-uniformly with increasing $\Gamma$. By decreasing the HPED capacity, more facilities are served with full efficiency. Next, combination 10 becomes optimal as soon as it becomes feasible, as we can serve both NClinic and HPED with full efficiency, spending the remaining on LPED. Combination 6 is optimal next as the LPED serves the small number of $s_3$ patients fully and allocates the remaining capacity to $s_2$. As there is now enough capacity to serve Clinics fully efficiently, combination 15 becomes optimal next. 
Combination 13 next briefly becomes optimal (although feasible earlier) when we can fully serve both the Clinics (stopping evolution flows) but sacrificing efficiency at HPED. 
Finally, combination 9 is optimal when we have enough capacity to serve both the Clinics and the HPED. As expected, combinations 1-5 become optimal when $\lambda = \Gamma$, i.e. when the capacity equals the total incoming flow, leading to an objective value (cost) of 0. By contrast, for Example 2, all combinations except 6, 7 and 8 become optimal at some capacity; and the preference order for the non-Covid majority is followed. Since Example 1 has Covid majority, the group of combinations (14, 10, 6) (favoring NClinic) appears before the group (15,11,7) (favoring Clinic).

\pfapptrue  
\def\pfsettingsonetable{
\begin{table}[]
\footnotesize	
\centering
\begin{tabular}{l|r|rrrr|rrrrr|rrrr|rrrr}
\toprule
{} &   Obj &  $\mu_{1}^{E}$ &  $\mu_{2,3}^{E}$ &   $\mu^{C}$ &   $\mu^{N}$ & $\alpha_{1}^{E}$ &  $\alpha_{2}^{E}$ &  $\alpha_{3}^{E}$ & $\alpha^{C}$ &  $\alpha^{N}$ &  $Q^{E}_{1}$ &    $Q^{E}_{2,3}$ &    $Q^{C}$ &    $Q^{N}$ &    $H_{1}$ &   $H_{2}^C$ &   $H_{2}^N$ &    $H_{3}$ \\
Soln &       &       &       &       &       &        &         &         &       &       &       &       &      &      &       &       &       &       \\
\midrule
16   &  0.82 &  0.30 &  0&  0&  0&   0.48 &     0&    0&  0&  0&  0.10 &  0&  0&  0&  1.12 &  1.54 &  1.03 &  1.01 \\
16   &  0.78 &  0.35 &  0&  0&  0&   0.58 &     0&    0&  0&  0&  0.12 &  0&  0&  0&  1.02 &  1.53 &  1.02 &  1.01 \\
16   &  0.74 &  0.40 &  0&  0&  0&   0.68 &     0&    0&  0&  0&  0.14 &  0&  0&  0&  0.92 &  1.52 &  1.01 &  1\\
16   &  0.70 &  0.45 &  0&  0&  0&   0.80 &     0&    0&  0&  0&  0.16 &  0&  0&  0&  0.82 &  1.51 &  1&  1\\
16   &  0.66 &  0.50 &  0&  0&  0&   0.92 &     0&    0&  0&  0&  0.18 &  0&  0&  0&  0.73 &  1.50 &  1&  1\\
12   &  0.62 &  0.54 &  0&  0&  0.01 &   1&     0&    0&  0&  0.02 &  0&  0&  0.01 &  0.01 &  0.65 &  1.48 &  0.97 &  0.99 \\
12   &  0.60 &  0.54 &  0&  0.01 &  0.05 &   1&     0&    0&  0.01 &  0.08 &  0&  0&  0.01 &  0.07 &  0.63 &  1.47 &  0.90 &  0.98 \\
12   &  0.58 &  0.55 &  0&  0.08 &  0.02 &   1&     0&    0&  0.08 &  0.04 &  0&  0&  0.10 &  0.03 &  0.61 &  1.34 &  0.94 &  0.97 \\
12   &  0.56 &  0.56 &  0&  0.12 &  0.03 &   1&     0&    0&  0.12 &  0.04 &  0&  0&  0.16 &  0.03 &  0.59 &  1.26 &  0.94 &  0.95 \\
12   &  0.54 &  0.56 &  0&  0.06 &  0.12 &   1&     0&    0&  0.06 &  0.19 &  0&  0&  0.09 &  0.16 &  0.56 &  1.36 &  0.76 &  0.94 \\
12   &  0.52 &  0.57 &  0&  0.15 &  0.08 &   1&     0&    0&  0.15 &  0.13 &  0&  0&  0.20 &  0.11 &  0.54 &  1.20 &  0.83 &  0.93 \\
14   &  0.50 &  0.35 &  0&  0&  0.50 &   0.67 &     0&    0&  0&  1&  0.12 &  0&  0&  0&  0.65 &  1.47 &  0.08 &  0.85 \\
14   &  0.46 &  0.40 &  0&  0&  0.50 &   0.79 &     0&    0&  0&  1&  0.14 &  0&  0&  0&  0.55 &  1.46 &  0.08 &  0.85 \\
14   &  0.42 &  0.46 &  0&  0&  0.49 &   0.93 &     0&    0&  0&  1&  0.16 &  0&  0&  0&  0.45 &  1.45 &  0.07 &  0.85 \\
10   &  0.39 &  0.48 &  0&  0.02 &  0.49 &   1&     0&    0&  0.02 &  1&  0&  0&  0.03 &  0&  0.39 &  1.40 &  0.07 &  0.84 \\
10   &  0.37 &  0.49 &  0&  0.07 &  0.49 &   1&     0&    0&  0.07 &  1&  0&  0&  0.09 &  0&  0.37 &  1.32 &  0.06 &  0.83 \\
10   &  0.35 &  0.50 &  0&  0.11 &  0.49 &   1&     0&    0&  0.11 &  1&  0&  0&  0.15 &  0&  0.35 &  1.24 &  0.06 &  0.82 \\
15   &  0.32 &  0.41 &  0&  0.74 &  0&   0.87 &     0&    0&  1&  0&  0.14 &  0&  0&  0&  0.36 &  0.09 &  0.96 &  0.78 \\
11   &  0.28 &  0.46 &  0&  0.74 &  0.01 &   1&     0&    0&  1&  0.01 &  0&  0&  0&  0.01 &  0.27 &  0.08 &  0.94 &  0.77 \\
11   &  0.26 &  0.46 &  0&  0.74 &  0.05 &   1&     0&    0&  1&  0.08 &  0&  0&  0&  0.07 &  0.25 &  0.08 &  0.85 &  0.76 \\
11   &  0.24 &  0.47 &  0&  0.74 &  0.09 &   1&     0&    0&  1&  0.15 &  0&  0&  0&  0.12 &  0.23 &  0.08 &  0.77 &  0.75 \\
11   &  0.22 &  0.48 &  0&  0.74 &  0.14 &   1&     0&    0&  1&  0.22 &  0&  0&  0&  0.18 &  0.20 &  0.08 &  0.69 &  0.73 \\
11   &  0.20 &  0.48 &  0&  0.73 &  0.18 &   1&     0&    0&  1&  0.30 &  0&  0&  0&  0.24 &  0.18 &  0.07 &  0.61 &  0.72 \\
11   &  0.18 &  0.49 &  0&  0.73 &  0.22 &   1&     0&    0&  1&  0.38 &  0&  0&  0&  0.30 &  0.16 &  0.07 &  0.53 &  0.71 \\
13   &  0.16 &  0.28 &  0&  0.73 &  0.49 &   0.61 &     0&    0&  1&  1&  0.10 &  0&  0&  0&  0.27 &  0.07 &  0.05 &  0.63 \\
13   &  0.12 &  0.33 &  0&  0.73 &  0.49 &   0.76 &     0&    0&  1&  1&  0.12 &  0&  0&  0&  0.17 &  0.06 &  0.04 &  0.63 \\
13   &  0.08 &  0.38 &  0&  0.73 &  0.49 &   0.92 &     0&    0&  1&  1&  0.13 &  0&  0&  0&  0.07 &  0.06 &  0.04 &  0.63 \\
9    &  0.06 &  0.40 &  0.03 &  0.73 &  0.49 &   1&     0&    0.06 &  1&  1&  0&  0.02 &  0&  0&  0.02 &  0.05 &  0.03 &  0.58 \\
9    &  0.05 &  0.40 &  0.08 &  0.73 &  0.49 &   1&     0&    0.16 &  1&  1&  0&  0.06 &  0&  0&  0.02 &  0.04 &  0.03 &  0.50 \\
9    &  0.04 &  0.40 &  0.14 &  0.73 &  0.48 &   1&     0&    0.28 &  1&  1&  0&  0.10 &  0&  0&  0.02 &  0.03 &  0.02 &  0.42 \\
9    &  0.03 &  0.40 &  0.19 &  0.73 &  0.48 &   1&     0&    0.40 &  1&  1&  0&  0.14 &  0&  0&  0.01 &  0.03 &  0.02 &  0.33 \\
9    &  0.02 &  0.40 &  0.24 &  0.72 &  0.48 &   1&     0&    0.54 &  1&  1&  0&  0.18 &  0&  0&  0.01 &  0.02 &  0.01 &  0.25 \\
9    &  0.02 &  0.40 &  0.29 &  0.72 &  0.48 &   1&     0&    0.68 &  1&  1&  0&  0.22 &  0&  0&  0.01 &  0.01 &  0.01 &  0.17 \\
9    &  0.01 &  0.40 &  0.35 &  0.72 &  0.48 &   1&     0&    0.83 &  1&  1&  0&  0.26 &  0&  0&  0&  0.01 &  0&  0.08 \\
\bottomrule
\end{tabular}
   \caption{Example 1}
    \label{tab: settings1table}
\end{table}
\pfappfalse  
}

\begin{figure}
\centering
\begin{subfigure}[b]{0.3\textwidth}
         \centering
         \includegraphics[width=6cm]{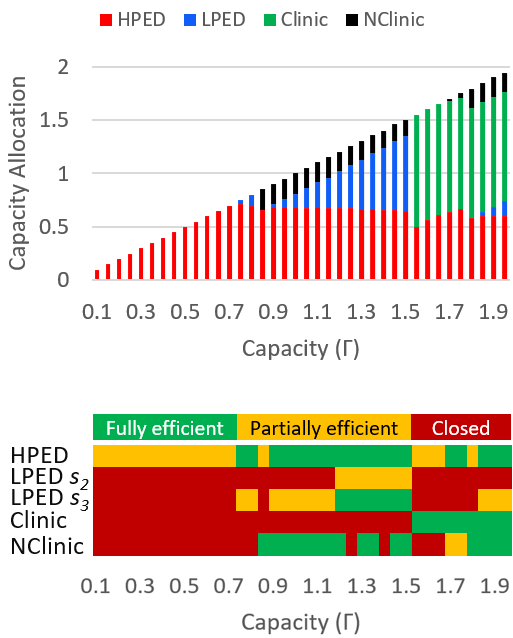}
         \caption{Example 1 (pre-vaccination, Covid majority)}
         \label{fig: sec6comparisons1_1}
     \end{subfigure}
     \hspace{1cm}
     \begin{subfigure}[b]{0.3\textwidth}
         \centering
         \includegraphics[width=6cm]{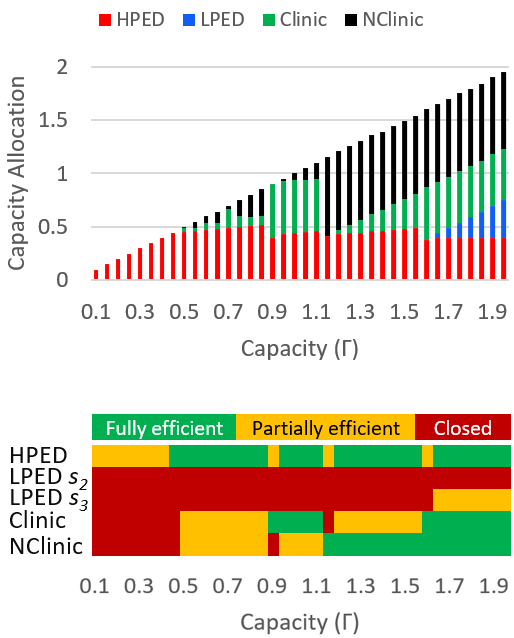}
         \caption{Example 2 (many vaccinated, non-Covid majority)}
         \label{fig: sec6comparisons1_2}
     \end{subfigure}
\caption{Optimal Allocations and Efficiencies with varying capacity} \label{fig: sec6comparisonsPD1}
\end{figure}

\pfapptrue  
\def\pfsecsixcomparisonsAppndx{
\begin{figure}
\centering
\includegraphics[width=6cm]{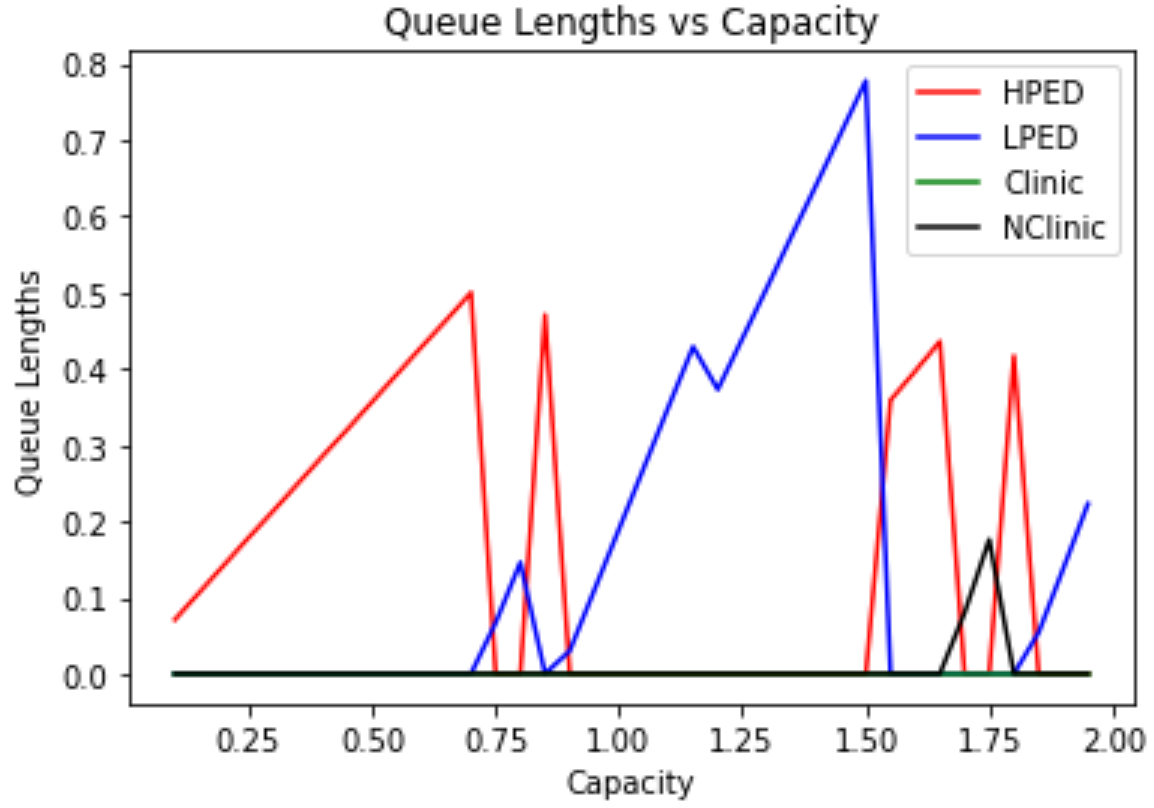}
\includegraphics[width=6cm]{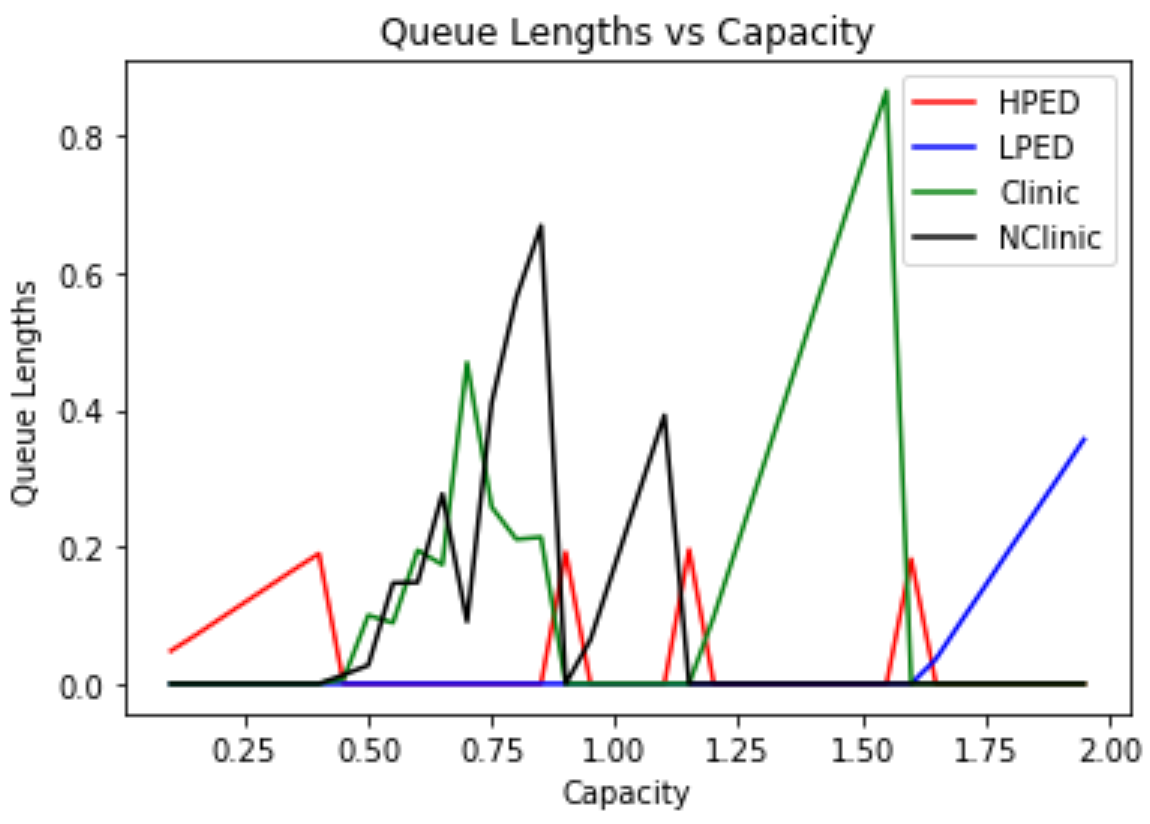}
\includegraphics[width=6cm]{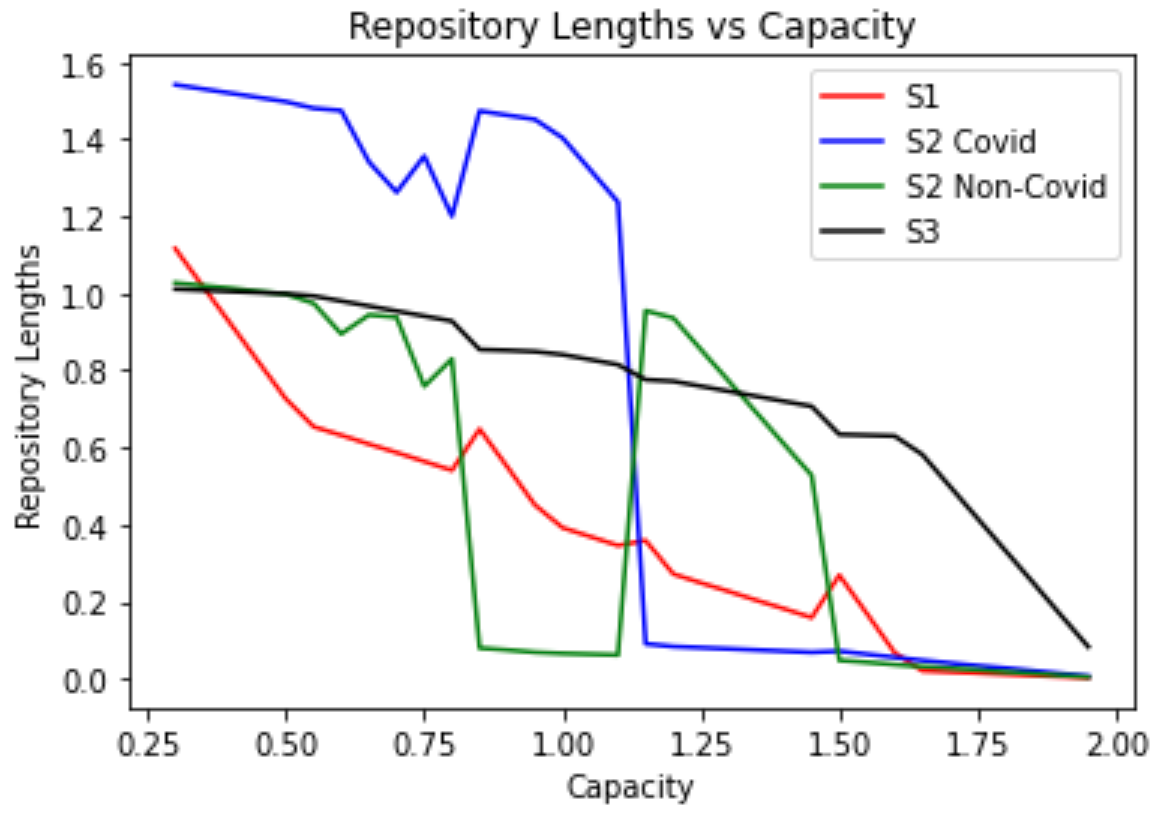}
\includegraphics[width=6cm]{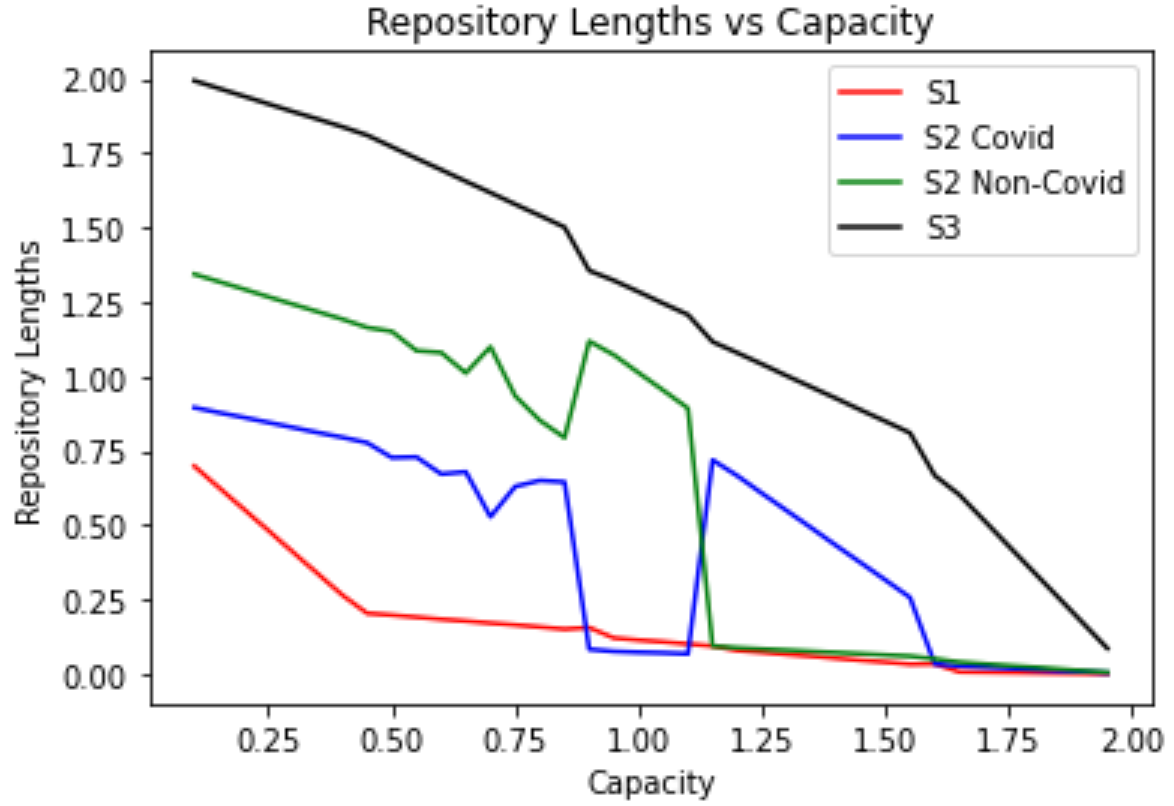}
\caption{The progression of optimal solutions with varying capacity, for Example 1 and 2} \label{fig: sec6comparisonsAppndx}
\end{figure}
\pfappfalse  
}

\pfapptrue  
\def\pfQLengths{
\textbf{Optimal capacity allocation:} As seen in case of both Examples 1 and 2, the capacity allocation at all the facilities  changes non-uniformly with increasing $\Gamma$. For instance, if we track the capacity allocated to HPED in Example 1, it starts with increasing linearly when combination 16 is optimal, and the slope changes when we switch to combination 12. The optimal policy sometimes indicates to decrease the capacity currently allocated to facility $f$ as $\Gamma$ increases, so as to accommodate more facilities with full efficiency. This phenomenon is observed, for example, when combination 14 becomes optimal, decreasing the capacity allocated to HPED (causing a dip at $\Gamma = 0.85$) in order to serve the NClinic with full efficiency (thereby reducing its major $s_2$ evolution flow).

Since Example 1 has Covid majority, the group of combinations (14, 10, 6) (favoring NClinic) appears before the group (15,11,7) (favoring Clinic). This ordering is reversed for Example 2, consistent with the non-Covid majority order.  Figure \ref{fig: sec6comparisonsPD1} shows this change in groups at $\Gamma = 1.5$ for Example 1 and at $\Gamma = 1.1$ for Example 2 as the optimal policy  shifts capacities between the two clinics. 
\\
\textbf{Efficiencies:} Similar to the capacities, the efficiencies also follow a non-uniform pattern with respect to $\Gamma$, and these can be understood with reference to the preference order in Figure \ref{fig: sec6ex2fig1}. As the allocation changes, we also see facilities fluctuate between being non-functional, partially efficient and fully efficient. This pattern may be very different from that of the capacities, as the capacity allocation as a whole determines the flows entering each facility. As discussed for Example 1, the capacity at the Clinics influences the evolution flow at the ED, as well as the repository flows from the $H_i$'s, thereby influencing the total incoming flow at the HPED (and its efficiency $\alpha_1^E$). Owing to these feedback flows, there are 4 ranges of $\Gamma$ where $\alpha_1^E=1$ while the capacity needed to do so decreases as $\Gamma$ increases. For Example 2, however,  the majority of $s_1$ population bags the most capacity, with the optimal policy  making comparatively smaller deviations in $\alpha_1^E$.  Near $\Gamma = 2$ in both the examples, we see combination 9 being optimal with LPED gaining efficiency while all the other facilities work with full efficiency, i.e., the LPED in both cases is the last facility to receive capacity. 


\textbf{Queue Lengths}:  We also track the facility queue and repository lengths as the optimal solution changes, as shown in Figure \ref{fig: sec6comparisonsAppndx}. The facility queue lengths largely follow the same pattern as that of the capacities, as those are linear functions of the corresponding $\mu_i^f$'s, for each combination. The repository lengths, however, are linear functions of each other as well as multiple capacity allocations, following an entirely different pattern. The repository lengths depict the general severity distribution of the public within hospital's area of influence and are also used to compute our total objective, which is piecewise linear. 
\pfappfalse  
}

\subsection{Three-period problems}

 \pfapptrue  
\def\pfthreeperiodsettings{
\begin{table}[]
\footnotesize	
\centering
\begin{tabular}{|c|c|c|}
\hline
 Parameters                                                                    & 3 Period Example 3                                                                                                                                                        & 3 Period Example 4                                                                                                                                                        \\ \hline
 $\lambda$                                                                     & 2.5; 4; 3.5                                                                                                                                                               & 4; 4; 3.15                                                                                                                                                                \\ \hline
 $s_{1}, s_{2}, s_{3}$                                                         & \begin{tabular}[c]{@{}c@{}}(0.75, 0.5, 0.25);  (0.75, 0.5, 0.25);  \\ (0.75, 0.5, 0.25)\end{tabular}                                                                      & \begin{tabular}[c]{@{}c@{}}(0.75, 0.5, 0.25);  (0.75, 0.5, 0.25); \\  (0.75, 0.5, 0.25)\end{tabular}                                                                      \\ \hline
 $\lambda_{1}, \lambda_{2}, \lambda_{3}$                                       & \begin{tabular}[c]{@{}c@{}}(0.5, 0.5, 1.5); (2, 1.4, 0.6); \\ (1.4, 1.05, 1.05)\end{tabular}                                                                              & \begin{tabular}[c]{@{}c@{}}(1.8, 1.6, 0.6); (0.8, 2.4, 0.8); \\ (0.63, 0.63, 1.89)\end{tabular}                                                                           \\ \hline
 $p$                                                                           & 0.25; 0.7; 0.7                                                                                                                                                           & 0.25; 0.7; 0.7                                                                                                                                                            \\ \hline
$\mathds{P}(covid)$                                                           & 0.2; 0.8; 0.6                                                                                                                                                             & 0.85; 0.7; 0.6                                                                                                                                                            \\ \hline
 $r$                                                                           & 0.2; 0.4; 0.4                                                                                                                                                             & 0.6; 0.4; 0.2                                                                                                                                                             \\ \hline
 $\delta_{10},  \delta_{21}, \delta_{32},\delta_{12},\delta_{23}, \delta_{34}$ & \begin{tabular}[c]{@{}c@{}}(0.2, 0.2, 0.2, 0.1, 0.1, 0.1); \\ (0.1, 0.1, 0.1, 0.1, 0.1, 0.1); \\ (0.1, 0.1, 0.1, 0.2, 0.2, 0.2)\end{tabular}                              & \begin{tabular}[c]{@{}c@{}}(0.2, 0.2, 0.2, 0.1, 0.1, 0.1); \\ (0.1, 0.1, 0.1, 0.2, 0.2, 0.2); \\ (0.1, 0.1, 0.1, 0.2, 0.2, 0.2)\end{tabular}                              \\ \hline
 $\beta_{1}, \beta_{2}, \beta_{3}$                                             & \begin{tabular}[c]{@{}c@{}}$(0.125, 0.125, 0.125);  (0.125, 0.125, 0.125)$; \\ ($0.125, 0.125, 0.125)$\end{tabular} & \begin{tabular}[c]{@{}c@{}}$(0.125, 0.125, 0.125);  (0.125, 0.125, 0.125)$; \\ ($0.125, 0.125, 0.125)$\end{tabular} \\ \hline
 $\sigma_{1}, \sigma_{2}, \sigma_{3}$                                          & \begin{tabular}[c]{@{}c@{}}$(0.2, 0.2, 0.2); (0.2, 0.2, 0.2)$; \\ ($0.2, 0.2, 0.2) $\end{tabular} & \begin{tabular}[c]{@{}c@{}}$(0.2, 0.2, 0.2); (0.2, 0.2, 0.2)$; \\ ($0.2, 0.2, 0.2) $\end{tabular}  \\ \hline
 $t_1, t_2, t_3$                                                                      & 5, 5, 5                                                                                                                                                                 & 5, 5, 5                                                                                       \\ \hline
$\Gamma$                                                                      & 1.75; 2; 3                                                                                                                                                                & 2; 2.5; 3                                                                                                                                                                 \\ \hline
\end{tabular}
\caption{Parameter settings for 3 period problem instances} 
\label{tab: paramsforthreeperiod}
\end{table}

\pfappfalse  
}
\begin{figure}
\centering
\begin{subfigure}[b]{0.3\textwidth}
         \centering
         \includegraphics[width=5.5cm]{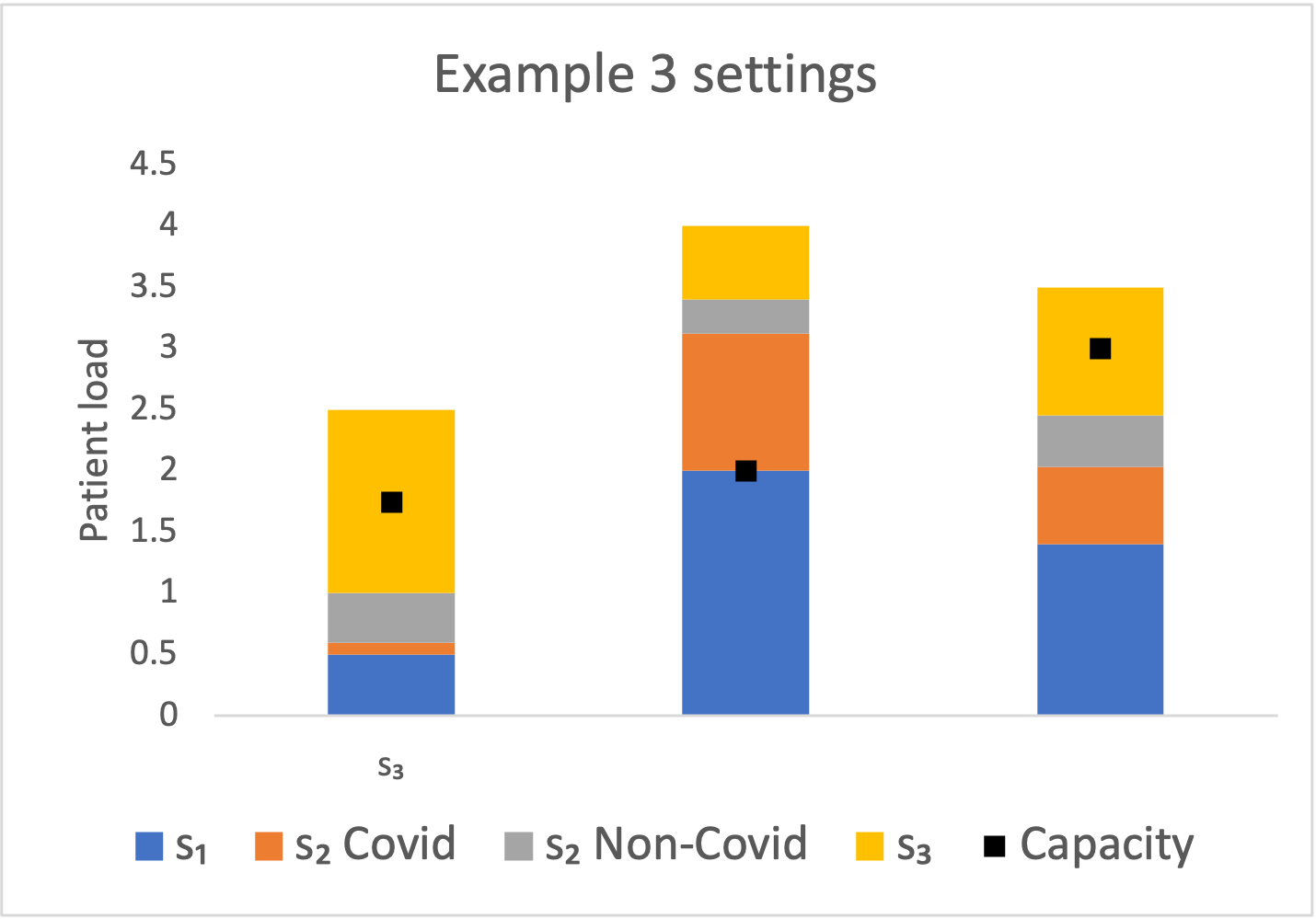}
         \caption{Example 3 (Early waves of pandemic, pre-vaccine)}
         \label{fig: set3}
     \end{subfigure}
     \hspace{1cm}
     \begin{subfigure}[b]{0.3\textwidth}
         \centering
         \includegraphics[width=5.5cm]{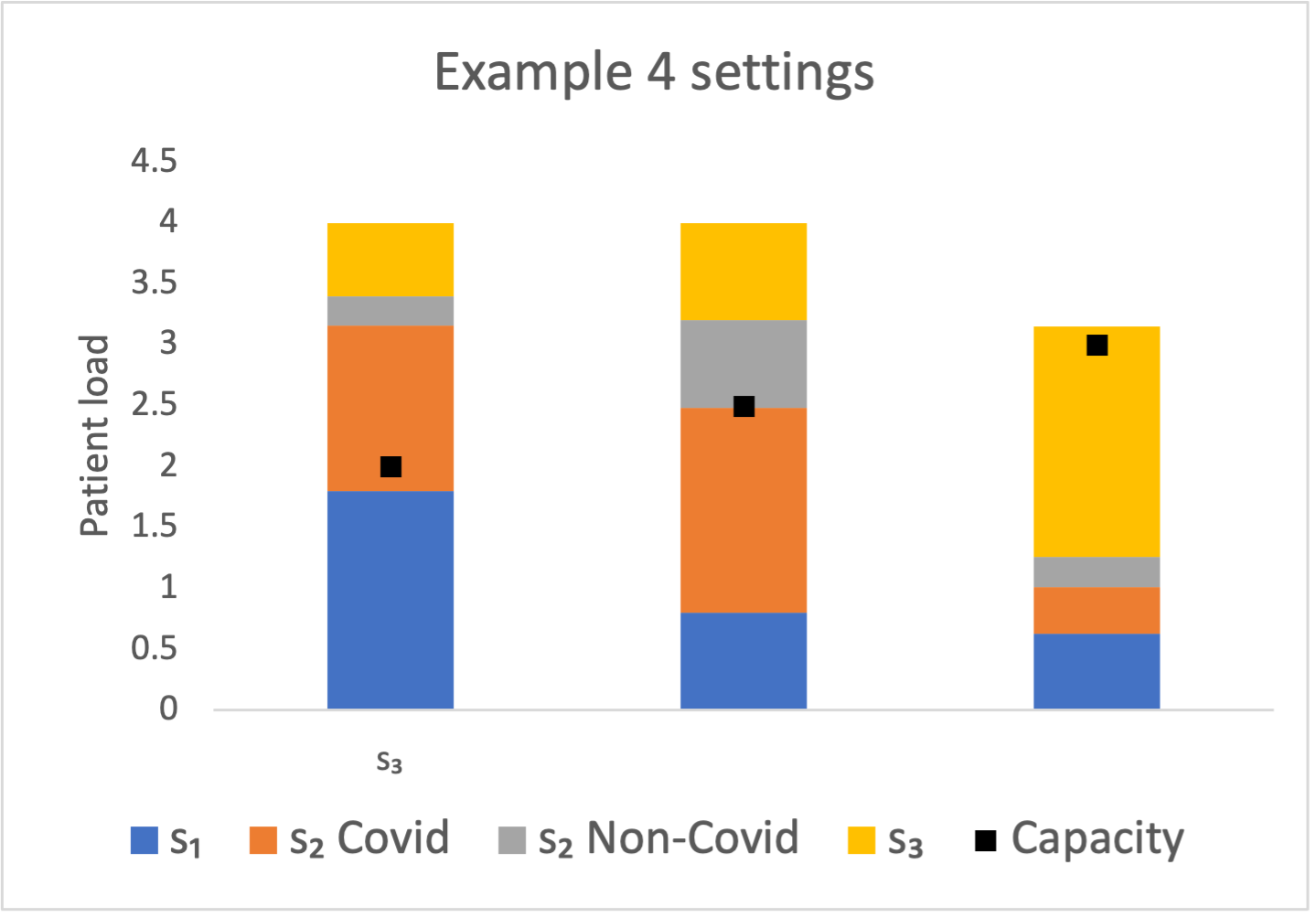}
         \caption{Example 4 (Later waves of pandemic with vaccines)}
         \label{fig: set4}
     \end{subfigure}
\caption{Multi-period pandemic load and capacity profiles} \label{fig: set3set4}
\end{figure}

We now study the multi-period problem under two different  pandemic scenarios, whose settings are described in Table \ref{tab: paramsforthreeperiod}, Appendix \ref{appndx: ForTwoPeriodSection}. The first setting (Example 3; see Figure~\ref{fig: set3set4}), represents the early stages of a pandemic, with patient load starting at a low level, increasing in Period 2 (with higher total patient load, and a much higher proportion of severity $s_1$ patients), and then gently tapering off in Period 3, as the disease spread reduces. On the other hand, the second setting (Example 4; see Figure~\ref{fig: set3set4}) represents a later stage wave, starting with a peak and with good vaccination coverage and effective treatments reducing the number of severity $s_1$ patients starting from Period 2. In this setting, patient load remains at a high level for two periods (with the split of severity $s_1$ patients reducing), with total load tapering off in the third period. In both examples, the total capacity $\Gamma$ increases with time progressing, as the hospital system prepares for higher demand. Parameters like the rates of disease evolution $\delta_{ij}$, risk $r$, the percentage of Covid patients $\mathds{P}(covid)$, and the severity distribution $\lambda_i$ are chosen in line with these pandemic profiles (Appendix \ref{appndx: ForNumericalSection}). 

Recall that our multi-period solution algorithm enumerates all feasible extreme points for Period 1, and for each of these extreme points, enumerates the feasible extreme points in Period 2 (factoring in Period 1 carryovers), and next finds the best extreme point in Period 3, which is now uniquely determined. Tables \ref{tab: 3periodSettings3} and \ref{tab: 3periodsettings4} provide details of all solutions for Examples 3 and 4 respectively in Appendix \ref{appndx: ForNumericalSection}, with relevant snapshots in Tables \ref{tab:3periodSettings3_compact} and \ref{tab:3periodsettings4_compact}. The values in the cells under the ``ExtPt'' columns are labels for the extreme points (note that a combination has at most 3 extreme points, as shown in Table \ref{tab: simplifiedcombs}; we refer to them using subscripts $a,b$ and $c$). 
The objective value in each period $n$ is shown in columns `Pd n Obj' for the $n$th period. `Global Objective' is the weighted sum of the objective values for the three periods, in line with Equation \eqref{eqn: 2periodobjective} (with each period of length 5 units). The optimal policy for the three-period problem is shown in bold font. The greedy policy that myopically finds the best extreme point in each period accounting for carryovers (i.e., the one corresponding to the minimum objective value within the column), is italicized. As seen from Tables \ref{tab: 3periodSettings3} and  \ref{tab: 3periodsettings4}, the optimal policies for Examples 3 and 4 choose extreme points in order $(9, 15_a, 12_c)$ and $(13, 15_a, 10_b)$ respectively.  
 Figure \ref{fig: effectiveLambdas} shows the effective $\lambda$ including carryovers from the previous period and also the optimal capacity allocation in each period, by severity level. 

\begin{table}[]
\footnotesize	
    \centering
    \begin{tabular}{c|c|c|c|l|l|l|l|}
\toprule
Index & Pd 1 ExtPt & Pd 2 ExtPt & Pd 3 ExtPt &  Pd 1 Obj &  Pd 2 Obj &  Pd 3 Obj &  Global Objective \\
\midrule
\midrule
: &        :  &       :  &       :  &     : &   : &    : &            : \\
3  &         9  &       $14_b$  &       $16_a$  &     0.209 &     1.691 &     0.849 &            1.243 \\
\textbf{4}  &         \textbf{9}  &       \textbf{15}$_a$  &       \textbf{12}$_c$  &     \textbf{0.209} &     \textbf{1.479} &     \textbf{0.776} &            \textbf{1.076 (Optimal)} \\
5  &         9  &       $15_b$  &       $12_a$  &     0.209 &     1.644 &     0.767 &            1.119 \\
\textit{6}  &         \textit{9 } &       \textit{16}$_a$  &       \textit{12}$_c$  &     \textit{0.209} &     \textit{1.118} &    \textit{ 0.889} &            \textit{1.093 (Greedy)} \\
: &        :  &       :  &       :  &     : &   : &    : &            : \\
\bottomrule
\end{tabular}
    \caption{Snapshot of solutions: Example 3 (Early waves of pandemic, pre-vaccine)}
    \label{tab:3periodSettings3_compact}
\end{table}

\begin{table}[]
\footnotesize	
    \centering
    \begin{tabular}{c|c|c|c|l|l|l|l|}
\toprule
Index & Pd 1 ExtPt & Pd 2 ExtPt & Pd 3 ExtPt &  Pd 1 Obj &  Pd 2 Obj &  Pd 3 Obj &  Total Objective \\
\midrule
: &        :  &       :  &       :  &     : &   : &    : &            : \\
5  &        13  &       $14_b$  &       $10_b$  &     1.450 &     1.382 &     0.544 &            1.395 \\
\textbf{6}  &        \textbf{13}  &      \textbf{15}$_a$  &       \textbf{10}$_b$  &     \textbf{1.450} &     \textbf{1.236} &    \textbf{ 0.437} &            \textbf{1.224 (Optimal)} \\
7  &        13  &       $15_b$  &       $10_b$  &     1.450 &     1.327 &     0.449 &            1.262 \\
: &        :  &       :  &       :  &     : &   : &    : &            : \\
\textit{20} &       \textit{16}$_a$  &       \textit{12}$_b$  &       \textit{10}$_b$  &     \textit{1.060} &     \textit{1.170} &     \textit{0.765} &           \textit{ 1.428 (Greedy)} \\
: &        :  &       :  &       :  &     : &   : &    : &            : \\
\bottomrule
\end{tabular}
    \caption{Snapshot of solutions: Example 4 (Later waves of pandemic with vaccines)}
    \label{tab:3periodsettings4_compact}
\end{table}

 \begin{figure}
\centering
     \begin{subfigure}[b]{0.3\textwidth}
         \centering
         \includegraphics[width=5.5cm]{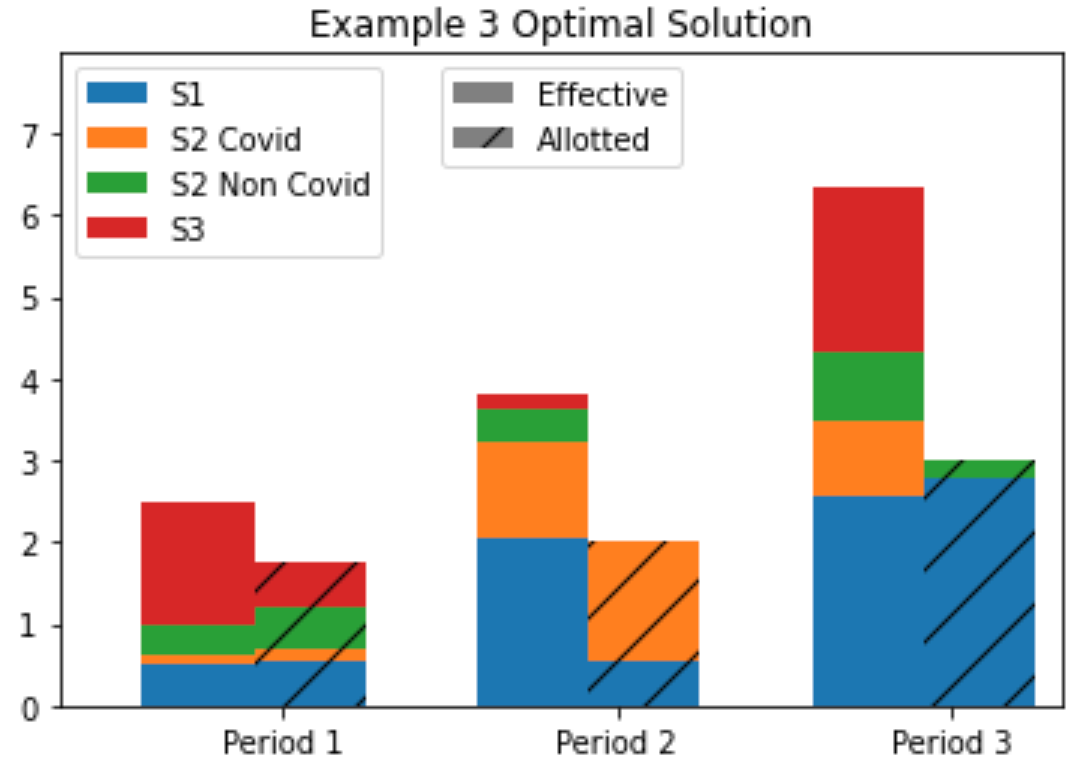}
         \caption{Example 3 (Early waves of pandemic, pre-vaccine)}
         \label{fig: Effset3}
     \end{subfigure}
\hspace{1cm}
     \begin{subfigure}[b]{0.3\textwidth}
         \centering
         \includegraphics[width=5.5cm]{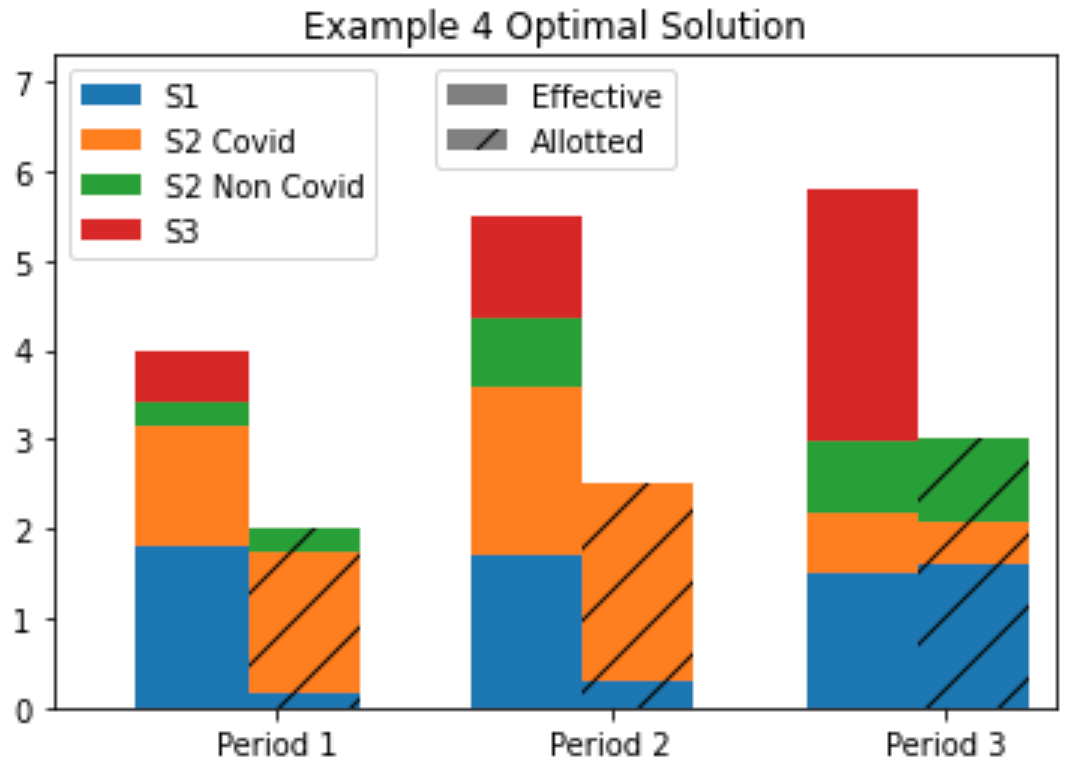}
         \caption{Example 4 (Later waves of pandemic with vaccines)}
         \label{fig: Effset4}
     \end{subfigure}
\caption{Effective load (after adding carryovers) vs optimal capacity allocation} \label{fig: effectiveLambdas}
\end{figure}


\textbf{Solution characteristics.} The optimal policy for Example 3 (see Figure \ref{fig: Effset3}) serves patients of all severities in the first period (partially myopically), but rapidly shifts its priorities in the next periods. It prioritizes the Covid $s_2$ population in the second period, to reduce the fast evolving $s_2$ population, and shifts to prioritizing $s_1$ in the third period, where the $s_2$ load is minimal. The policy also allocates some capacity on non-Covid $s_2$ and $s_3$ in Period 1, later favoring non-Covid to Covid with the decreasing spread and carrying over $s_3$ patients to the end of Period 3.

For Example 4 (see Figure \ref{fig: Effset4}), observe the trajectories of $s_1$, $s_2$ non-Covid patients in the optimal allocation. In contrast to Example 3, there is a significant carryover of $s_1$ patients in the first period, but almost none from the second to the third period. 
Because of the high evolution rates from $s_2$, the optimal policy prioritizes the $s_2$ Covid population in the first two periods and later prioritize the $s_1$ population. The decreasing rate of demands, particularly for $s_1$ (as shown in Figure \ref{fig: set3set4}) in this pandemic profile makes a different prioritization optimal, allowing the pandemic to end with only $s_3$ patients remaining at the end of Period 3. 

\textbf{The value of being forward looking:} Tables \ref{tab:3periodSettings3_compact} and \ref{tab:3periodsettings4_compact} compare the optimal solution profiles with the myopic or greedy solution profiles. For Example 3, the optimal policy, i.e. $(9, 15_a,  12_c)$ partially overlaps with the greedy policy, i.e. $(9, 16_a,  12_c)$. In Period 1, combination 9 is the best choice (as we have capacity to serve all $s_2$, $s_3$ patients), but extreme point $15_a$ is worse than extreme point $14_a$ in Period 2. The greedy policy prioritizes $s_1$ in the second period, allocating the remaining to $s_3$, oblivious to the decreasing load in Period 3 (Figure \ref{fig: set3}), although reaching a fairly close global objective value to the optimal policy. 

For Example 4, the greedy policy, i.e., $(16_a, 12_b, 10_b)$ diverges from the optimal in the very first period -- Combination 13 (the optimal choice for Period 1 that fully serves both the clinics) is seemingly the worst solution for a one-period problem, yet the carryovers from this combination achieve the optimal solution to the multi-period problem. Even over the first two periods, solutions indexed  19, 20, 21 offer better objectives than $(13,15_a)$, the components of the optimal policy. Although combination 13 prioritizes Clinics ($s_2$) over HPED ($s_1$) leading to a higher objective in Period 1, the later periods have lower $s_1$ load (Figure \ref{fig: set4}) and thus using extreme point $10_b$ in Period 3 addresses all $s_1$ carryovers. The non-myopic optimal policy offers a significantly improved approach, as the decreasing pandemic load allows it to finally result in only $s_3$ patients remaining.

For multi-period problems, we thus see that optimal policies are  non-trivial, with changing priorities in serving patients of various severities and disease types from one period to the next. This is because the optimal policy accounts not just for evolution and feedback flows (explaining why it may allocate more capacity than the corresponding load), but also for the fact that current policies endogenously shape future effective load greatly. We also observe that optimal and myopic policies become closer in behavior as the period lengths increase, because  carryovers modify the effective arrival rates in longer periods to smaller extents. 

\section{Conclusions and Extensions} \label{sec:conclusion}

In this work, we studied a hospital system that operates both an Emergency Department (ED) and a medical clinic in the context of the COVID-19 pandemic. Patients contact the provider through a phone call or may present directly at the ED; patients can be COVID (suspected/confirmed) or non-COVID, and have different severities.  Depending on severity, patients who contact the provider may be directed to the ED (to be served in a few hours), be offered an appointment at the Clinic (to be seen in a few days), or be treated via phone as severity is low. Patients  make decisions to join a facility by comparing their risk perceptions versus their expected service times and choose to enter a facility only if it is beneficial. Moreover, patient severities may evolve if they wait over days, prompting them to change the facility they choose. The hospital system aims to allocate service capacity across facilities to minimize costs from patients deaths or defections.

Our approach is three pronged. First, we provide a fluid model framework to solve for optimal capacity allocation in a multi-facility healthcare system, accounting for endogeneous patient behavior and evolution. Second, we analytically characterize the stationary one-period problem by decomposing the solution space into mutually exclusive and collectively exhaustive solution spaces, each of which exhibit physically meaningful structures and can be solved tractably. Thus the global optimal is achieved through a simple enumeration. We further prove a strong ordering of progression of optimal solutions as the available capacity increases. Third, we model an evolving pandemic as a sequence of one-period problems with carryovers, and current decisions affecting the future. Algorithmically, we establish a parsimonious and provably efficient way of computing the stationary optimal solution for the multi-period problem by leveraging the one-period solution structures by enumerating only extreme points, without solving a dynamic programming problem.

We find both analytical and computational insights. Our first insight is that even in single period problems, the endogeneity due to patient choices and evolving severities results in non-greedy capacity allocations, that is, the prioritization of highest severity patients depends on the relative number of medium severity patients and their evolution rates. Second, this result is further reinforced in the multi-period setting. Optimal policies in multi-periods also account for carryovers and future demands and thus make capacity allocations based on effective loads. Third, depending on the trajectory of the pandemic, greedy solutions may be near-optimal or very far from optimal. Cases in which carryovers are low relative to fresh arrival rates in a period may have greedy solutions behave closer to optimal.

Our approach allows for several natural extensions. First, multiple facilities and severity levels can be modeled using our decomposition and extreme point enumeration framework. The combinations can similarly be defined as a cross product of facility efficiency vectors. Second, multiple types of capacities, such as doctors and nurses or specific equipment, may also be captured using our framework. Third, a variety of linear objectives can be incorporated, such as penalizing queue lengths for some or all facilities, deleting appointments due to patient evolution, and penalizing patient transfers between facilities. In these cases as well, it suffices to examine only the extreme points and combinations thereof over multiple periods.

\bibliographystyle{informs2014}
\bibliography{biblio}

\newpage

\begin{APPENDICES}
\section{Details from Section \ref{sec: FluidModel}} 
\pfnotationtable
\label{appndx: ForFluidFormulations}
\pfClinics
\section{Proofs of the 1-period Problem} \label{appndx: ForOnePeriodSection}
\pfapptrue
\proof{Details of the Proof to Theorem \ref{thm: existence}.}

\pfexhaustive
\pfcontiguityoffeasible
\pfexistencelemma
\pfOnetoFivefeasible

\pffullcapacity

\pfthmtwo

\pfthmthree

\pfthmfour

\section{Proofs for the multi period Problem} \label{appndx: ForTwoPeriodSection}
\pfapptrue
\pftwoperiodlemma
\pfextptperiodtwo
\pflinearperiodone
\pfcandidateoptsolnspdone
\pfthefinallemma

\section{Numerical Insights Details} \label{appndx: ForNumericalSection}

\pfsecsixcomparisonsAppndx
\pfQLengths
\pfthreeperiodsettings
\begin{table}[]
\footnotesize	
    \centering
    \begin{tabular}{c|c|c|c|l|l|l|l|}
\toprule
Index & Pd 1 ExtPt & Pd 2 ExtPt & Pd 3 ExtPt &  Pd 1 Obj &  Pd 2 Obj &  Pd 3 Obj &  Global Objective \\
\midrule
\midrule
1  &         9  &        13  &       $12_c$ &     0.209 &     1.616 &     0.733 &            1.085 \\
2  &         9  &       $14_a$  &       $10_b$  &     0.209 &     1.244 &     0.837 &            1.083 \\
3  &         9  &       $14_b$  &       $16_a$  &     0.209 &     1.691 &     0.849 &            1.243 \\
\textbf{4}  &         \textbf{9}  &       \textbf{15}$_a$  &       \textbf{12}$_c$  &     \textbf{0.209} &     \textbf{1.479} &     \textbf{0.776} &            \textbf{1.076 (Optimal)} \\
5  &         9  &       $15_b$  &       $12_a$  &     0.209 &     1.644 &     0.767 &            1.119 \\
\textit{6}  &         \textit{9 } &       \textit{16}$_a$  &       \textit{12}$_c$  &     \textit{0.209} &     \textit{1.118} &    \textit{ 0.889} &            \textit{1.093 (Greedy)} \\
7  &       $10_a$  &       $14_a$  &       $12_b$  &     0.239 &     1.321 &     0.888 &            1.215 \\
8  &       $10_a$  &       $14_b$  &       $16_a$  &     0.239 &     1.772 &     0.911 &            1.340 \\
9  &       $10_a$  &       $15_a$  &       $12_b$  &     0.239 &     1.578 &     0.808 &            1.167 \\
10  &       $10_a$  &       $15_b$  &       $16_a$  &     0.239 &     1.720 &     0.823 &            1.253 \\
11 &       $10_a$  &       $16_a$  &       $12_b$  &     0.239 &     1.198 &     0.927 &            1.215 \\
12 &       $12_a$  &       $14_a$  &       $12_c$  &     0.348 &     1.522 &     0.948 &            1.323 \\
13 &       $12_a$  &       $14_b$  &       $16_a$  &     0.348 &     1.905 &     1.006 &            1.480 \\
14 &       $12_a$  &       $15_a$  &       $12_b$  &     0.348 &     1.713 &     0.888 &            1.293 \\
15 &       $12_a$  &       $15_b$  &       $16_a$  &     0.348 &     1.867 &     0.941 &            1.410 \\
16 &       $12_a$  &       $16_a$  &       $12_b$  &     0.348 &     1.342 &     1.004 &            1.339 \\
\bottomrule
\end{tabular}
    \caption{Example 3 (Early waves of pandemic, pre-vaccine): Complete Analysis}
    \label{tab: 3periodSettings3}
\end{table}

\begin{table}[]
\footnotesize	
    \centering
    \begin{tabular}{c|c|c|c|l|l|l|l|}
\toprule
Index & Pd 1 ExtPt & Pd 2 ExtPt & Pd 3 ExtPt &  Pd 1 Obj &  Pd 2 Obj &  Pd 3 Obj &  Total Objective \\
\midrule
1  &        13  &       $12_a$  &       11b  &     1.450 &     0.919 &     0.515 &            1.331 \\
2  &        13  &       $12_b$  &       11b  &     1.450 &     0.829 &     0.463 &            1.239 \\
3  &        13  &       $12_c$  &       11b  &     1.450 &     0.829 &     0.458 &            1.258 \\
4  &        13  &       $14_a$  &       11b  &     1.450 &     0.884 &     0.451 &            1.269 \\
5  &        13  &       $14_b$  &       $10_b$  &     1.450 &     1.382 &     0.544 &            1.395 \\
\textbf{6}  &        \textbf{13}  &      \textbf{15}$_a$  &       \textbf{10}$_b$  &     \textbf{1.450} &     \textbf{1.236} &    \textbf{ 0.437} &            \textbf{1.224 (Optimal)} \\
7  &        13  &       $15_b$  &       $10_b$  &     1.450 &     1.327 &     0.449 &            1.262 \\
8  &       $14_a$  &       $12_a$  &       $10_b$  &     1.118 &     1.209 &     0.785 &            1.490 \\
9  &       $14_a$  &       $12_b$  &       $10_b$  &     1.118 &     1.162 &     0.755 &            1.447 \\
10  &       $14_a$  &       $12_c$  &       $10_b$  &     1.118 &     1.162 &     0.757 &            1.462 \\
11 &       $14_a$  &       $14_a$  &       $10_b$  &     1.118 &     1.295 &     0.773 &            1.517 \\
12 &       $14_a$  &       $14_b$  &       $10_b$  &     1.118 &     1.792 &     0.836 &            1.658 \\
13 &       $14_a$  &       $16_b$  &       $14_a$  &     1.118 &     1.829 &     0.894 &            1.697 \\
14 &       $15_a$  &       $14_a$  &       $10_b$  &     1.391 &     1.473 &     0.741 &            1.582 \\
15 &       $15_a$  &       $14_b$  &       $10_b$  &     1.391 &     1.912 &     0.795 &            1.704 \\
16 &       $15_a$  &       $15_a$  &       $15_a$  &     1.391 &     1.801 &     0.715 &            1.673 \\
17 &       $15_a$  &       $15_b$  &       $12_b$  &     1.391 &     1.860 &     0.727 &            1.654 \\
18 &       $15_a$  &       $16_a$  &       $10_b$  &     1.391 &     1.204 &     0.756 &            1.490 \\
19 &       $16_a$  &       $12_a$  &       $10_b$  &     1.060 &     1.236 &     0.807 &            1.488 \\
\textit{20} &       \textit{16}$_a$  &       \textit{12}$_b$  &       \textit{10}$_b$  &     \textit{1.060} &     \textit{1.170} &     \textit{0.765} &           \textit{ 1.428 (Greedy)} \\
21 &       $16_a$  &       $12_c$  &       $10_b$  &     1.060 &     1.170 &     0.769 &            1.448 \\
22 &       $16_a$  &       $14_a$  &       $10_b$  &     1.060 &     1.305 &     0.775 &            1.500 \\
23 &       $16_a$  &       $14_b$  &       $10_b$  &     1.060 &     1.759 &     0.832 &            1.627 \\
24 &       $16_a$  &       $16_b$  &       $12_c$  &     1.060 &     1.802 &     0.909 &            1.663 \\
\bottomrule
\end{tabular}
    \caption{Example 4 (Later waves of pandemic with vaccines): Complete Analysis}
    \label{tab: 3periodsettings4}
\end{table}
\end{APPENDICES}

\end{document}